\documentclass[universe,article,accept,moreauthors,pdftex,10pt,a4paper]{mdpi}

\providecommand{\aap}{Astron. Astrophys.}
\providecommand{\apjs}{Astrophys. J. Suppl.}
\providecommand{\apjl}{Astrophys. J.}
\providecommand{\apj}{Astrophys. J.}
\providecommand{\nat}{Nature}

\providecommand{\npa}{Nucl. Phys. A}

\providecommand{\epja}{Eur. Phys. J. A}

\providecommand{\prc}{Phys. Rev. C}
\providecommand{\prd}{Phys. Rev. D}
\providecommand{\pr}{Phys. Rev.}
\providecommand{\prl}{Phys. Rev. Lett.}

\firstpage{1} 
\pubvolume{4}
\issuenum{0}
\articlenumber{00}
\pubyear{2018}
\copyrightyear{2018}
\history{Received: 26 April 2018; Accepted: 16 May 2018; Published: 25 May 2018}
\issuenum{0} 

\usepackage{sidecap}
\usepackage{amssymb}
\usepackage{upgreek}

\Title{Towards a Unified Quark-Hadron-Matter Equation of~State for Applications in Astrophysics and Heavy-Ion~Collisions}

\Author{Niels-Uwe F. Bastian $^{1,}$*\orcidA{}, David Blaschke $^{1,2,3}$\orcidB{}, Tobias Fischer $^{1}$\orcidC{} and Gerd R\"opke $^{3,4}$}
\AuthorNames{Niels-Uwe F. Bastian, David Blaschke, Tobias Fischer and Gerd R\"opke}

\address{%
$^{1}$ \quad Institute of Theoretical Physics, University of Wroclaw, 50-137 Wroclaw, Poland; david.blaschke@gmail.com~(D.B.); tobias.fischer@ift.uni.wroc.pl (T.F.); \\
$^{2}$ \quad Bogoliubov Laboratory for Theoretical Physics, Joint Institute for Nuclear Research, 141980 Dubna, Russia\\
$^{3}$ \quad National Research Nuclear University (MEPhI), 115409 Moscow, Russia\\
$^{4}$ \quad Institute of Physics, University of Rostock, 18051 Rostock, Germany; gerd.roepke@uni-rostock.de%
}
\corres{Correspondence: niels-uwe@bastian.science}

\abstract{
We outline an approach to a unified equation of state for quark-hadron matter on the basis of a $\Phi-$derivable approach to the generalized Beth-Uhlenbeck equation of state for a cluster decomposition of thermodynamic quantities like the density.
To this end we summarize the cluster virial expansion for nuclear matter and demonstrate the equivalence of the Green's function approach and the $\Phi-$derivable formulation.
As an example, the formation and dissociation of deuterons in nuclear matter is discussed. 
We formulate the cluster $\Phi-$derivable approach to quark-hadron matter which allows to take into account the specifics of chiral symmetry restoration and deconfinement in triggering the Mott-dissociation of hadrons. This approach unifies the description of a strongly coupled quark-gluon plasma with that of a medium-modified hadron resonance gas description which are contained as limiting cases. 
The developed formalism shall replace the common two-phase approach to the description of the deconfinement and chiral phase transition that requires a phase transition construction between separately developed equations of state for hadronic and quark matter phases. 
Applications to the phenomenology of heavy-ion collisions and astrophysics are outlined.  
}


\begin{document}

\section{Introduction}

The development of a unified equation of state (EoS) for quark-hadron matter, where the hadrons are not elementary degrees of freedom but rather appear as composites of their quark constituents (bound and scattering states of effective quark interactions) is a formidable task because it requires the implementation of dynamical mechanisms of quark confinement as well as chiral symmetry breaking which characterise the hadronic phase of matter and mechanisms for deconfinement and chiral symmetry restoration which determine the transition to the quark(-gluon) phase of strongly interacting matter.

The aspect of bound state formation and dissociation has been well studied in warm dense nuclear matter, where a cluster virial expansion has been developed as a most effective means of description of a system that contains clusters of different sizes with internal quantum numbers like in the nuclear statistical equilibrium picture, but generalizes it by including residual binary interactions among them (second cluster virial coefficient) and their dissociation due to compression and heating within the Mott effect. 
The consistent description of bound and scattering states on the same footing is provided by the generalized Beth-Uhlenbeck equation of state that uses phase shifts in order to describe correlations and their modifications in a hot, dense environment. 
It turns out that for the short-ranged strong interactions it is not the screening of the interaction which drives the dissociation of the bound states but rather the Pauli blocking effect that inhibits cluster formation when the phase space is densely populated. 
Being based on pure symmetry arguments, this effect is sufficiently general to apply to any fermionic many-particle system where---as a function of the density---the formation of bound state gets first favored (principle of Le Chatelier and Brown) until the Pauli principle inhibits scattering processes that would lead to cluster formation and a homogeneous phase of fermionic quasiparticles emerges: The dense nuclear matter.

Since nucleons themselves can be considered as clusters of quarks, the more fundamental degrees of freedom of quantum chromodynamics (QCD) as the underlying gauge field theory of strong interactions, we want to develop in this work the basics of a corresponding approach that will describe the transition from hadronic to quark matter as a Mott transition driven by the Pauli blocking effect.
Despite this striking analogy there are also specifics for the quark degrees of freedom which need to be taken into account.
These are mainly the internal quantum numbers (color, flavor, spin) that cause the confinement of colored states (quarks, gluons, diquarks etc.), the dynamical chiral symmetry breaking and combined symmetry requirements as well as relativistic kinematics.  

\begin{figure}[H] 
	\centerline{\includegraphics[width=0.65\textwidth]{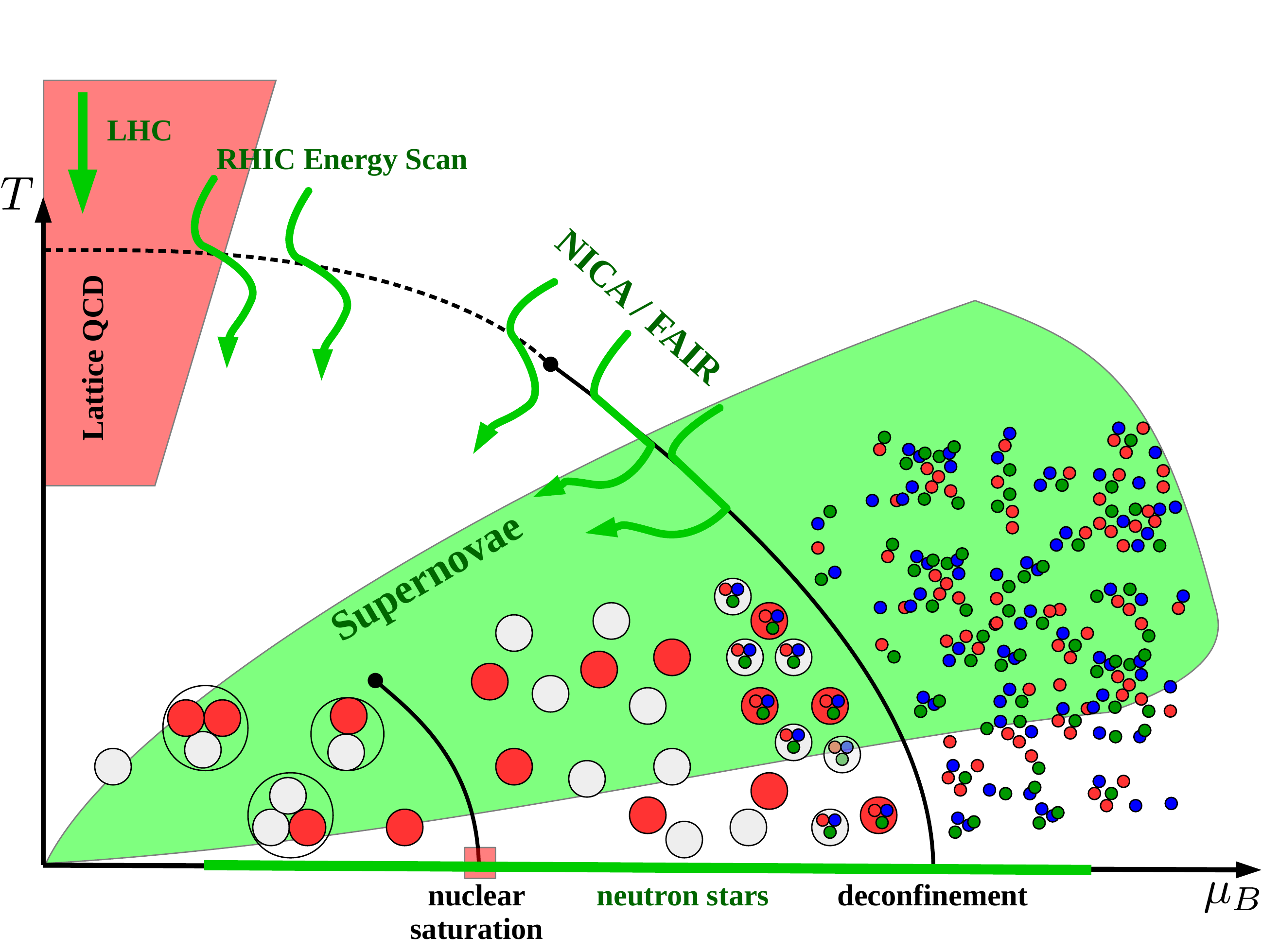}} 
	\caption{%
		Schematic phase diagram of strongly interacting matter with the solid lines indicating first order liquid-gas and hadronic-quark matter phase transitions ending in dots symbolizing the critical endpoints. 
		While for the nuclear liquid-gas phase transition the critical point is at a temperature of $T_c^{LG} = 16.6\pm 0.9$ MeV \cite{Natowitz:2002nw}, the position of the critical endpoint of the deconfinement transition is not known.
		The dashed line indicates the pseudocritical temperature $T_c(\mu_B)$ of the crossover transition from hadronic to quark matter which has been determined at vanishing baryochemical potential $\mu_B=0$ in lattice QCD simulations to $T_c(0)=154 \pm 9$ MeV \cite{Bazavov:2011nk}.
		The red and white colored circles stand for neutrons and protons which at low densities can form nuclear clusters that get dissociated in the liquid phase.
		Similar to that case, nucleons themselves can be viewed as clusters of quarks symbolized by colored dots that get dissociated in the quark matter phase.
		The green shaded area is the domain of the phase diagram accessible by supernova explosions.
		The red shaded area is the region accessible by ab-initio simulations of QCD at finite temperatures and chemical potentials on space-time lattices.
		The~green arrowed lines symbolize trajectories of heavy-ion collisions at different center of mass energies ranging form LHC over RHIC to the planned NICA/FAIR experiments.
		The~green line on the chemical potential axis depicts the range of values accessible in compact star interiors.
		In this case, the third dimension of the charge (or isospin) chemical potential relevant for isospin asymmetric systems in supernovae, compact stars and their mergers is suppressed.} 
	\label{schematic_pd}  
\end{figure}

In Figure \ref{schematic_pd} we show the phase diagram of QCD according to the present state of knowledge. In~particular it addresses the aspect of cluster formation and dissociation in dense quark-hadron matter, which is the main goal of the approach to be outlined in this work.

To this end in Section \ref{sec:qs-nucl} we first review the cluster virial approach for nuclear matter in the form of a generalized Beth-Uhlenbeck EoS. 
Then in Section \ref{sec:phi-nucl} we suggest that it might be obtained from the $\Phi-$derivable approach in its field theoretic formulation when for the $\Phi$ functional the class of all two-loop Feynman diagramns is chosen that can be constructed from cluster Green's functions and cluster T-matrices and consequently take the form of generalized ``sunset'' type diagrams. 

In Section \ref{sec:phi-quarks} we develop the approach for quark matter with mesons and baryons as clusters dominating the hadronic phase of matter since quarks and diquarks are suppressed by confining interactions that give rise to diverging selfenergies for those states in the low-density (confinement) phase. 
This mechanism is realized here within the so-called string-flip model that is generalized in relativistic form.
As a first step, a selfconsistent mean-field approximation is performed which results in a new quark matter equation of state with confining features, thus superior to previous NJL model approaches and their Polyakov-loop generalized versions.
We demonstrate how the quark Pauli-blocking effect for hadrons is already inherent in the $\Phi-$derivable approach and can be made apparent by a perturbation expansion with respect to cluster selfenergies on top of the quasiparticle approximation. 
The Pauli-blocking effect can be mimicked by a hadronic excluded volume which should therefore be taken into account when a hadronic EoS is extrapolated from a low-density limit (hadron resonance gas, nuclear statistical equilibrium) to the vicinity of the deconfinement transition. 
In the quark matter phase, we implement higher order repulsive interactions which may be justified by multi-pomeron exchange interactions, thus non-perturbative effects of the gluon sector which we are not treating dynamically at this stage.
These effects cause a stiffening of the high-density quark matter EoS that are essential for hybrid compact star applications.

In Section \ref{sec:appl-astro} we discuss the role of nuclear clusters for the astrophysics of supernovae and the deconfinement transition in compact stars while in Section \ref{sec:appl-hic} we consider the potential of a unified approach to quark-hadron matter for applications in heavy-ion collisions. We point out that nuclear cluster formation may occur directly at the hadronisation transition and thus result in their (sudden) chemical freeze-out together with other hadronic species as indicated by the ALICE experiment at the CERN LHC. 

In Section \ref{sec:conclusions} we draw conclusions for the further development of the approach towards a unified description of quark-hadron matter and its phenomenology in heavy-ion collisions and astrophysics.      
\section{Quantum Statistical Approach and  the Cluster Virial Expansion}\label{sec:qs-nucl}
A systematic approach to the cluster expansion of thermodynamic
properties is obtained from quantum statistics. We consider a system consisting of species $i$ with the (conserved) particle number $N_i$ and the corresponding chemical potential $\mu_i$, described by the hamiltonian $H$ at equilibrium with temperature $T$. The index $i$ will be dropped in the following.
The grand canonical thermodynamic potential is given by
\begin{equation}
	 \Omega=-P V= - T \ln {\rm Tr}\,\, {\rm e}^{-(H-\mu N)/T},
\end{equation}
where $P$ is the pressure and $V$ the volume.
It can be represented by diagrams within a perturbation expansion \cite{Baym:1961zz,Baym:1962sx}, see also \cite{KKER1986}. 
We have
\begin{equation}
	P= \frac{1}{V} {\rm Tr} \ln[-G_1^{(0)}]-\frac{1}{2 V} \int_0^1 
	\frac{d \lambda}{\lambda}{\rm Tr} \Sigma_\lambda G_\lambda,
\end{equation}
or
\begin{equation}
\label{Eq:3}
	\parbox[h][0.80cm][c]{0.7\linewidth}{
	\includegraphics[scale=0.45]{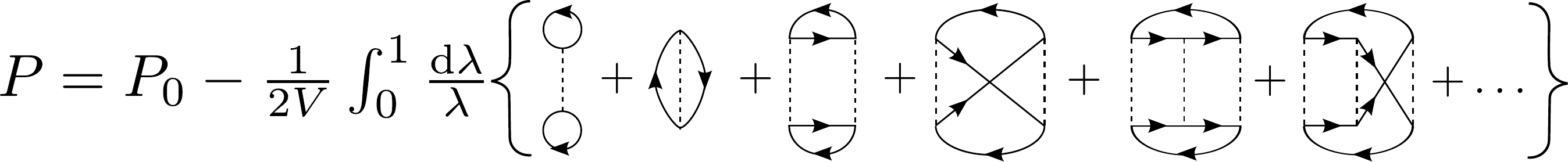}
    }
\end{equation}
\noindent
where $\lambda$ is a scaling factor substituting the two-particle interaction $V_2$ by $\lambda~V_2$. $G_1^{(0)}$ is the free single-particle propagator that gives the ideal part of the pressure $P_0$.
The full single-particle Green function $G_\lambda$ and the self-energy $\Sigma_\lambda$ are taken with the coupling constant $\lambda$.
Depending on the selected diagrams, different approximations can be found. 
In particular, the second virial coefficient for charged particle systems has been investigated, see Ref.~\cite{KKER1986}.\footnote{Note that the interaction in nuclear systems is strong. 
	However, the perturbation expansion is performed with respect to the imaginary 
	part of the self-energy that is assumed to be small. 
	Most of the interaction is already taken into account in the self-consistent 
	determination of the quasiparticle energies. 
	With increasing density, the Fermi energy will dominate the potential energy 
	so that the correlations are suppressed.
	A quasiparticle description can be used to calculate the nuclear structure.
}

An alternative way to derive the equation of state is to start from the expression for the total nucleon density
\begin{equation}
	n_{\tau_1}(T,\mu_p,\mu_n) = \frac{2 }{V} \sum_{p_1} \int_{-\infty}^\infty\frac{d \omega}{2 \pi} f_{1}(\omega) S_1(1,\omega)\,, 
	\label{eosspec}
\end{equation}
where $V$ is the system volume, the variable $1=\{{\bf p_1},\sigma_1,\tau_1\}$ denotes the single particle states (here nucleons in momentum representation), $\tau_1 =n,\,p$, and summation over spin direction is collected in the factor 2. 
Both the Fermi distribution function and the spectral function depend on the temperature and the chemical potentials $\mu_p, \mu_n$, which are not given explicitly.
The spectral function $S_1(1,\omega)$ of the single-particle Green function $G_1(1,i z_\nu)$ is related to the single-particle self-energy $\Sigma(1,z)$ according to
\begin{equation}
	\label{spectral}
	S_1(1,\omega) = \frac{2 {\rm Im}\,\Sigma_1(1,\omega-i0)}{(\omega - E^{(0)}(1)- {\rm Re}\, \Sigma_1(1,\omega))^2 + 
	({\rm Im}\,\Sigma_1(1,\omega-i0))^2 }\,,
\end{equation}
where the imaginary part has to be taken for a small negative imaginary part in the frequency; $E^{(0)}(1)=p_1^2/(2m_1)$.

Both approaches are equivalent. The perturbation expansion can be represented
by Feynman diagrams, see Equation (\ref{Eq:3}). However, a finite order perturbation
theory will not produce bound states, and partial summations of an infinite number of
e.g., ladder diagrams must be performed to get bound states.
As shown by Baym and Kadanoff \cite{Baym:1961zz,Baym:1962sx}, 
self-consistent approximations to the one-particle Green function can be given 
based on a functional $\Phi$ so that
\begin{equation}
\Sigma_1(1,1') = \frac{\delta  \Phi}{\delta G_1(1,1')}.
\end{equation}

Different approximations for the generating functional $\Phi$ are discussed in the following Sections~\ref{sec:phi-nucl} and \ref{sec:phi-quarks}.
The self-consistent $\Phi-$derivable approximations not only lead to a fully-conserving transport theory.
In the equilibrium case they also have the property that different methods to obtain the grand partition function such as integrating the expectation value of the potential energy with respect to the coupling constant $\lambda$, Equation (\ref{Eq:3}), or integrating the density $n$ with respect to the chemical potential $\mu$, lead to the same result \cite{Weinhold:1997ig,Weinhold:1998tha}. 
In particular, with
\begin{equation}
	\Omega = -{\rm Tr}\,\, \ln (-G_1) -{\rm Tr} \Sigma_1 G_1 +\Phi
\end{equation}
also 
\begin{equation}
	n = - \frac{\partial \Omega}{\partial \mu}
\end{equation}
holds in the considered approximation.

\begin{figure}[H] 
	\centerline{\includegraphics[scale=0.4]{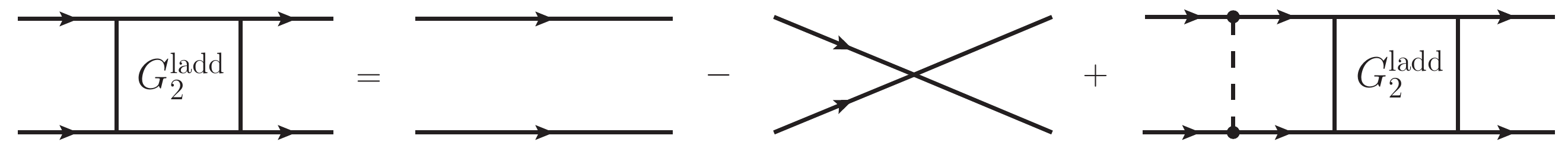}} 
	\caption{Ladder approximation $G_{A}^{\rm ladder}$ for $A=2$. Iteration gives the infinite sum of ladder diagrams.}
	
	\label{Fig:G2a} 
\end{figure}

The latter approach (using Equation \eqref{eosspec}) has been extensively used in many-particle systems~\cite{Zimmermann:1985ji,Ropke:1982vzx,Ropke:1982ino,Ropke:1983lbc,Schmidt:1990zz}, in particular in connection with the chemical picture. 
The main idea of the chemical picture is to treat bound states on the same footing as ``elementary'' single particles.
This way one describes correctly the low-temperature, low-density region of many-body systems where bound states dominate.
Within a quantum statistical approach, the chemical picture is realized considering  the $A$-particle propagator.
In ladder approximation, see Figure \ref{Fig:G2a}, the Bethe-Salpeter equation (BSE)
\begin{equation}
\begin{array}{ll}
	G_{A}^{\rm ladder}(1\dots A;1'\dots A';z_A)=G_{A}^{0}(1\dots A;z_A)\delta_{\rm ex}(1\dots A;1'\dots A')
\\	+ \sum_{1"\dots A"} G_{A}^{0}(1\dots A;;z_A)V_A(1\dots A;1"\dots A")G_{A}^{\rm ladder}(1"\dots A";1'\dots A';z_A)\,,
	\label{BSE}
\end{array}	
\end{equation}
is obtained, where $z_A$ is the $A$-particle Matsubara frequency, $G_{A}^{0}$ is the product of
single-particle propagators, $V_A(1\dots A;1"\dots A")=\frac{1}{2} \sum_{i,j}V_2(ij,i"j")\prod_{k\neq i,j}\delta(k,k")$
is the $A$-particle interaction. $\delta_{\rm ex}(1\dots A;1'\dots A')$ describes the antisymmetrization of the $A$-particle 
state.

The BSE is equivalent to the $A$-particle wave equation. Neglecting all medium effects 
we have
\begin{equation}
\begin{array}{ll}
	[E^{(0)}(1)+ \dots+E^{(0)}(A)] \psi_{A\nu P}(1\dots A)+\sum_{1'\dots A'} V_A(1\dots A;1'\dots A') \Psi_{A\nu P}(1'\dots A')
\\	= E_{A,\nu}^{(0)}( P) \Psi_{A\nu P}(1\dots A)\,,
\end{array}	
\end{equation}
where $\nu$ indicates the internal quantum number of the $A$-particle eigen states, including the channel $c$
describing spin and isospin state. Different excitations are possible in each channel $c$, in particular bound states and scattering states.
The chemical picture uses the eigen-representation of the  $A$-particle propagator. In ladder approximation,
neglecting medium effects,
\begin{eqnarray}
\label{GAladder}
	G_{A}^{\rm ladder,0}(1\dots A;1'\dots A';z_A)&=&\sum_{\nu}
	\langle 1\dots A |\psi_{A \nu P} \rangle \frac{1}{z_A-E^{(0)}_{A,\nu}(P)} 
	\langle \psi_{A \nu P} |1'\dots A' \rangle\,\,.
\end{eqnarray}

In particular, it contains the contribution of $A$-particle bound states (nuclei) similar
to the ``elementary'' single particle propagator
\begin{equation}
	\label{G1}
	G_1^{(0)}(1,z) = \frac{1}{z-E^{(0)}_1(p_1)}\,.
\end{equation}

Within the representation of the perturbative expansion by Feynman diagrams, the chemical
picture implies that diagrams containing single-particle propagators should be completed by 
adding $A$-particle propagators, at least for the bound states $\nu$. However, also the scattering 
parts $\nu$ have to be considered for a full description.

Coming back to Equation (\ref{spectral}), different approximations for the self-energy $\Sigma_1(1,z)$
can be considered. The main issues are the implementation of bound state formation which is 
of relevance in the low-density region, and the account for density effect if considering higher densities,
such as mean-field approximations. To give some systematics with respect to the different approximations
calculating the self-energy and the corresponding approximation for the EoS, in Ref. \cite{Ropke:2017own} 
the following overview was presented.

As seen in Table \ref{Tab:2}, to consider density effects, the ideal gas approximation 
is improved by the mean-field approximation. Quasiparticle shifts are introduced
in a semi-empirical way within the relativistic mean-field (RMF) approximation\footnote{A Pad{\'e} approximation of the nucleon quasiparticle shifts, applicable in a wide temperature range, can be found in Ref.  \cite{Roepke:2017ohb}.} or
considering Skyrme parametrizations. This~approximation is assumed to give an adequate 
description of dense matter near and above the saturation density but fails to 
describe correlations, in particular bound state formation, in the low density region.

\begin{table}[H]
	\caption{Aspects of the quantum statistical  description of a many-particle system with bound and scattering states in the low-density limit (left column) and its modification due to medium effects at high densities (right column).}
	\centering
	{\begin{tabular}{ll}
			{\it low density limit }& {\it high density modification (medium effects)}
		\\	\midrule
			\multicolumn{2}{c}{\bf(1) elementary particles}
		\\	\midrule
			{\it Ideal Fermi gas: }
		&	{\it Quasiparticle quantum liquid:}
		\\	neutrons, protons
		&	mean-field approximation
		\\	(electrons, neutrinos,\dots)
		&	Skyrme, Gogny, RMF
		\\	\midrule
			\multicolumn{2}{c}{\bf(2) bound state formation}
		\\	\midrule
			{\it Nuclear statistical equilibrium:}
		&	{\it Chemical equilibrium of quasiparticle clusters:}
		\\	ideal mixture of all bound states
		&	medium modified bound state energies
		\\	chemical equilibrium, mass action law
		&	self-energy and Pauli blocking
		\\	\midrule
			\multicolumn{2}{c}{\bf(3) continuum contributions}
		\\	\midrule
			{\it Second virial coefficient:}
		&	{\it Generalized Beth-Uhlenbeck formula:}
		\\	account of continuum correlations ($A=2$)
		&	medium modified binding energies,
		\\	scattering phase shifts, Beth-Uhlenbeck Eq.
		&	medium modified scattering phase shifts
		\\	\midrule
			\multicolumn{2}{c}{\bf(4) chemical \& physical picture}
		\\	\midrule
			{\it Cluster virial approach:}
		&	{\it Correlated medium:}
		\\	all bound states (clusters)
		&	phase space occupation by all bound states
		\\	scattering phase shifts of all pairs
		&	in-medium correlations, quantum condensates
		\\	\bottomrule
	\end{tabular}}
	\label{Tab:2}
\end{table}

For this, the chemical picture is necessary, which is realized by a cluster decomposition 
of the self-energy, see \cite{Ropke:1982ino}. In particular, considering only the bound state part
of the free $A$-particle  propagator, the nuclear statistical equilibrium (NSE) is obtained.
However, also the scattering part of the free $A$-particle  propagator must be 
considered to obtain the correct second virial coefficient in the case $A=2$,
as given by the Beth-Uhlenbeck formula \cite{Beth:1937zz}. A cluster Beth-Uhlenbeck formula has been
worked out \cite{Ropke:2012qv} which considers the chemical picture, where scattering states between
two arbitrary clusters $A$ and $B$ are taken into account.

The inclusion of density effects for these approximations, as collected in   Table \ref{Tab:2},
is not simple. A generalized Beth-Uhlenbeck formula has been worked out in Ref. \cite{Schmidt:1990zz}.
In this approach, the Pauli blocking and the dissolution of bound states with increasing density is 
shown. 

A more detailed approach should also take into account that the medium is not considered 
as uncorrelated, but correlations in the medium have to be taken into account. 
The cluster-mean-field (CMF) approximation has been worked out \cite{Ropke:1983lbc}
where also clustering of the medium is treated. This is of relevance in the region where
correlations and bound state formation dominate.

These effects have been considered recently \cite{Ropke:2014fia} in warm dense matter.
Improving the NSE, medium modifications of the bound states are discussed. 
For a consistent description, in particular of the correct second virial coefficient, scattering phase shifts have been implemented.
If clustering is relevant, the medium cannot be approximated by an uncorrelated, ideal Fermion gas of quasiparticles, but correlations must be taken into account.

Here we focus on the cluster virial expansion \cite{Ropke:2012qv}.~We consider density effects such as 
single-particle quasiparticle shifts and Pauli blocking, which are responsible for the dissolution of 
bound states and the appearance of a new state of matter. In the present section we treat nuclear clusters 
which are dissolved to form a nuclear Fermi liquid when density increases. Later on we go a step forward to investigate the 
dissolution of hadrons by medium effects to form the quark-gluon~plasma.

As shown in \cite{Ropke:2014fia}, using the cluster decomposition of the self-energy which takes into account, in~ particular, cluster formation,
we obtain
\begin{eqnarray}
	\label{eos}
	&&  n^{\rm tot}_n(T,\mu_n,\mu_p)= \frac{1 }{V} \sum_{A,\nu,P}N 
	f_{A,Z}[E_{A,\nu}(P;T,\mu_n,\mu_p)] , \nonumber\\ 
	&&  n^{\rm tot}_p(T,\mu_n,\mu_p)= \frac{1 }{V} \sum_{A,\nu,P}Z 
	f_{A,Z}[E_{A,\nu}(P;T,\mu_n,\mu_p)] \, ,
	\label{quasigas}
\end{eqnarray}
where $\bf P$ denotes the center of mass (c.o.m.) momentum of the cluster (or, for $A=1$, the momentum of the nucleon). 
The internal quantum state $\nu$ contains the proton number $Z$ and neutron number $N=A-Z$ of the cluster, 
\begin{equation}
	f_{A,Z}(\omega;T,\mu_n,\mu_p)=\frac{1}{ \exp [(\omega - N \mu_n - Z \mu_p)/T]- (-1)^A}
	\label{vert}
\end{equation}
is the Bose or Fermi distribution function for even or odd $A$,
respectively, that is depending on $\{T,\mu_n,\mu_p\}$. 
The integral over $\omega$ is performed within the quasiparticle approach, the quasiparticle energies $E_{A,\nu}(P;T,\mu_n,\mu_p)$
are depending on the medium characterized by $\{T,\mu_n,\mu_p\}$. Expressions for the in-medium modifications are given in \cite{Ropke:2014fia}.

In Equation (\ref{eos}) the sum is to be taken over the mass number $A$ of the cluster, the center-of-mass
momentum $\bf P$, and the intrinsic quantum number $\nu$. The summation over $\nu$ concerns the bound states as far as they exist, as well as the continuum of scattering states. 
Solving the few-body problem what is behind the calculation of the $A$-nucleon T matrices in the Green function approach, 
we can introduce different channels ($c$) characterized, e.g., by spin and isospin quantum numbers. 
We assume that these channels decouple. In contrast to the angular momentum which is not conserved,
e.g., for tensor forces, the contribution of different channels to Equation (\ref{eos}) is additive.
The remaining intrinsic quantum numbers will be denoted by $\nu_c$, 
it concerns the bound states as far as they exist (ground states and excited states), as well as the continuum of scattering states.  

We analyze the contributions of the clusters ($A\geq 2$), suppressing the thermodynamic variables $\{T,\mu_n,\mu_p\}$.
We have to perform the integral over the c.o.m. momentum $\bf P$ what, in general, must be done numerically since the dependence of the in-medium quasiparticle energies $E_{A,\nu}(P;T,\mu_n,\mu_p)$ on  $\bf P$ is complicated.
We have in the non-degenerate case $\left[ \,\sum_P \to V/(2 \pi)^3 \int d^3P \,\right]$
\begin{eqnarray}
	\label{components}
	&&\frac{1 }{V} \sum_{\nu,P}f_{A,Z}[E_{A,\nu}(P)]=\sum_{c} e^{\left(N\mu_n+Z\mu_p\right)/T} 
	\!\!\int\!\! \frac{d^3 P}{(2 \pi)^3}\sum_{\nu_c}g_{A,\nu_c} e^{-E_{A,\nu_c}(P)/T} =\sum_{c}
	\!\int\!\! \frac{d^3 P}{(2 \pi)^3} z_{A,c}(P)
\end{eqnarray}
with  $g_{A,c} =2 s_{A,c} +1$  the degeneration factor in the channel $c$.
The partial density of the channel $c$ at $\bf P$ contains the intrinsic partition function
\begin{equation}
\label{zpart}
	z_{A,c}(P;T,\mu_n,\mu_p)=e^{\left(N\mu_n+Z\mu_p\right)/T}
	\!\left\{\sum_{\nu_c}^\mathrm{bound} \!\! g_{A,\nu_c} e^{-E_{A,\nu_c}(P)/T} \Theta\!\left[-E_{A,\nu_c}(P)+E_{A,c}^{\rm cont}(P)\right]\!\right\} 
	+ z^{\rm cont}_{A,c}(P).
\end{equation}

It can be decomposed in the bound state contribution and the contribution of scattering states $z^{\rm cont}_{A,c}(P;T,\mu_n,\mu_p) $.
We emphasize that the subdivision of the intrinsic partition function into a bound and a scattering contribution is artificial and not of physical relevance.

The summations over $A, c$ and $\bf P$  of (\ref{zpart}) remain to be done for the EOS (\ref{eos}), and $Z$ may be included in $c$.
The region in the parameter space, in particular $\bf P$, where bound states exist, may be restricted what is expressed by 
the step function $\Theta(x)=1, x \geq 0;\,\,=0$ else. The continuum edge of scattering states is denoted by 
$E_{A,c}^{\rm cont}(P;T,\mu_n,\mu_p)$.
The intrinsic partition function $z^{\rm cont}_{A,c}( P)$  contains the scattering state contribution (non degenerate case)
\begin{equation}
	\label{zcont}
	z^{\rm cont}_{A,c}(P) =\int_0^\infty \frac{dE}{2\pi}e^{\left(-E-P^2/(2 M_A)+N\mu_n+Z\mu_p\right)/T}  
	2 \sin^2\delta_c(E) \frac{d\delta_c (E)}{dE}~.
\end{equation}

Going beyond the ordinary Beth-Uhlenbeck formula \cite{Beth:1937zz},  for nuclear matter the generalized Beth-Uhlenbeck formula has been worked out in Ref.~\cite{Schmidt:1990zz}. 
Here, the single-particle contribution is described by quasiparticles, and to avoid double counting, 
the corresponding mean-field term must be subtracted from the contribution of scattering states what leads to the
appearance of the term $  \sin^2\delta_c(E) $.

The NSE follows as a simple approximation where the sum is performed only over the bound states, 
and medium effects are neglected.~We obtain the model of a mixture of non-interacting bound clusters,
which can react so that chemical equilibrium is established. This approximation may be applicable for the low-density, low-temperature region where the components are nearly freely moving, and intrinsic excitations are not of relevance. However, the continuum of scattering states (the intrinsic quantum number $\nu_c$ may contain the relative momentum) are of relevance at increasing temperature, 
which is clearly seen in the exact Beth-Uhlenbeck expression \cite{Beth:1937zz} 
for the second virial coefficient. Another argument to take the continuum of scattering states 
into account is the dissolution of bound states at increasing density. Here, the thermodynamic properties behave smoothly because the contribution of the bound state to the intrinsic partition function is replaced by the corresponding contribution of the scattering states.

It turns out advantageous for analyses of the thermodynamics of the Mott transition, to avoid the separation into a bound and a scattering state part of the spectrum and to include the discrete part of the spectrum into the definition of the phase shifts, so that these are merely parameters of a polar representation of the complex $A-$particle cluster propagator. The partition function then takes the~form 
\begin{equation}
\label{z}
	z_{A,c}(P;T,\mu_n,\mu_p) =\int_{-\infty}^\infty \frac{dE}{2\pi}~e^{\left(-E-P^2/(2 M_A)+N\mu_n+Z\mu_p\right)/T}  
2 \sin^2\delta_c(E) \frac{d\delta_c (E)}{dE}~.
\end{equation}

\begin{figure}[H] 
	\centerline{\includegraphics[scale=0.4]{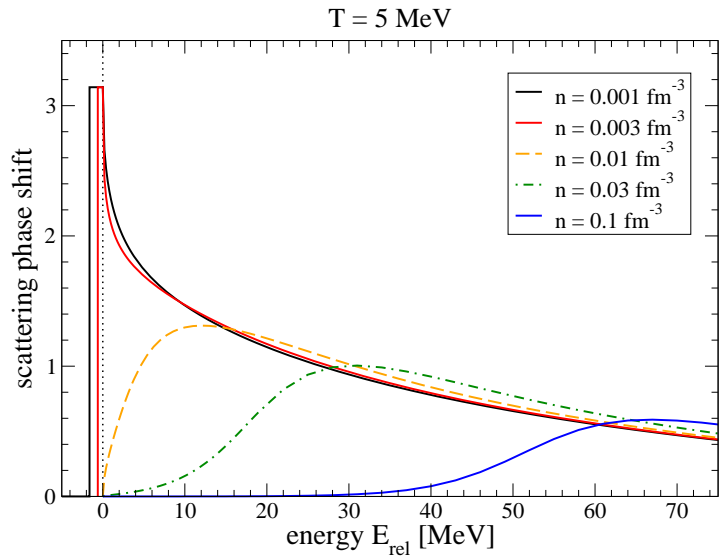}} 
	\caption{Integrand of the intrinsic partition function as function of the intrinsic energy in the 
		deuteron channel. Different densities of the medium are considered, the temperature is $T=5$ MeV. 
		From \cite{Ropke:2014mwa}.
	} 
	\label{Phaseshifts} 
\end{figure}

As an example let us consider the case $A=2$. A consistent description of the medium effects should contain not only the mean-field shift of the quasiparticle energies but also the Pauli blocking, as shown in this work for conserving approximations.  
Within the generalized Beth-Uhlenbeck approach~\cite{Schmidt:1990zz}, Pauli blocking modifies the binding energy of the bound state (deuteron) 
in the isospin-singlet channel as well as the scattering phase shifts. 
The integrand of the intrinsic partition function is shown in Figure \ref{Phaseshifts}.

The intrinsic partition function $z^{\rm cont}_{A,c}( P)$ cannot be divided unambiguously into a bound state contribution
and a contribution of scattering states.~As seen from Figure \ref{Phaseshifts}, the sum of both contributions 
behaves smoothly when with increasing density the bound state is dissolved. 
Therefore, we emphasize 
that one should consider only the total  intrinsic partition function including both, bound and scattering contribution,
as expressed by the generalized phase shifts as shown in Figure \ref{Phaseshifts}.

Whereas the two-body problem can be solved, e.g., using separable interaction potentials, and the account of 
medium effects has been investigated \cite{Schmidt:1990zz}, 
the evaluation of the contribution of clusters with mass numbers $A>2$ to the EoS is challenging. 
The medium modification of the cluster binding energies has been
calculated, e.g., using a variational approach \cite{Ropke:2014fia}. 
Problematic is the inclusion of scattering states, in particular the treatment of different  channels describing 
the decomposition of the $A$-particle cluster.

In this work, we discuss the concept of a cluster-virial expansion and corresponding 
generalized cluster-Beth Uhlenbeck approaches. 
This concept is based on the chemical picture 
where bound states are considered as new components and can be treated in the same way as 
``elementary'' particles. It is generally accepted that a second virial coefficient can be introduced for systems consisting of atoms, but also with molecules as components. 
The problem is the introduction of an effective 
interaction between the components (including the quantum symmetry postulates for fermions or bosons), 
and intrinsic excitations of the bound components are described in some approximations.
Within the NSE, we describe a nuclear system as a mixture of  ``free'' nucleons in single-particle states
as well as of clusters (nuclei with $A>1$). Taking into account the interaction between these components
of a nuclear system, from measured scattering phase shifts (for instance $\alpha - \alpha$ scattering)
a virial EoS can be derived. Results have been presented in Ref. \cite{Horowitz:2005nd}.


On a more microscopic level, we consider here interacting quarks which can form bound states (hadrons),
and the general approach should include both cases, the region of the quark-gluon plasma
and, after the confinement transition, the region of well established hadrons. A main difficulty is the introduction
of an effective interaction which can be made by fitting empirical data. However, a systematic quantum statistical 
approach is needed to derive such effective interaction from a fundamental Lagrangian, and to introduce the 
cluster states performing consistent approximations and avoiding double counting of the contributions to 
the EoS and other physical properties.

\begin{figure}[H] 
	\centerline{\includegraphics[scale=.45]{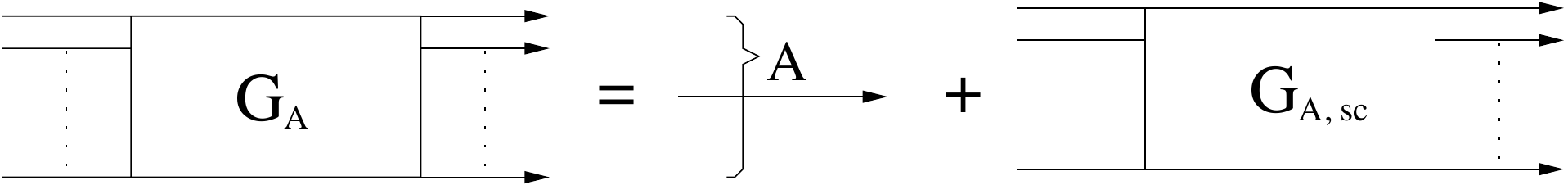}} 
	\caption{Splitting of the $A$-particle cluster propagator into a 
		bound and scattering contribution. Note~that the internal quantum number has
		been dropped.
	} 
	\label{Fig:2} 
\end{figure}

The Green function technique as well as the path-integral approach are such systematic quantum statistical 
approaches. Different contributions to the EoS are represented by Feynman diagrams, and double counting 
is clearly excluded. A selection of diagram classes can be performed to recover the chemical picture.
As seen from Equation (\ref{GAladder}), after separation of the center-of-mass 
momentum $\vec P$,
%
the~propagator for the $A$-particle bound states is
\begin{eqnarray}
G_{A,\nu}^{\rm bound}(1\dots A;1'\dots A';z_A)&=&
\langle 1\dots A |\psi_{A \nu P} \rangle \frac{1}{z_A-E^{(0)}_{A,\nu}(P)} 
\langle \psi_{A \nu P} |1'\dots A' \rangle\,
\label{Apropagator}
\end{eqnarray}
where $\nu$ covers only the bound state part of the internal quantum state of the $A$-particle cluster.
As~a new element, the bound state propagator is introduced as indicated in 
Figure \ref{Fig:2}. 
This bound state propagator has the same analytical form like the single 
particle propagator (\ref{G1}), besides the appearance of the internal wave 
function that determines the vertex function.

As discussed above, different approximations are obtained such as 
the nuclear statistical equilibrium (NSE) and the cluster mean-field (CMF) 
approximation using the chemical picture.
These~approximations are based on the bound state part of the $A$-particle 
propagator.
They give leading contributions in the low-density, low temperature range where
bound states dominate the composition of the many-particle system. 

To improve the approximation, the remaining part of continuum correlations has also been taken into account.
From the point of view of the physical picture, these contributions arise in 
higher orders of the virial expansion of the equation of state. 
As example, the formation of the $A$-particle bound state 
is seen in the $A$-th virial coefficient, the mean-field shift due to a 
cluster $B$ in the $(A+B)$-th
virial coefficient. The chemical picture indicates which high-order virial 
coefficients of the virial expansion are essential, if the many-particle 
system is strongly correlated so that bound states are formed.


In an improved approximation, the scattering part of the $A$-particle 
propagator has to be considered.
It contributes also to the $A$-th virial coefficient. 
The scattering processes within the $A$-particle system can have different 
channels.
As an example we discuss here binary elastic scattering processes between 
sub-clusters $A_1$ and $A_2$ of the system of $A$ particles, $A=A_1+A_2$. 
Binary phase shifts $\delta_{A_1,A_2}(E)$ are introduced that describe the 
corresponding scattering experiments. 
They can also be calculated within few-body theory. 
Besides the effective interaction between the sub-clusters that are depending 
on the internal wave function of the sub-clusters, also virtual transitions to 
excited states may be taken into account.
In general, the effective interaction is non-local in space and time, i.e., 
momentum and frequency dependent.

A generalized  cluster Beth-Uhlenbeck formula, see Equation (\ref{zpart}), 
is obtained 
when in particle loops not the free propagator, but 
quasiparticle Green's functions are used.
If the quasiparticle shift is calculated in Hartree-Fock
approximation,
the first order term of the interaction must be excluded from the ladder
$T_2^{\rm ladder}$ 
matrix to avoid double counting. The bound state part is not affected, 
it is determined by an infinite number of diagrams. The scattering part 
is reduced subtracting the Born contribution as shown in
Equation  (\ref{z}) 
by the $2 \left[\sin (\delta_{c})\right]^{2}$ term; for the derivation
see Ref.~\cite{Schmidt:1990zz}.

The continuum correlations that are not considered in the NSE give a 
contribution to the second virial coefficient in the chemical picture. 
We can extract from the continuum part two contributions~\cite{Ropke:2014fia}: 
resonances that can be treated like new particles in the law of mass action, 
and the quasiparticle shift of the different components contributing
to the law of mass action. 
Both processes are expected to represent significant contributions of the 
continuum. 
After projecting out these effects, the residual contribution 
of the two-nucleon continuum is assumed to be reduced. 
One can try to parametrize the residual part, using the ambiguity in 
defining the bound state contribution. 
Eventually the residual part of the continuum correlations can be neglected.

The correct formulation of the cluster-virial expansion  \cite{Ropke:2012qv}
is considered to be a main ingredient towards a unified EoS describing quark matter 
as well as
nuclear matter. One has to introduce the interaction between the constituents
in a systematic way.~The account of correlations in the continuum is 
essential near the confinement phase transition where the hadronic bound states 
disappear. Conserving approximations may lead to acceptable results for the EoS.
An important issue is the account of correlation in the medium, in particular
when considering the Pauli blocking in the phase space of the elementary constituents.

\section{$\Phi$---Derivable Approach to the Cluster Virial Expansion for Nuclear Matter}\label{sec:phi-nucl}
Recently, it has been suggested \cite{Blaschke:2015bxa} that the cluster virial expansion for many-particle systems~\cite{Ropke:2012qv} can be formulated within the $\Phi$--derivable approach \cite{Baym:1961zz,Baym:1962sx}. 
This approach is straightforwardly generalized to $A$--particle correlations in a many-fermion system
\begin{align}\label{Phi-A}
	\Omega &= \sum_{l=1}^A \Omega_l = \sum_{l=1}^A \left\{ c_l \left[\operatorname{Tr} \ln \left(-G_l^{-1}\right) + \operatorname{Tr} \left(\Sigma_l~G_l \right) \right]+\sum_{\stackrel{i,j}{i+j=l}}\Phi[G_i,G_j,G_{i+j}]\right\}~,
\end{align}
where the full $A-$particle Green's function obeys the Dyson equation 
\begin{eqnarray}\label{G-A}
	G_{A}^{-1}&=&G_{A}^{(0)^{-1}} - \Sigma_A~,
\end{eqnarray}
where $G_{A}^{(0)}$ is the free  $A-$particle Green's function and the selfenergy is defined as a functional derivative of the two-cluster irreducible  $\Phi$ functional
\begin{eqnarray}\label{Sigma-A}
	\Sigma_A(1\dots A,1^\prime \dots A^\prime,z_A) = 
	\frac{\delta \Phi}{\delta G_A(1\dots A,1^\prime \dots A^\prime,z_A)}~.
\end{eqnarray}

This generalization of the $\Phi-$derivable approach fulfills by its construction the conditions of stationarity of the thermodynamical potential with respect to variations of the cluster Green's functions
\begin{eqnarray}\label{Omin}	
	\frac{\delta \Omega}{\delta G_A(1\dots A,1^\prime \dots A^\prime,z_A)} = 0~.
\end{eqnarray}

\begin{figure}[H]
	\begin{minipage}{0.5\textwidth}
		\centering
		\includegraphics[scale=0.2]{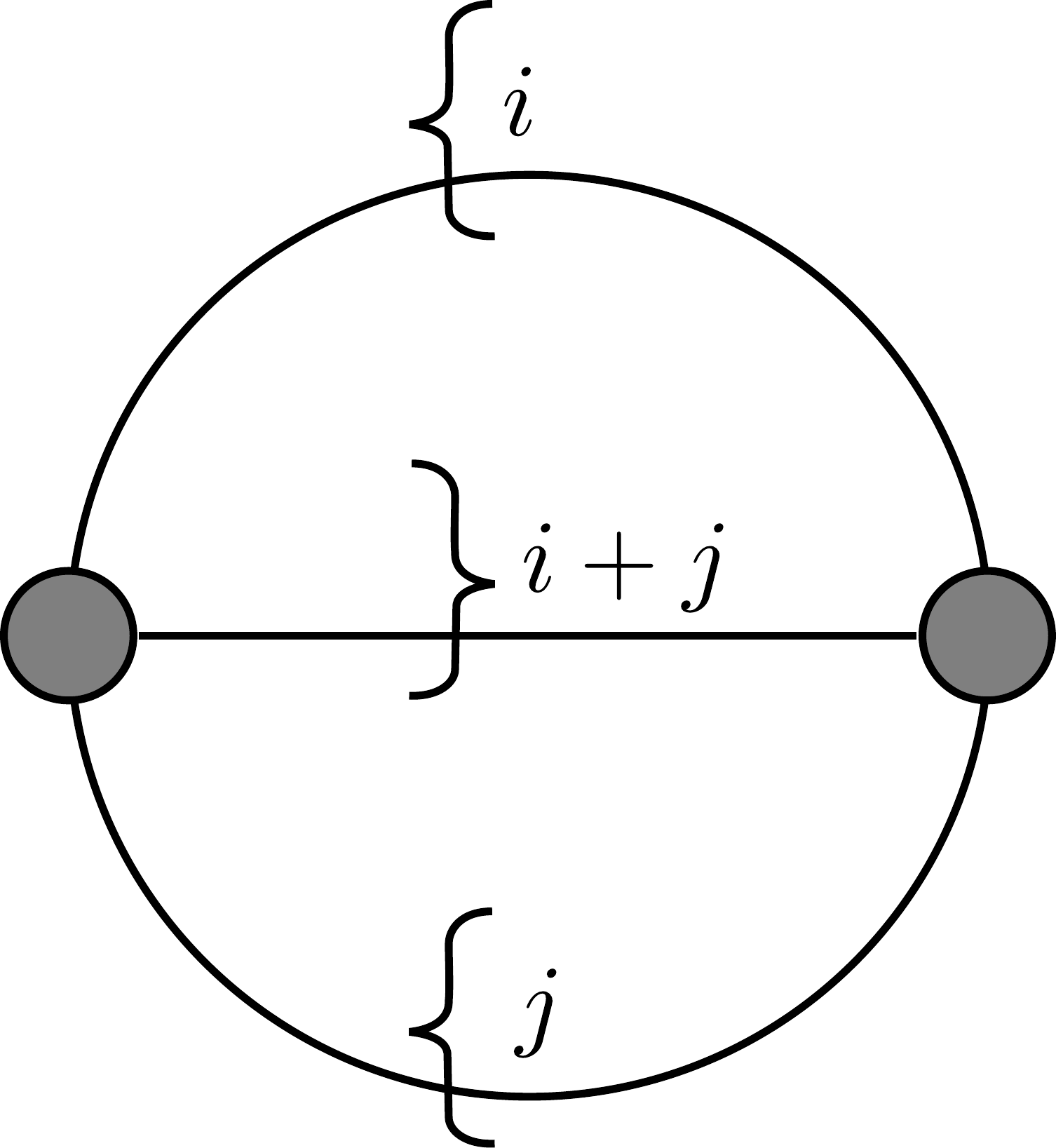}
	\end{minipage}
	\begin{minipage}{0.5\textwidth}
		\centering
		\includegraphics[scale=0.2]{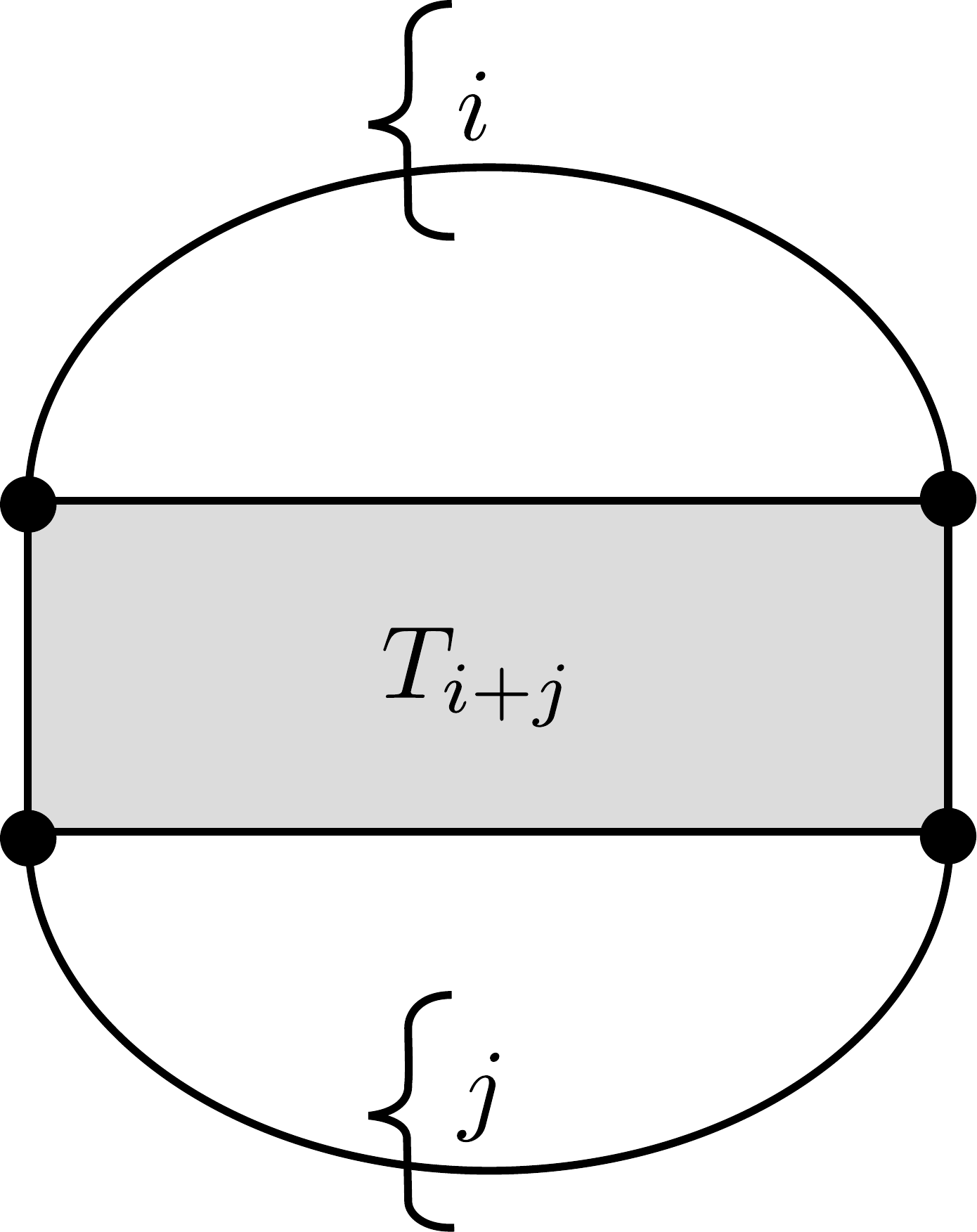}
	\end{minipage}
	\caption{Left panel: The $\Phi$ functional for the general case of $A-$particle correlations in a many-fermion system, whereby all bipartitions $A=i+j$ into lower order clusters of sizes $i$ and $j$
		shall be considered;
		Right panel: Equivalent representation of the diagram in the left panel with the highest order cluster Green's function and the vertex functions replaced by the cluster T matrix, see Figure \ref{fig:2}.} 
	\label{fig:1}
\end{figure}

The $\Phi$ functional for our purpose of defining a cluster decomposition of the system with inclusion of residual interactions among the clusters captured by a second virial coefficient is given by a sum of all two-loop diagrams that can be drawn with cluster Green's functions $G_{i+j}$ and their subcluster Green's functions $G_i$ and $G_j$ for a given bipartition with the appropriate vertex functions $\Gamma_{i+j;ij}$. This~generalization of the so-called ``sunset'' diagram case is depicted diagrammatically in Figure \ref{fig:1}.

\begin{figure}[H]
	\centering
	\includegraphics[scale=0.2]{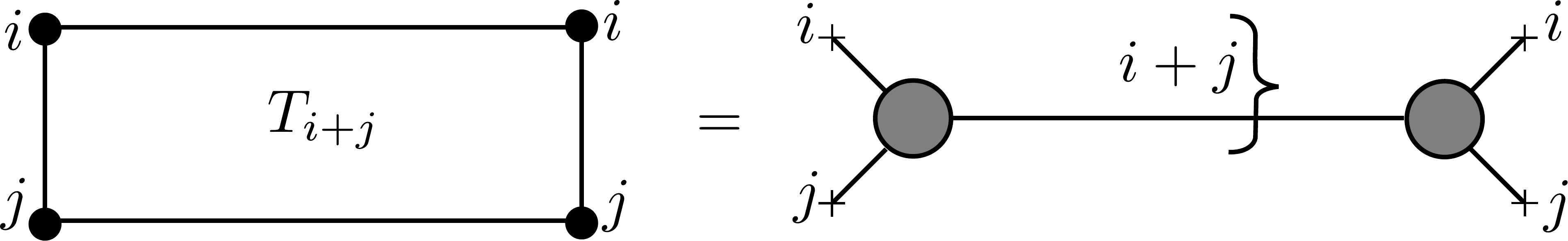}
	\caption{Diagrammatic representation for the replacement of the higher order Green's function $G_{i+j}$ and the corresponding vertex functions in the $\Phi$ functional for the cluster virial expansion by the $T_{i+j}$ matrix for binary collisions in the channel with the partition $i, j$.} 
	\label{fig:2}
\end{figure}

Herewith we have generalized the notion of the $\Phi-$derivable approach to that of a system where the hierarchy of higher order Green functions is built successively from the tower of all Greens functions starting with the fundamental one $G_1$.~The open question is how to define the vertex functions joining the cluster Greens functions. 
As a heuristic first step one may always introduce local coupling constants, like in the Lee model discussed in the context of a $\Phi-$derivable approach by Weinhold et al.~\cite{Weinhold:1997ig}. Introducing nonlocal formfactors at the vertices will correspond to a separable representation of the interaction. 
The definition of the interaction can be absorbed in the introduction of the cluster T-matrix in the corresponding channel which, in the ladder approximation, will reproduce the analytical properties encoded in the full cluster Green's function concerning bound state poles and scattering state continuum.  
This equivalence is depicted in Figure \ref{fig:2}.

The $T_{A}$ matrix fulfills the Bethe-Salpeter equation in ladder approximation
\begin{eqnarray}	
	T_{i+j}(1,2,\dots,A;1',2', \dots A'; z) = V_{i+j} + V_{i+j} G_{i+j}^{(0)} T_{i+j}~,
	\label{T_A}
\end{eqnarray}
which in the separable approximation for the interaction potential, 
\begin{eqnarray}
	V_{i+j}=\Gamma_{i+j}(1,2,\dots,i;i+1,i+2,\dots, i+j)\Gamma_{i+j}(1',2',\dots,i';(i+1)',(i+2)',\dots,(i+j)') ~,
	\label{V_A}
\end{eqnarray}
leads to the closed expression for the $T_{A}$ matrix
\begin{eqnarray}	
	T_{i+j}(1,2,\dots,i+j;1',2', \dots (i+j)'; z) = V_{i+j}\left\{1-\Pi_{i+j}\right\}^{-1} ~,
	\label{T_A-pole}
\end{eqnarray}
where the generalized polarization function 
\begin{eqnarray}
	\Pi_{i+j}={\rm Tr}\left\{\Gamma_{i+j} G_{i}^{(0)} \Gamma_{i+j} G_{j}^{(0)} \right\}
\label{Phi-loop}
\end{eqnarray}
has been introduced and the one-frequency free $i-$particle Green's function is defined by the
\mbox{$(i-1$)-}fold Matsubara sum
\begin{equation}
\begin{array}{lll}
	G_{i}^{(0)}(1,2,\dots, i; \Omega_i)&=&\sum_{\omega_1 \dots \omega_{i-1}}\frac{1}{\omega_1-E(1)}\frac{1}{\omega_2-E(2)}\dots\frac{1}{\Omega_i-(\omega_1+\dots \omega_{i-1})-E(i)}\\
	&=&\frac{(1-f_1)(1-f_2)\dots (1-f_i) - (-)^i f_1f_2\dots f_i}{\Omega_i - E(1)-E(2)-\dots E(i)}~.
\end{array}
\end{equation}

Note that for these Green's functions holds the relationship ($\Omega_{i+j}=\Omega_i+\Omega_j$)
\begin{eqnarray}
	G_{i+j}^{(0)}=G_{i+j}^{(0)}(1,2,\dots, i+j; \Omega_{i+j})&=&\sum_{\Omega_i}
	G_{i}^{(0)}(1,2,\dots, i; \Omega_i)G_{j}^{(0)}(i+1,i+2,\dots, i+j; \Omega_j)~.
\end{eqnarray}

Another set of useful relationships follows from the fact that in the ladder approximation both, the full two-cluster ($i+j$ particle) T matrix (\ref{T_A-pole}) and the corresponding Greens' function 
\begin{eqnarray}
G_{i+j}=G_{i+j}^{(0)}\left\{1-\Pi_{i+j}\right\}^{-1}
\label{Gij}
\end{eqnarray}
have similar analytic properties determined by the $i+j$ cluster polarization loop integral (\ref{Phi-loop}) and are related by the identity
\begin{eqnarray}	
T_{i+j} G_{i+j}^{(0)}  = V_{i+j} G_{i+j} ~,
\label{T-G}
\end{eqnarray}
which is straightforwardly proven by multiplying Equation (\ref{T_A-pole}) with $G_{i+j}^{(0)}$ and using Equation~(\ref{Gij}).
Since~these two equivalent expressions in Equation (\ref{T-G}) are at the same time equivalent to the two-cluster irreducible $\Phi$ functional introduced above in Equations~(\ref{Phi-A}) and Figure \ref{fig:1}, the~functional~relations
\begin{eqnarray}
T_{i+j}= \delta \Phi /\delta G_{i+j}^{(0)}~,\\
V_{i+j}= \delta \Phi /\delta G_{i+j}
\end{eqnarray}
follow and may become useful in proving cancellations that are essential for the relationship to the Generalized Beth-Uhlenbeck approach as discussed below. 
\subsection{Generalized Beth-Uhlenbeck EoS from the $\upPhi-$Derivable Approach}
Now we return to the question how the relation between the cluster $\Phi-$derivable approach to the partition Function (\ref{Phi-A}) and the generalized Beth-Uhlenbeck equation for the cluster density
may be established.
To this end we consider the partial density of the $A-$particle state defined as 
\begin{eqnarray}
	n_A(T,\mu) = -\frac{\partial \Omega_A}{\partial \mu}\,.
	\label{n_A}
\end{eqnarray} 

Taking into account that any analytic complex function $F(\omega)$ has the spectral representation
\begin{eqnarray}
	F(iz_n) = \int_{-\infty}^\infty\frac{d\omega}{2\pi}\frac{\mathrm{Im} F(\omega)}{\omega - iz_n}\,,
\end{eqnarray}
we perform the Matsubara summation\footnote{For odd $A$, $z_n=(2n+1)\pi T +\mu$ are the fermionic Matsubara frequencies and for even $A$, $z_n=2n\pi T + \mu$ the bosonic ones.} in Equation \eqref{Phi-A}
\begin{eqnarray}
	\sum_{z_n} \frac{c_A}{\omega - iz_n} = f_A(\omega)=\frac{1}{\exp[(\omega-\mu)/T] - (-1)^A}~.
\end{eqnarray}

Using the relation $\partial f_A(\omega)/\partial \mu = -\partial f_A(\omega)/\partial \omega$ we get for Equation \eqref{n_A} now
\begin{equation}
\begin{array}{l}
	n_A(T,\mu) = - d_A \int\frac{d^3 q}{(2\pi)^3}\int\frac{d\omega}{2\pi}f_A(\omega)
	\frac{\partial}{\partial \omega}\left[{\rm Im} \ln \left(-G_A^{-1}\right) 
	+ {\rm Im} \left(\Sigma_A~G_A \right) \right]
	+\sum_{\stackrel{i,j}{i+j=A}}\frac{\partial \Phi[G_i,G_j,G_A]}{\partial \mu}~,	
\end{array}
\label{n_A_} 	
\end{equation} 
where a partial integration over $\omega$ has been performed and the degeneracy factor $d_A$ for cluster state has been introduced, stemming from the trace operation in the internal spaces.

Now we use the fact that for two-loop diagrams of the sunset type a cancellation holds \cite{Vanderheyden:1998ph,Blaizot:2000fc} which we generalize here for cluster states
\begin{eqnarray}
	d_A \int\frac{d^3 q}{(2\pi)^3}\int\frac{d\omega}{2\pi}f_A(\omega)
	\frac{\partial}{\partial \omega}\left({\rm Re}\Sigma_A~ {\rm Im}G_A \right)
	-\sum_{\stackrel{i,j}{i+j=A}}\frac{\partial \Phi[G_i,G_j,G_A]}{\partial \mu} =0 ~.
	\label{cancel}
\end{eqnarray} 

Using generalized optical theorems \cite{Zimmermann:1985ji,Schmidt:1990zz} we can show that
\begin{eqnarray}
	\frac{\partial}{\partial \omega}\left[{\rm Im} \ln \left(-G_A^{-1}\right) 
	+ {\rm Im}\Sigma_A~ {\rm Re} G_A \right] = 2 {\rm Im} \left[G_A~{\rm Im}\Sigma_A~\frac{\partial}{\partial \omega} G_A^*~{\rm Im}\Sigma_A  \right]=-2\sin^2\delta_A \frac{\partial \delta_A}{\partial \omega}~, 
\end{eqnarray}
where the phase shifts $\delta_A$ have been introduced via the polar representation of the complex 
$A-$particle propagator $G_A=|G_A| \exp(i\delta_A)$.~With these ingredients follows from the cluster $\Phi-$derivable approach the cluster virial expansion for the density in the form of a generalized Beth-Uhlenbeck EoS

\begin{eqnarray}
	n(T,\mu) = \sum_{i=1}^A n_i(T,\mu) = \sum_{i=1}^A d_i \int\frac{d^3 q}{(2\pi)^3}\int\frac{d\omega}{2\pi}f_i(\omega) 2\sin^2\delta_i\frac{\partial \delta_i}{\partial \omega}~.
\end{eqnarray}

In this way we have drawn the connection between the cluster virial expansion of Ref.~\cite{Ropke:2012qv} reviewed in the previous section 
with the $\Phi-$derivable approach \cite{Baym:1961zz,Baym:1962sx}.
In the following subsection we consider the example of deuterons in nuclear matter in order to
elucidate the application of the~approach.
 
\subsection{Deuterons in Nuclear Matter}

\begin{figure}[H]
	\begin{minipage}{0.5\textwidth}
		\centering
		\includegraphics[scale=0.2]{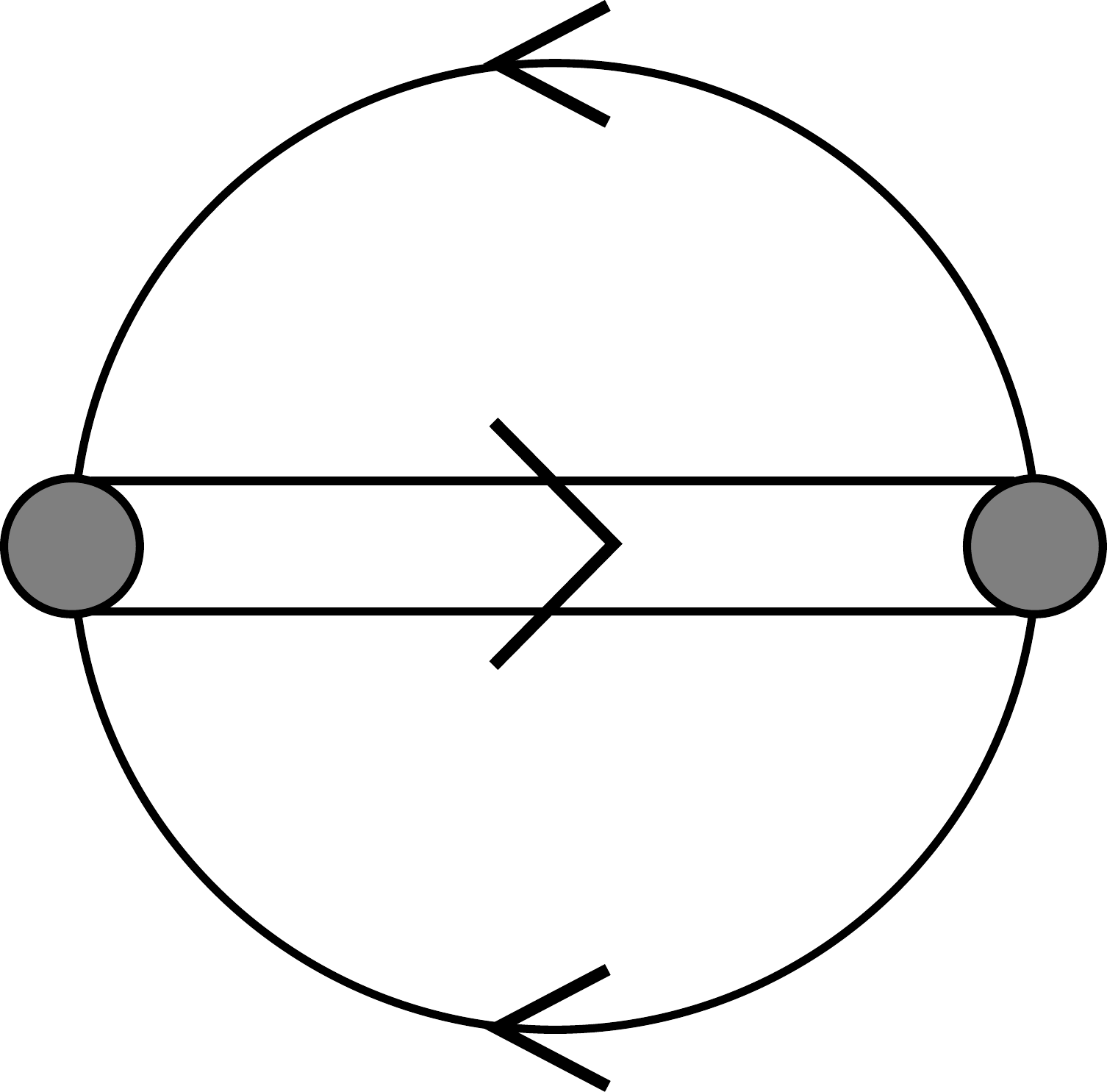}
	\end{minipage}
	\begin{minipage}{0.5\textwidth}
		\centering
		\includegraphics[scale=0.2]{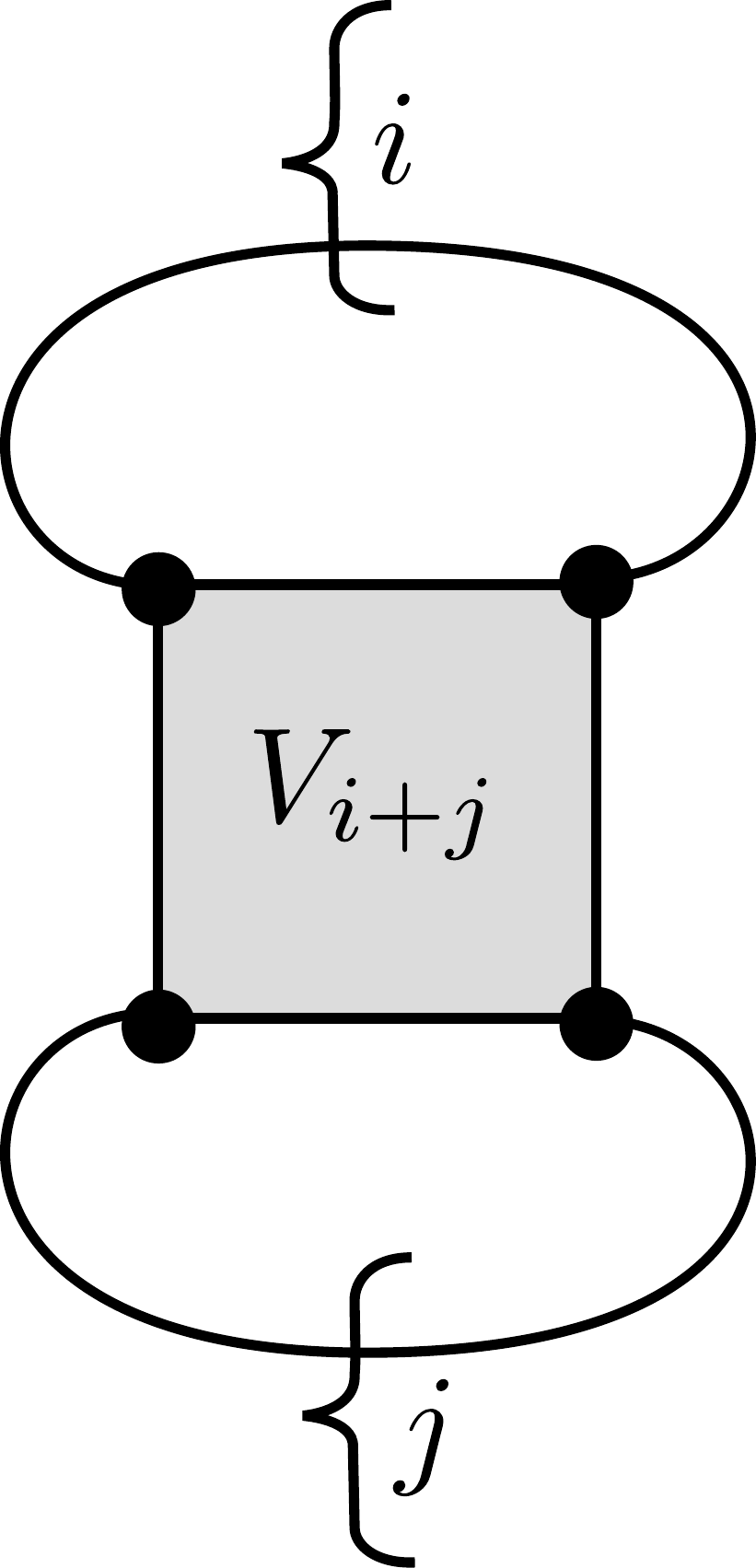}
	\end{minipage}
	\caption{(\textbf{Left}) panel: Two-particle irreducible  $\Phi$ functional describing two-particle correlations 
	(double line with arrow) of elementary fermions (single arrowed lines);
	(\textbf{Right}) panel: Cluster Hartree approximation, following from replacing the two-cluster T matrix in the $\Phi$ functional of Figure \ref{fig:1} with the two-cluster potential $V_{i+j}$. }
	\label{fig:3}
\end{figure}

Within the $\Phi-$derivable approach \cite{Baym:1961zz,Baym:1962sx} the grand canonical thermodynamic potential for a dense fermion system with two-particle correlations is given as
\begin{eqnarray}
	\label{2PI}
	\Omega = -{\rm Tr}\,\{ \ln (-G_1)\} -{\rm Tr} \{\Sigma_1 G_1\}
	+{\rm Tr}\,\{ \ln (-G_2)\} +{\rm Tr} \{\Sigma_2 G_2\} +\Phi[G_1,G_2]~,
\end{eqnarray}
where the full propagators obey the Dyson-Schwinger equations
\begin{eqnarray}
	G_1^{-1}(1,z)&=& z - E_1(p_1) - \Sigma_1(1,z);~
	G_2^{-1}(12,1'2',z) =z-E_1(p_1)-E_2(p_2)-\Sigma_2(12,1'2',z),
\end{eqnarray}
with selfenergies 
\begin{eqnarray}
	\Sigma_1(1,1') &=& \frac{\delta  \Phi}{\delta G_1(1,1')}~;~~
	\Sigma_2(12,1'2',z)=\frac{\delta  \Phi}{\delta G_2(12,1'2',z)}~,
\end{eqnarray}
which are defined by the choice for the $\Phi$ functional, a two-particle irreducible 
set of diagrams such the ones in Figure \ref{fig:3}.

The functional for the thermodynamic potential (\ref{2PI}) is constructed such that the requirement of its stationarity, 
\begin{eqnarray}
	\frac{\partial \Omega}{\partial G_1} = \frac{\partial \Omega}{\partial G_2} = 0~, 
\end{eqnarray}
in thermodynamic equilibrium is equivalent to Equation (\ref{eosspec})
\begin{eqnarray}
	\label{n}
	 n = - \frac{1}{V} \frac{\partial \Omega}{\partial \mu}
	= \frac{1 }{V} \sum_{1}
	\int_{-\infty}^\infty\frac{d \omega}{ \pi} f_{1}(\omega) S_1(1,\omega)\,, 
\end{eqnarray}
where $S_1(1,\omega)=2 \Im G_1(1,\omega+i\eta)$ is the fermion spectral function and Equation (\ref{n})
expresses particle number conservation in a system with volume $V$.  

Having introduced the notion of a cluster expansion of the $\Phi$ functional we want to suggest a definition which eliminates the unknown vertex functions in favour of the $T_{A+B}$ matrix which describes the nonperturbative binary collisions of $A-$ and $B-$ particle correlations in the channel $A+B$, see Figure \ref{fig:2}.
The application of this scheme to the simplest case of two-particle correlations in the deuteron channel in nuclear matter results in the selfenergy \cite{Schmidt:1990zz}
\begin{equation}
\label{cluster-SE}
\Sigma(1,z)=\sum_2\int\frac{d\omega}{2\pi} S(2,\omega)\bigg\{ 
f(\omega)V(12,12) -\int\frac{dE}{\pi} \Im T(12,12;E+i\eta) \frac{f(\omega)+g(E)}{E-z-\omega}\bigg\}~,
\end{equation}
where $f(\omega)=[\exp(\omega/T)+1]^{-1}$ is the Fermi function and 
$g(\omega)=[\exp(\omega/T)-1]^{-1}$ the Bose function.
The decomposition (\ref{cluster-SE}) corresponds to a cluster decomposition of the nucleon density 
\begin{equation}
n(\mu,T)=n_{\rm qu}(\mu,T) + 2n_{\rm corr}(\mu,T)~,
\end{equation}
where the correlation density  
\begin{equation}
\label{ncont}
n_{\rm corr}=\int \frac{dE}{2\pi} g(E) 2 \sin^2\delta(E) \frac{d\delta(E)}{dE}~,
\end{equation}
contains besides the bound state a scattering state contribution as can be seen from examining the derivative of the phase shift shown in Figure \ref{Phaseshifts}.~The one-particle density of free quasiparticle nucleons $n_qu$ is reduced in order to fulfil the baryon number conservation in the presence of deuteron correlations and contains a selfenergy contribution due to the deuteron correlations in the medium.
This~improvement of the quasiparticle picture due to the correlated medium accounted for by the consistent definition of the selfenergy as a derivative of the $\Phi$ Functional (\ref{Phi-A}) is the reason the continuum 
correlations (\ref{ncont}) are reduced by the factor $2\sin^2 \delta$ as compared to the traditional Beth-Uhlenbeck formula \cite{UHLENBECK1936729,Beth:1937zz}.~For details, see  \cite{Zimmermann:1985ji,Schmidt:1990zz}.~With the definition of the $\Phi$ functional via the T matrix in Figure \ref{fig:2} we were able to show the correspondence between the generalized Beth-Uhlenbeck approach and the $\Phi-$derivable approach
for the nonrelativistic potential model approach to two-particle correlations in a warm, dense Fermion system \cite{Schmidt:1990zz,Ropke:2012qv}.
Now we would like to discuss its application to a relativistic model for correlations in quark matter:
mesons, diquarks and~baryons.

\section{Cluster Virial Expansion for Quark-Hadron Matter within the $\Phi$ Derivable Approach}\label{sec:phi-quarks}

\begin{figure}[H]
	\begin{minipage}{0.2\textwidth}
		\centering
		\includegraphics[scale=0.18]{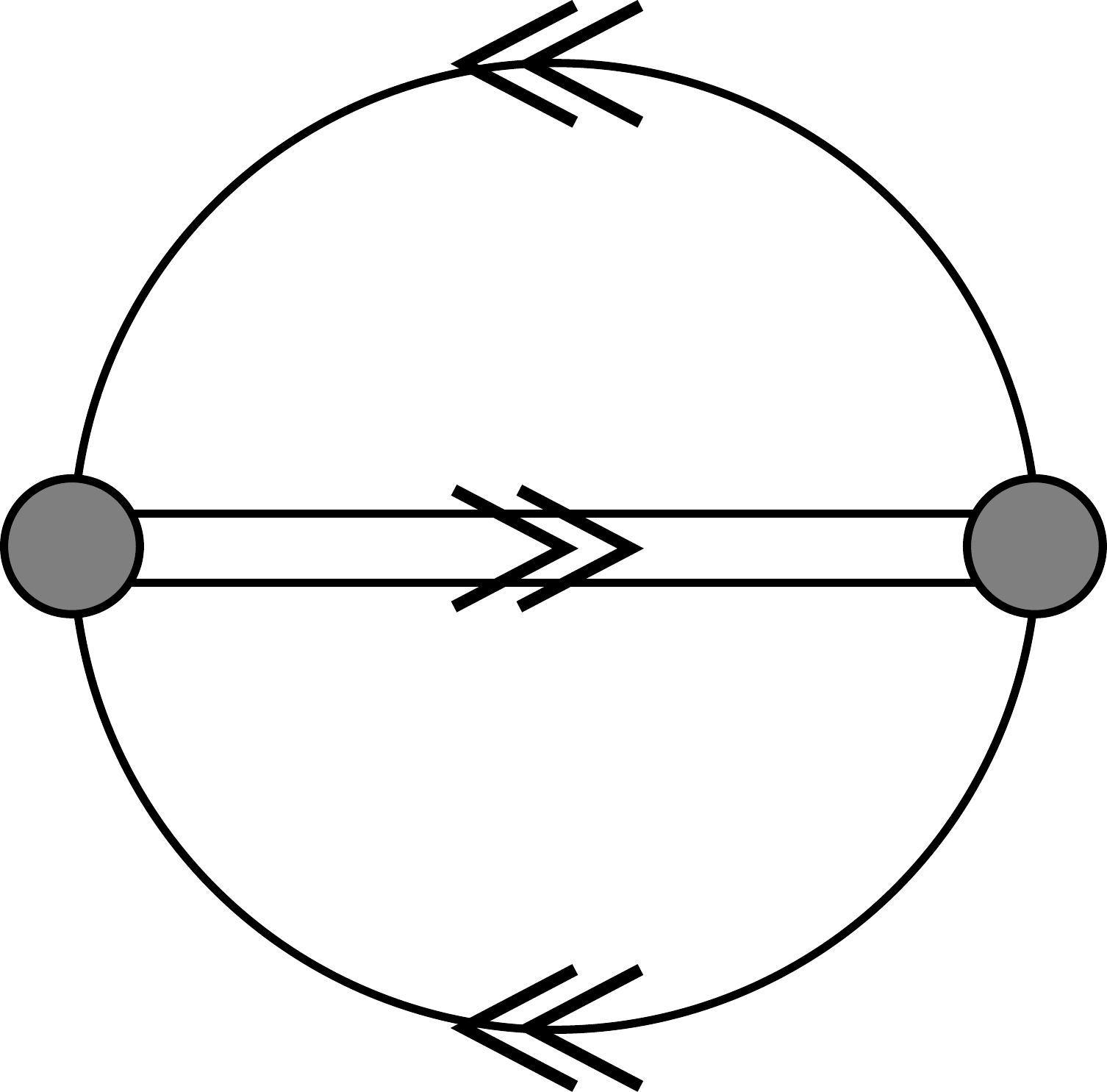}
	\end{minipage}%
	\begin{minipage}{0.2\textwidth}
		\centering
		\includegraphics[scale=0.18]{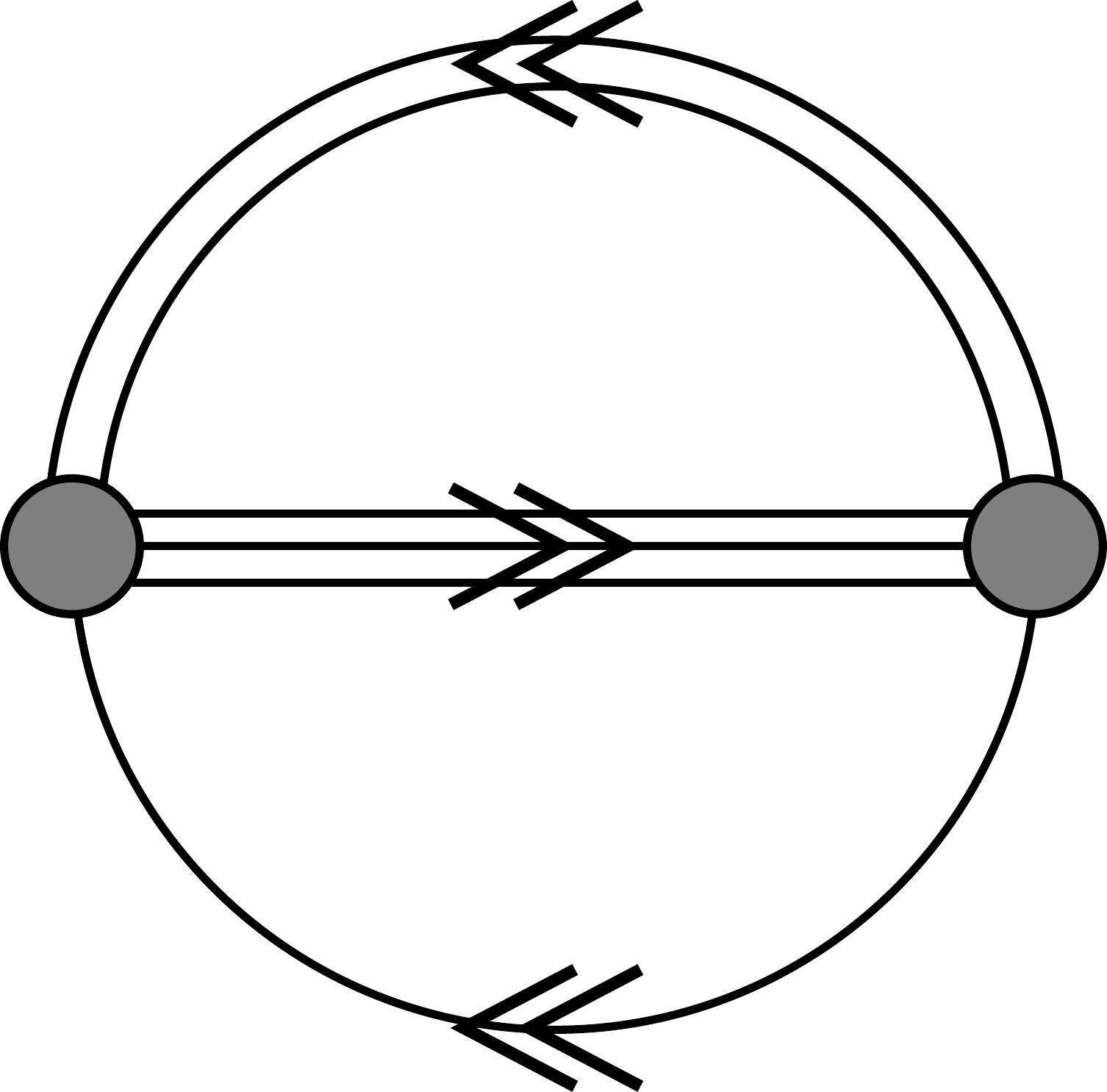}
	\end{minipage}%
	\begin{minipage}{0.2\textwidth}
		\centering
		\includegraphics[scale=0.18]{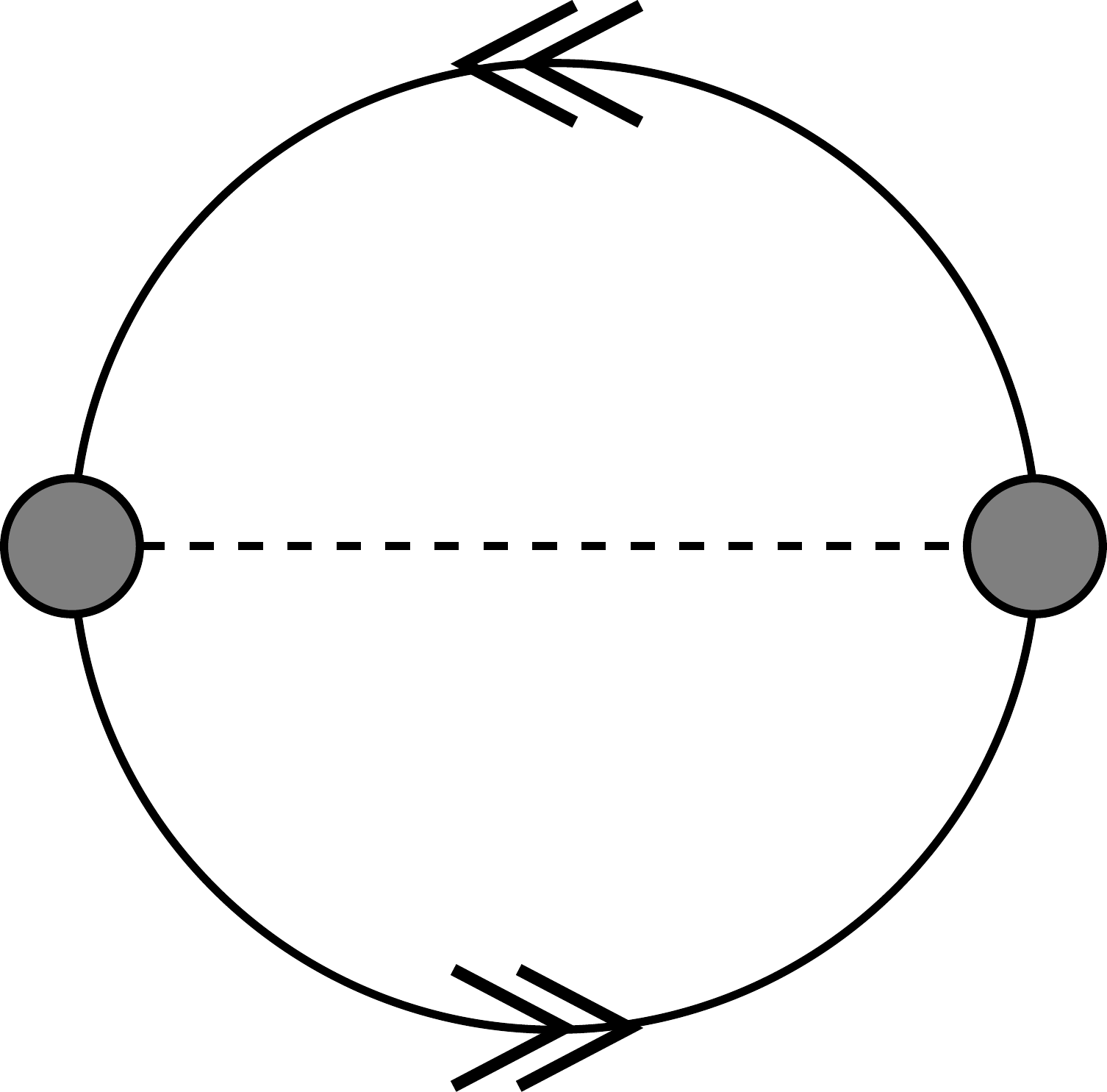}
	\end{minipage}%
	\begin{minipage}{0.2\textwidth}
		\centering
		\includegraphics[scale=0.18]{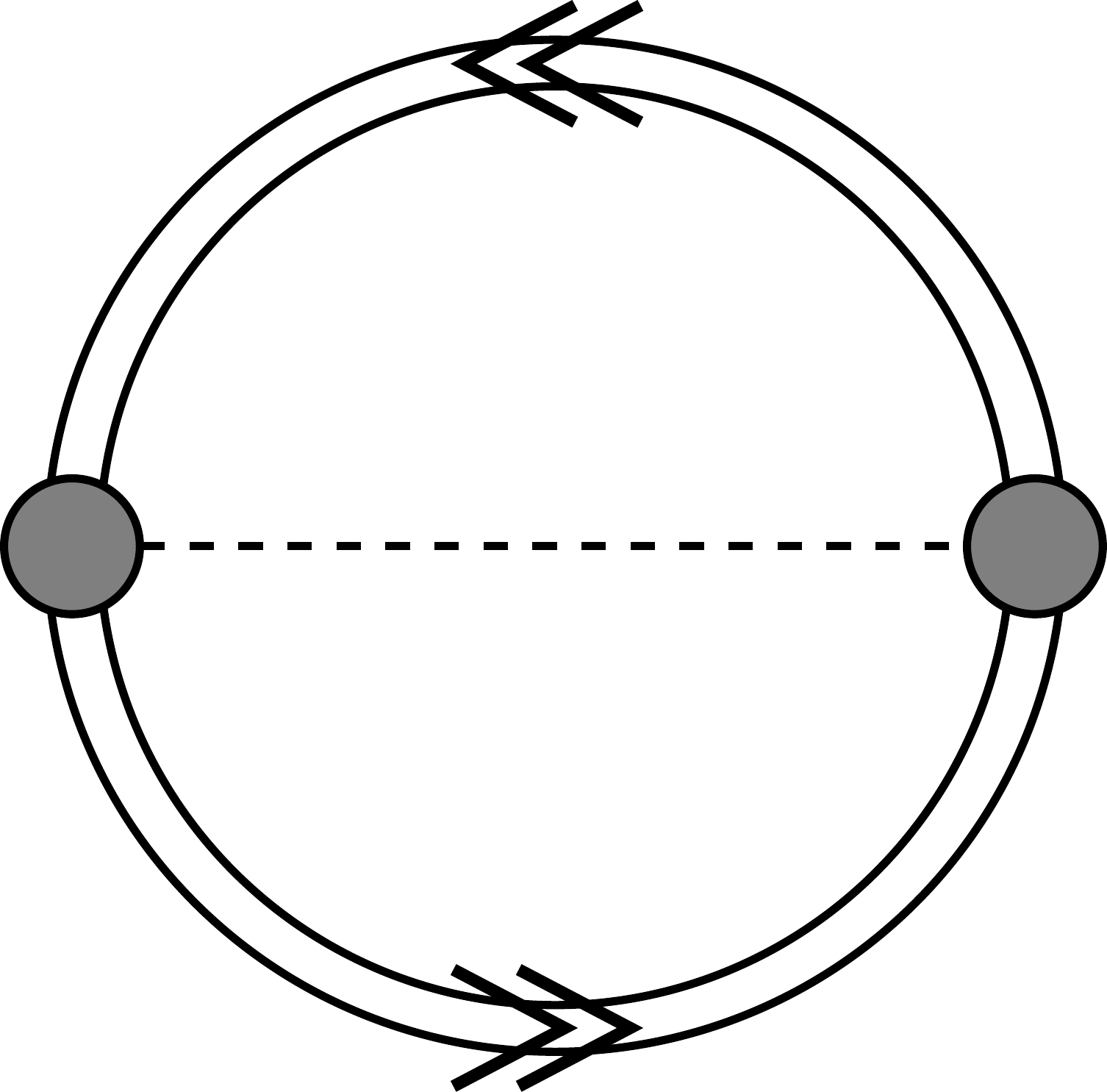}
	\end{minipage}%
	\begin{minipage}{0.2\textwidth}
		\centering
		\includegraphics[scale=0.18]{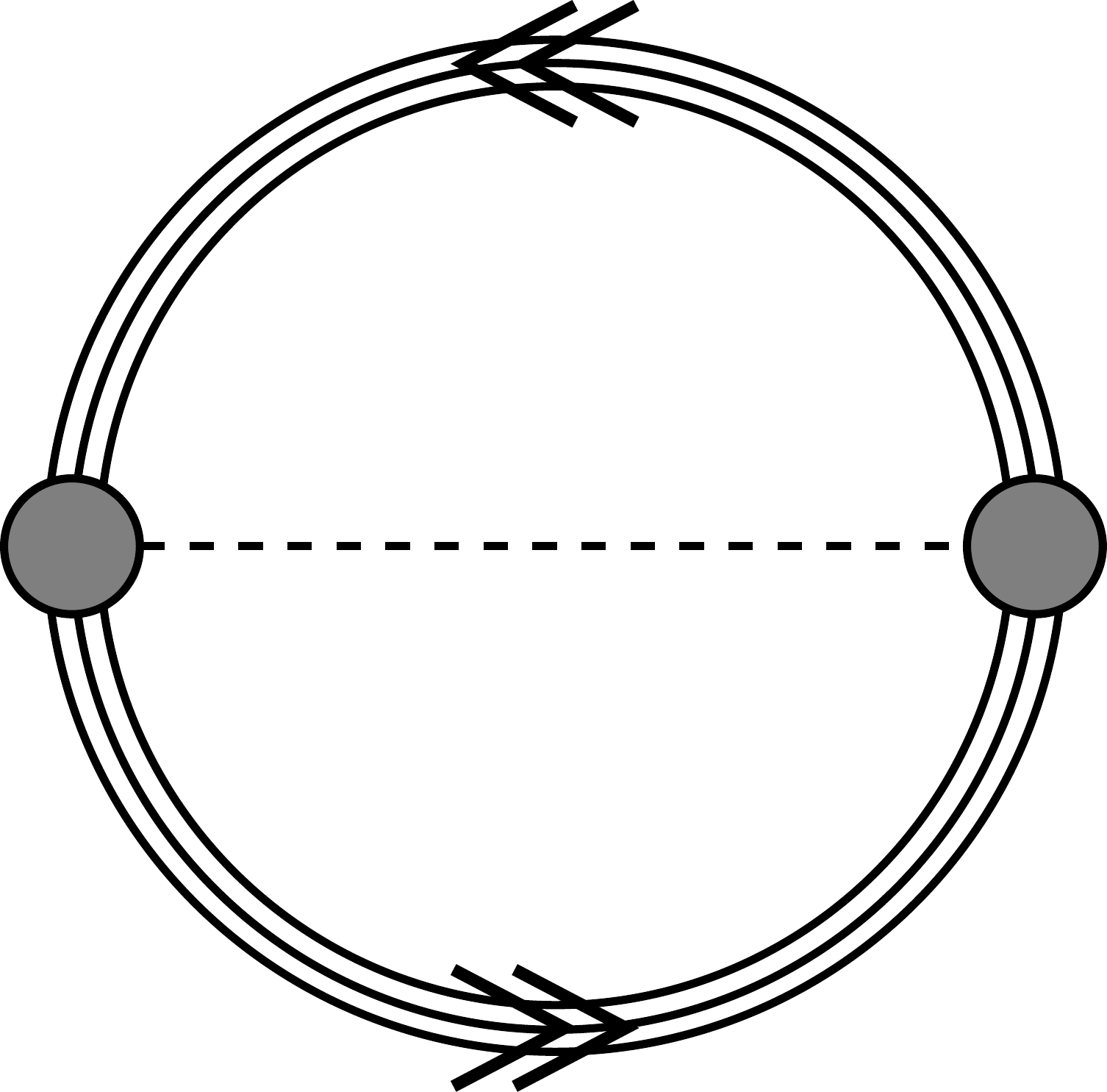}
	\end{minipage}%
	\caption{The contributions to the $\Phi$ functional for the quark-meson-diquark-baryon system.} 
	\label{fig:4}
\end{figure}

Finally, we would like to sketch how the $\Phi$ derivable approach can be employed to define a cluster virial expansion for quark-hadron matter consisting of quarks (\textit{Q}), mesons (\textit{M}), diquarks (\textit{D}) and baryons (\textit{B}) that can represent a unified quark-hadron matter EoS. The thermodynamical potential for this system obtains the form very similar to the case of clustered nuclear matter, i.e.
\begin{eqnarray}
	\label{Om1}
	\Omega &=& \sum_{i=Q,M,D,B} {c_i} \left[\mathrm{Tr} \ln \left(-G_i^{-1}\right) 
	+ \mathrm{Tr} \left(\Sigma_i~G_i \right) \right]+\Phi\left[G_Q,G_M,G_D,G_B\right]~,\\
	&=& \sum_{i=Q,M,D,B} d_i \int\frac{d^3 q}{(2\pi)^3}\int\frac{d\omega}{2\pi}
	\left\{ \omega + 2 T \ln \left[1 - {\mathrm e}^{-\omega/T} \right]\right\} 
	2\sin^2\delta_i\frac{\partial \delta_i}{\partial \omega}~.
	\label{Om2}
\end{eqnarray}
where $c_i=1/2$ ($c_i=-1/2$) for bosonic (fermionic) states and $d_i$ are the degeneracy factors that stem from the trace operation in the internal spaces of the quark, meson, diquark and baryon states. 
We~suggest that in going from Equation (\ref{Om1}) to (\ref{Om2}) the same cancellations apply that were 
used above for the density formula and that are known to apply also for the entropy \cite{Blaizot:2000fc,Vanderheyden:1998ph} would allow to derive this generalized Beth-Uhlenbeck equation of state for the thermodynamic potential, i.e., the negative pressure, once we restrict ourselves to the minimal set of two-particle irreducible diagrams in defining the $\Phi$ functional 
by the class of sunset type diagrams only, as given in Figure \ref{fig:4}.
From this $\Phi$ functional follow the selfenergies defining the full Greens functions of the system by functional derivation
\begin{equation}
\Sigma_i = \frac{\delta~ \Phi\left[G_Q,G_M,G_D,G_B\right]}{\delta~G_i }~.
\end{equation}

\begin{figure}[H]
	\begin{minipage}{0.33\textwidth}
		\centering
		\includegraphics[scale=0.18]{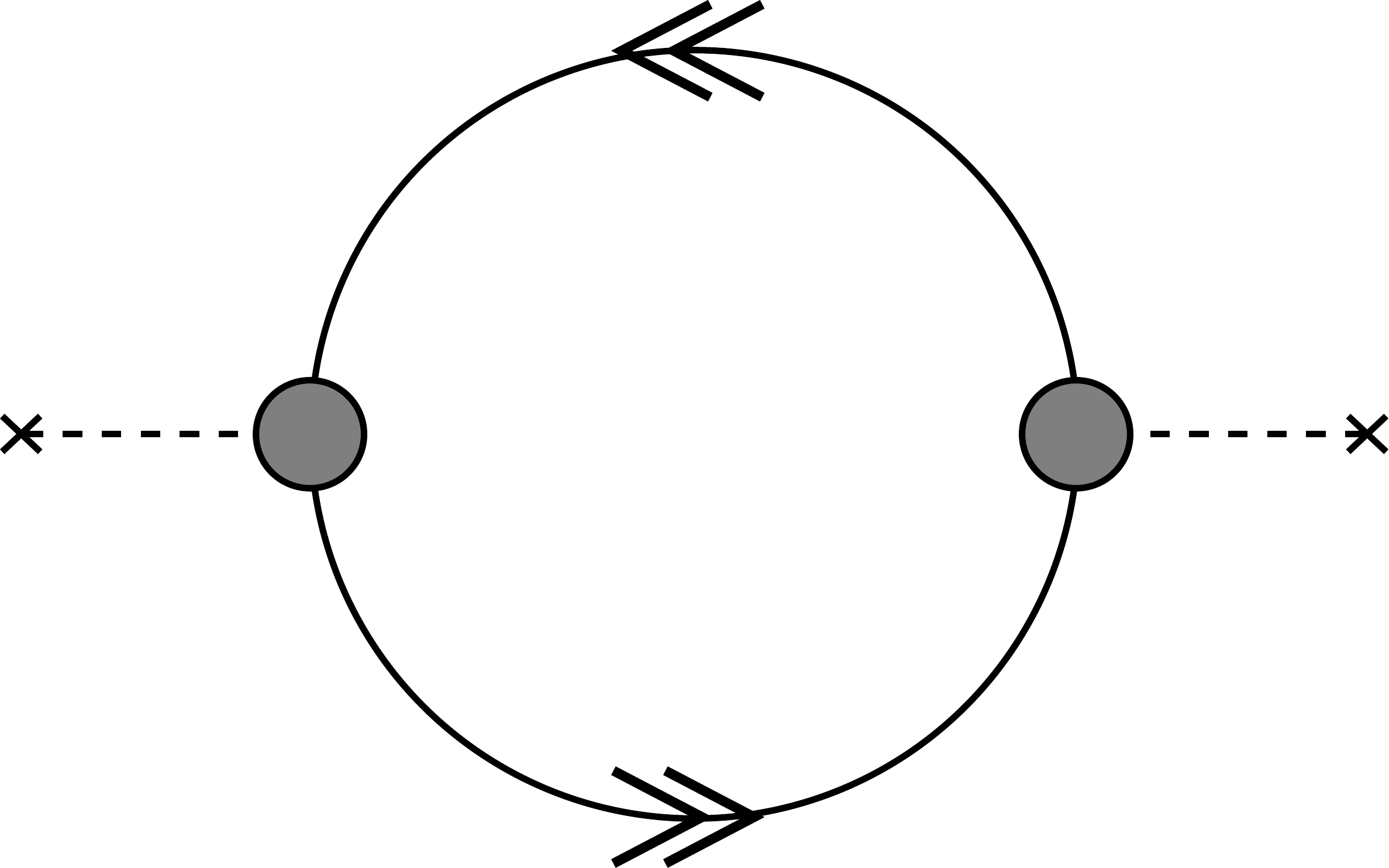}
	\end{minipage}%
	\begin{minipage}{0.33\textwidth}
		\centering
		\includegraphics[scale=0.18]{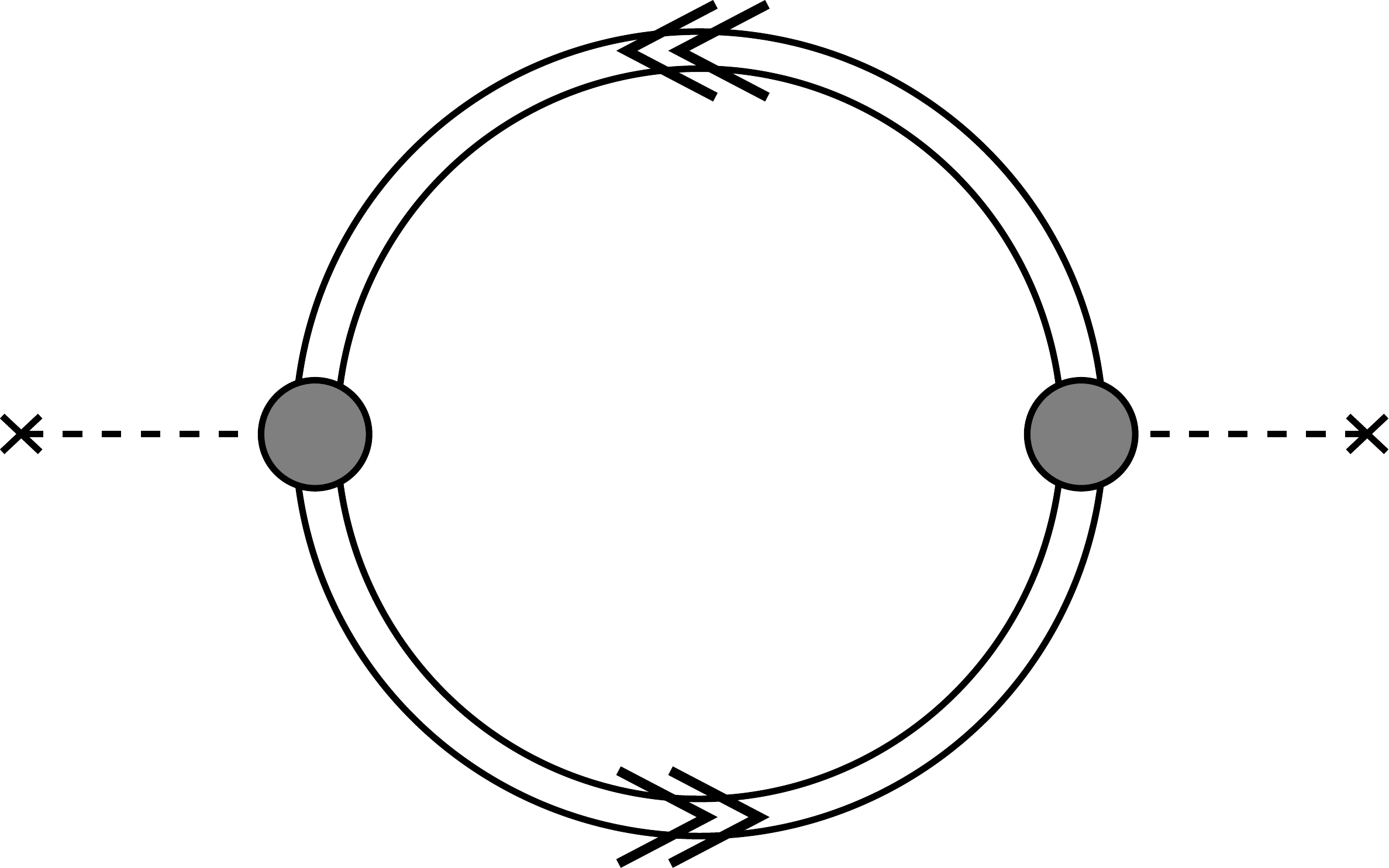}
	\end{minipage}%
	\begin{minipage}{0.33\textwidth}
		\centering
		\includegraphics[scale=0.18]{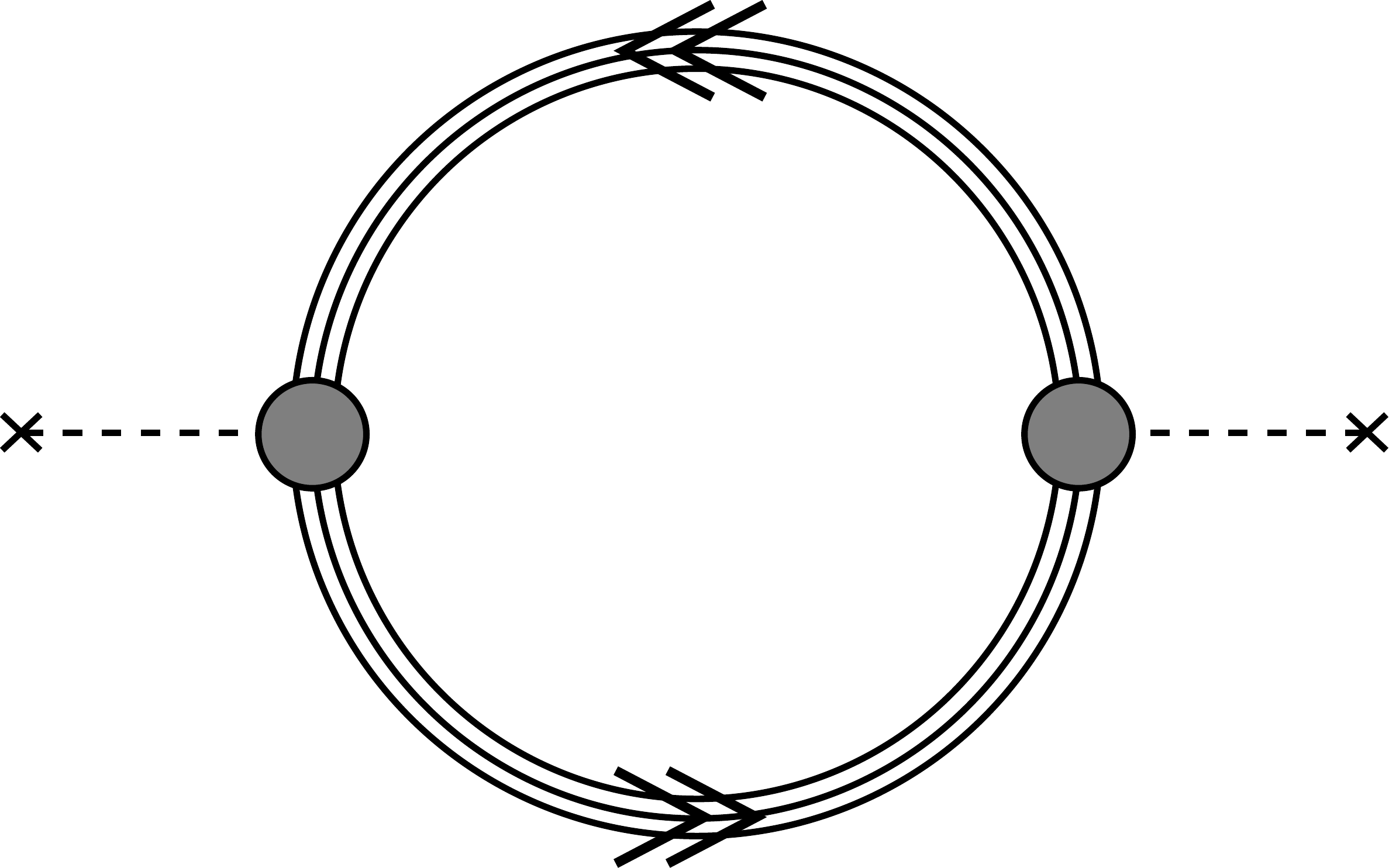}
	\end{minipage}
	\begin{minipage}{0.33\textwidth}
		\centering
		\includegraphics[scale=0.18]{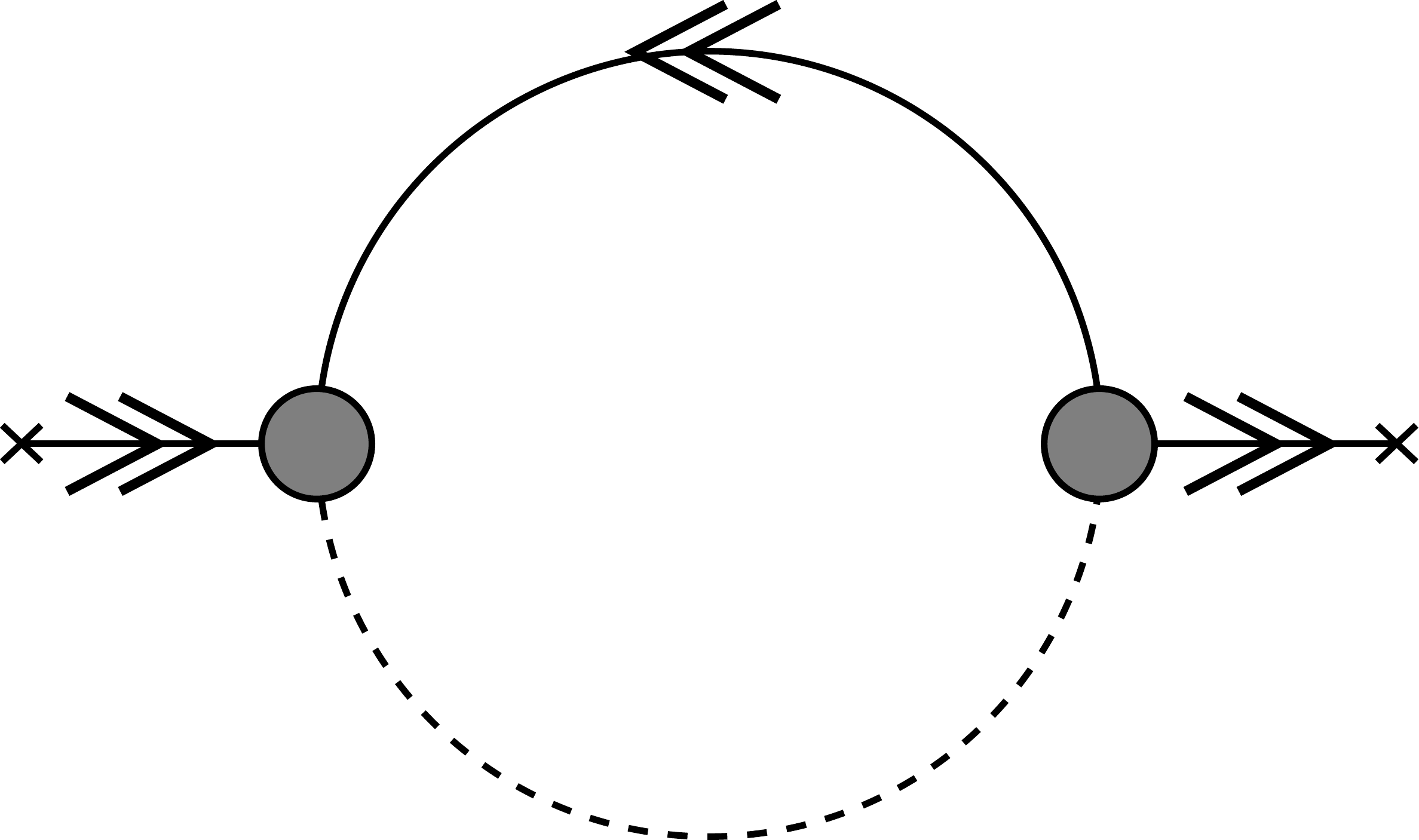}
	\end{minipage}%
	\begin{minipage}{0.33\textwidth}
		\centering
		\includegraphics[scale=0.18]{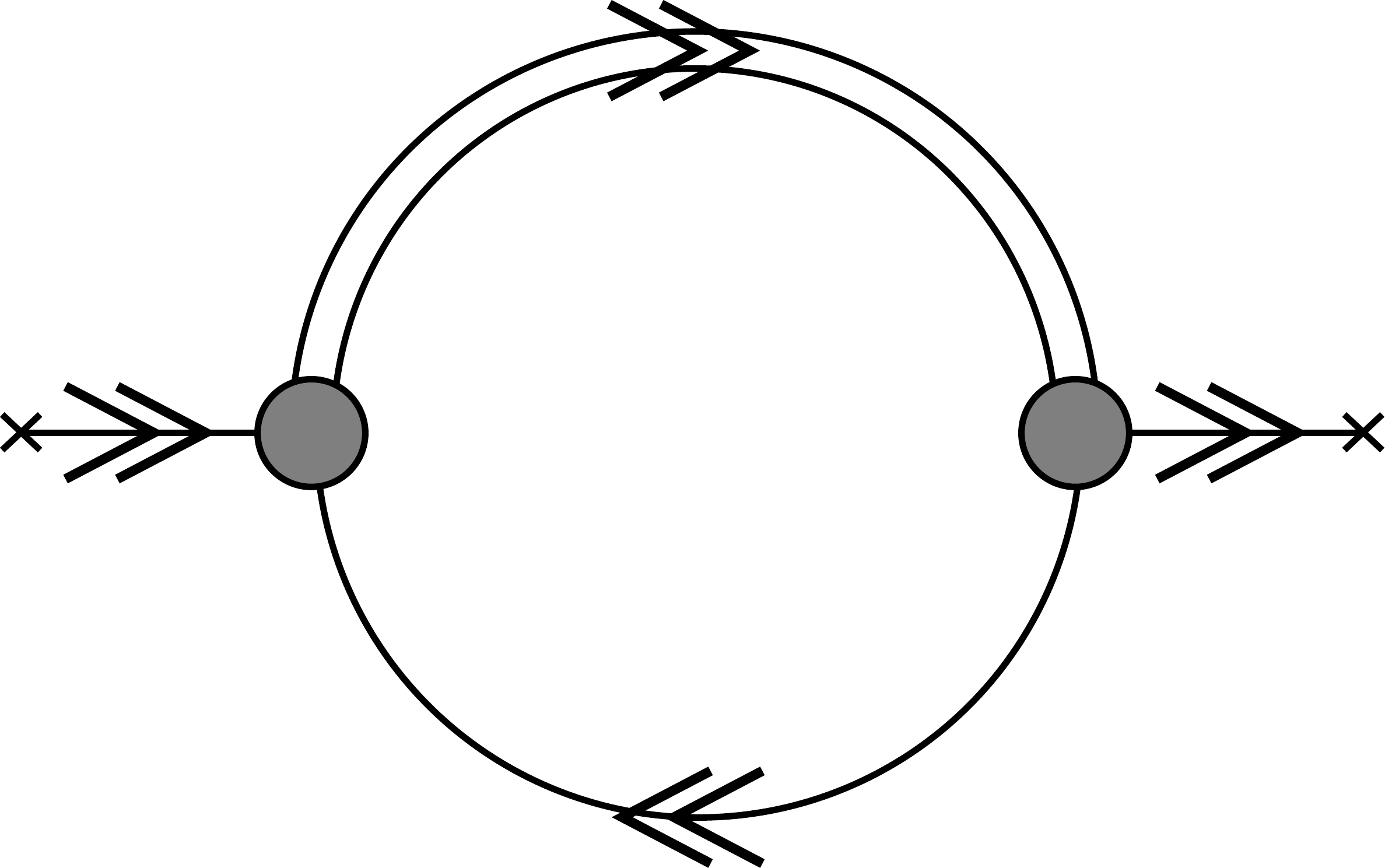}
	\end{minipage}%
	\begin{minipage}{0.33\textwidth}
		\centering
		\includegraphics[scale=0.18]{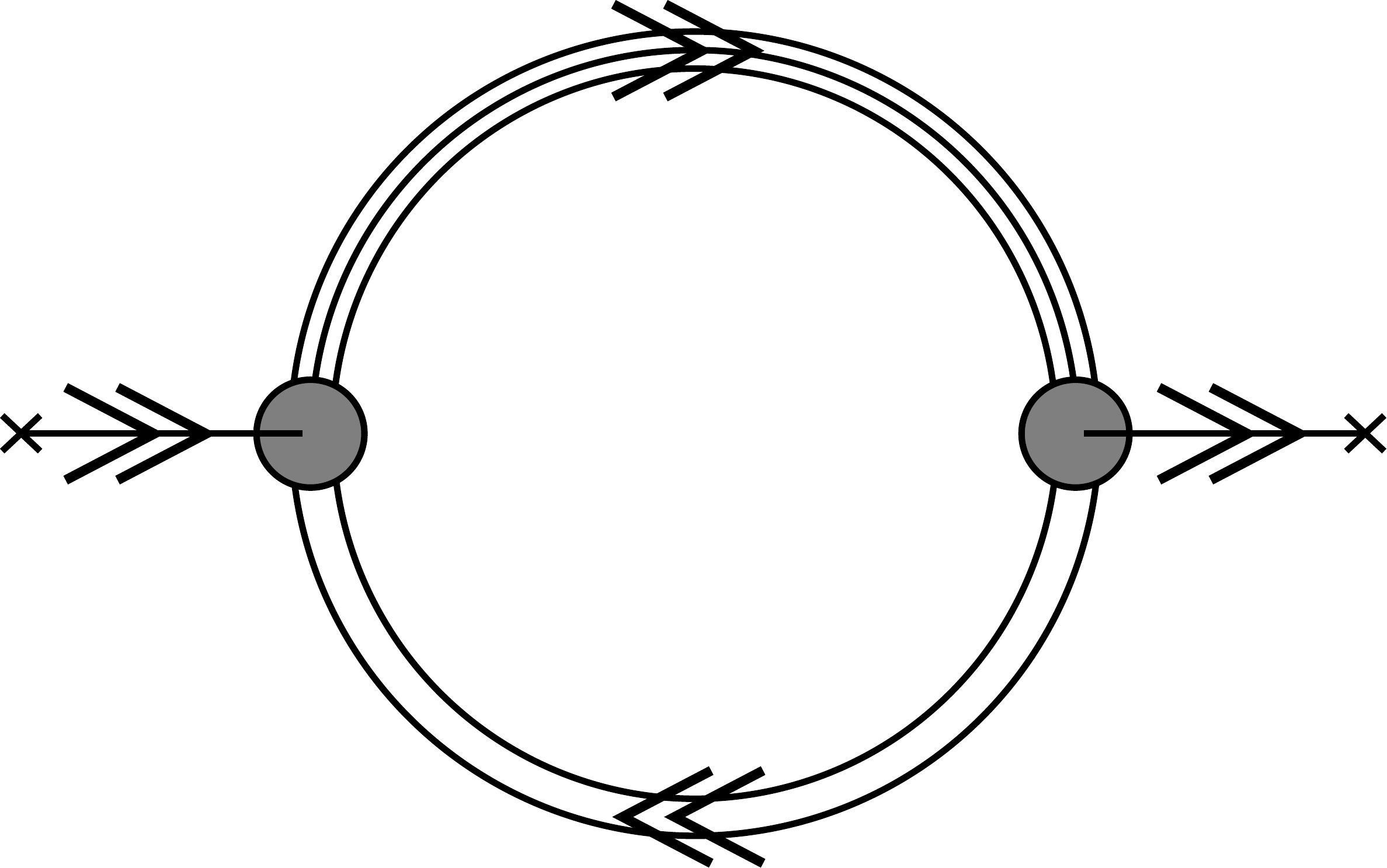}
	\end{minipage}
	\begin{minipage}{0.33\textwidth}
		\centering
		\includegraphics[scale=0.18]{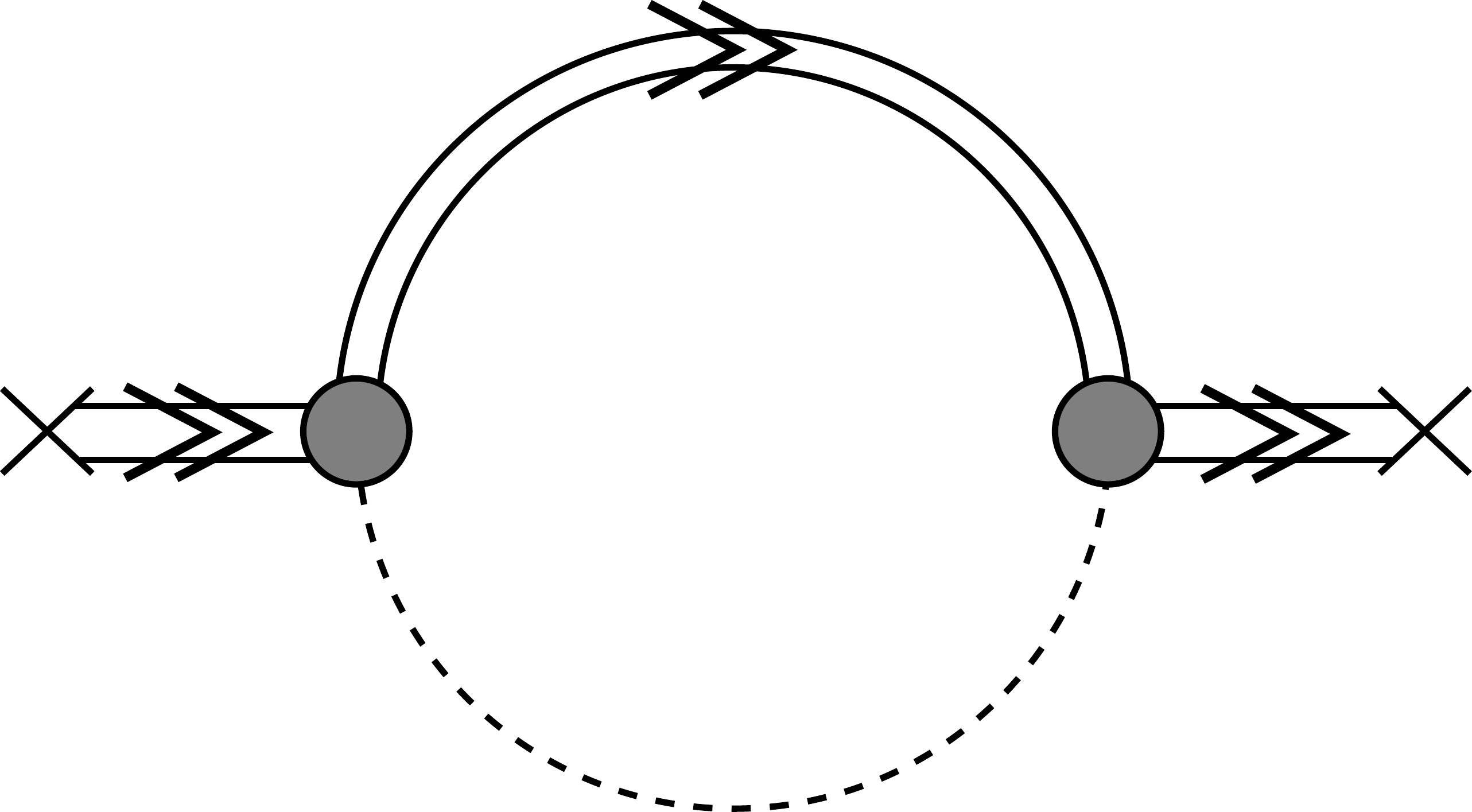}
	\end{minipage}%
	\begin{minipage}{0.33\textwidth}
		\centering
		\includegraphics[scale=0.18]{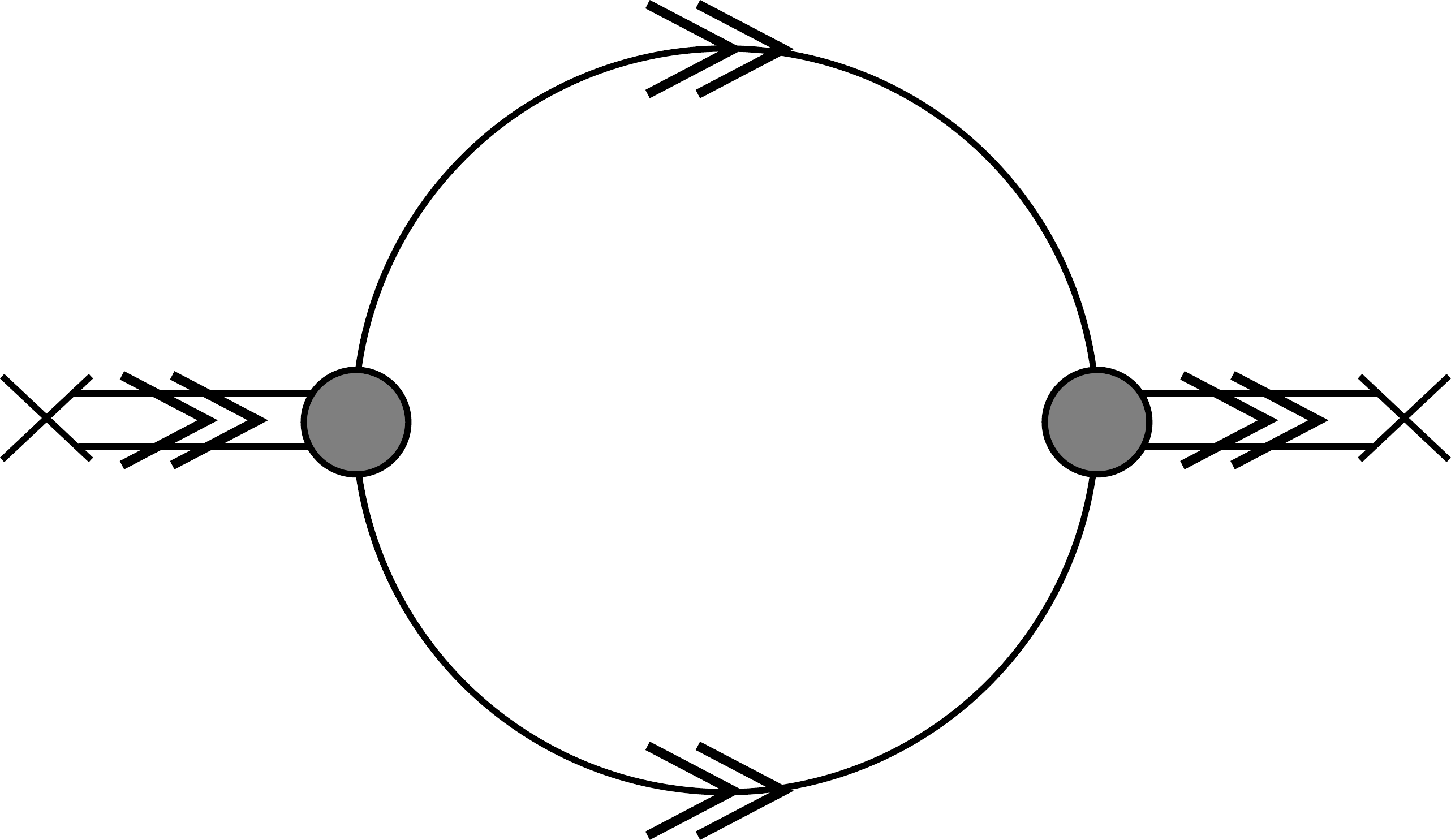}
	\end{minipage}%
	\begin{minipage}{0.33\textwidth}
		\centering
		\includegraphics[scale=0.18]{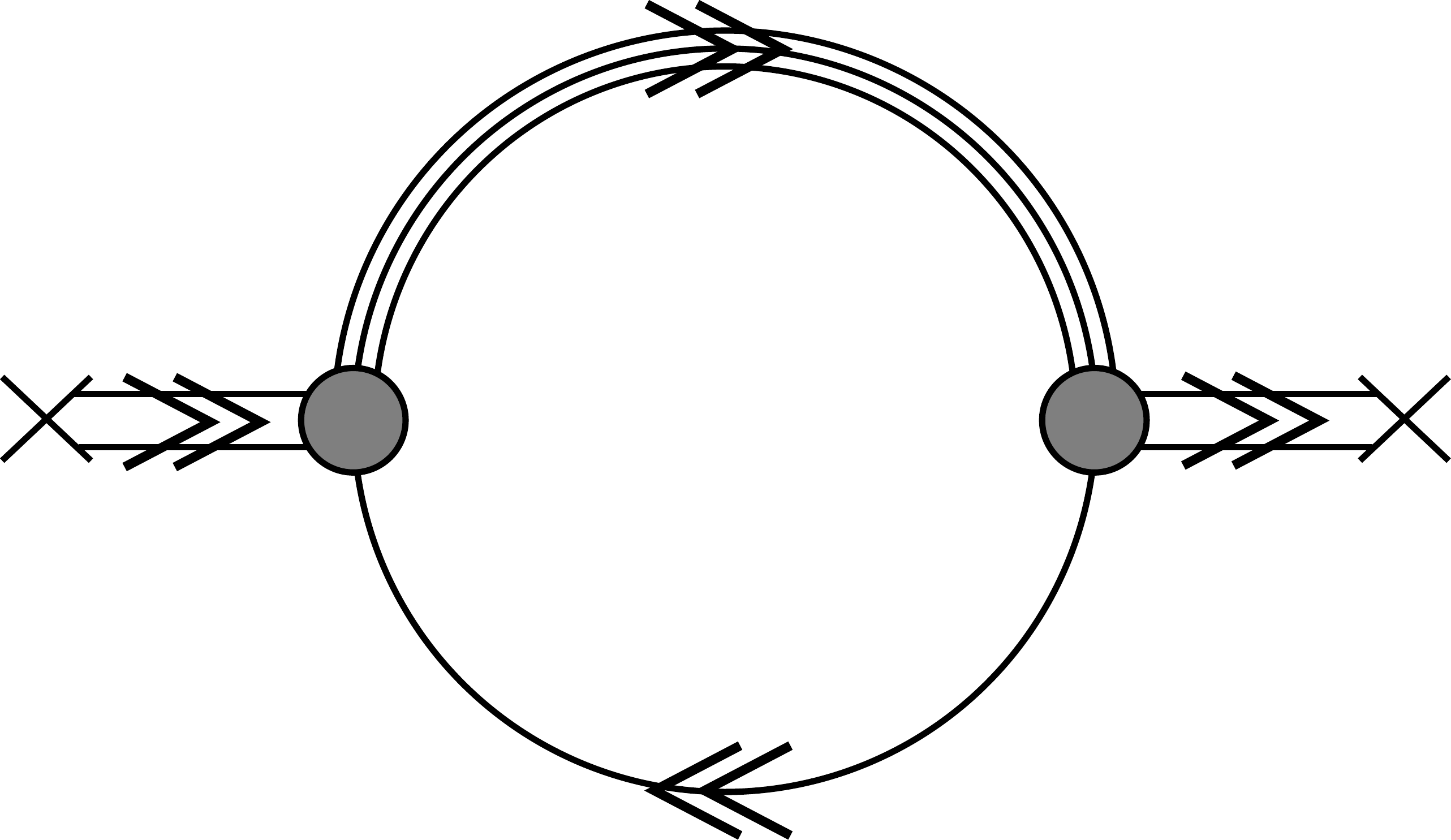}
	\end{minipage}
	\begin{minipage}{0.5\textwidth}
		\centering
		\includegraphics[scale=0.18]{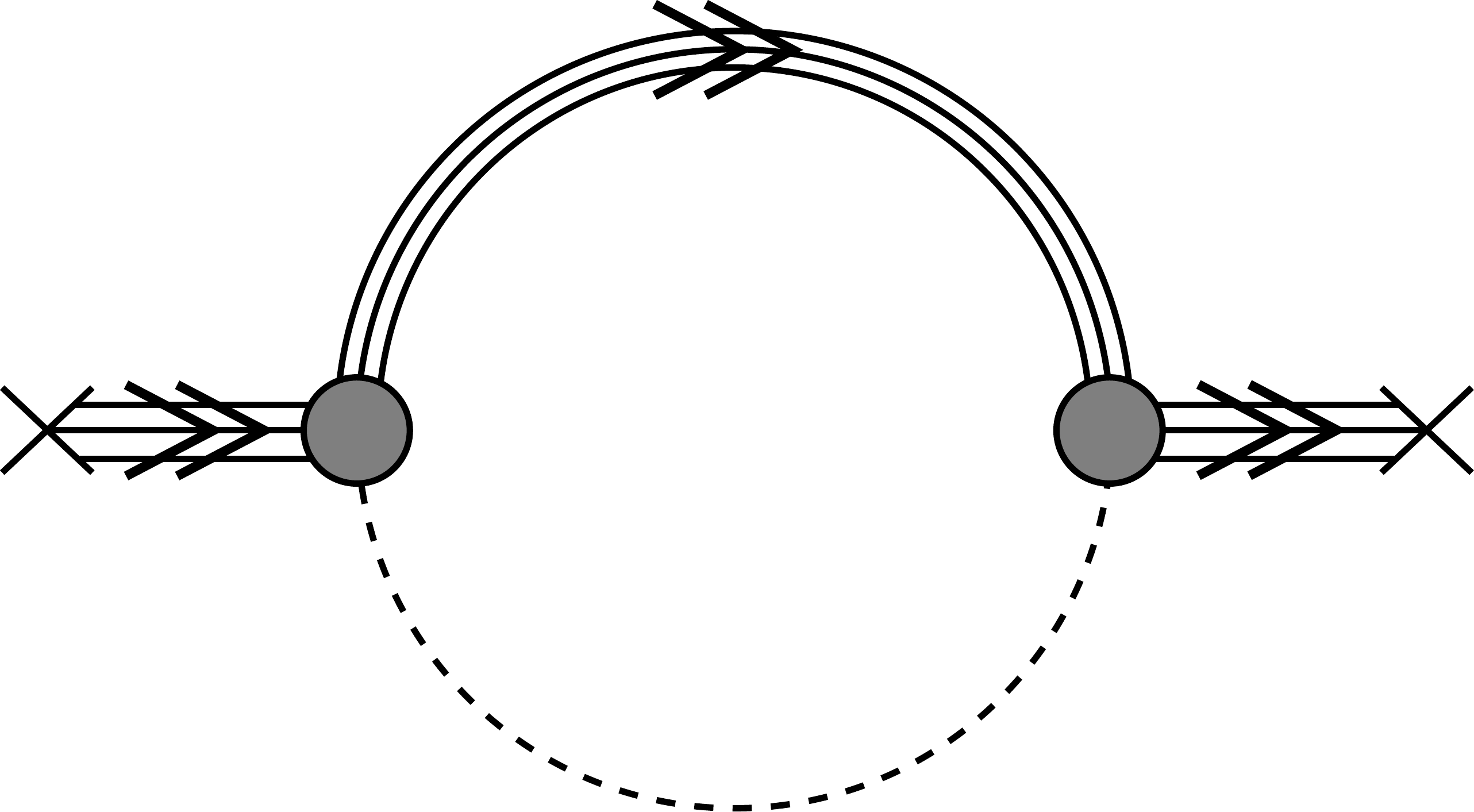}
	\end{minipage}%
	\begin{minipage}{0.5\textwidth}
		\centering
		\includegraphics[scale=0.18]{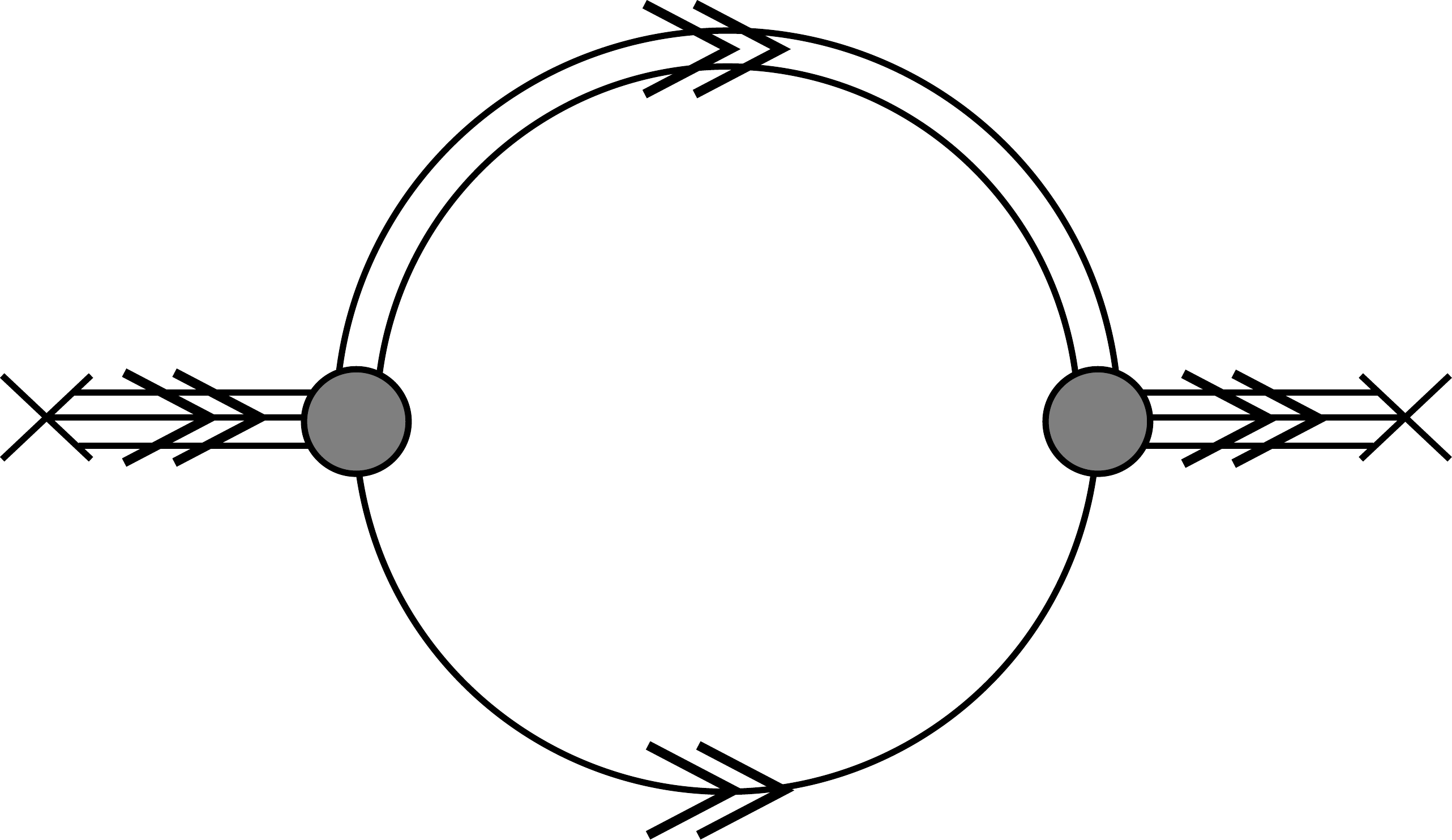}
	\end{minipage}
	\caption{The selfenergy contributions for the Greens functions of the quark-meson-diquark-baryon system,
		defined by the $\Phi$ functional contributions shown in Figure \ref{Fig:2}. 
		From top to down the four rows of diagrams show the selfenergies for the full propagators of mesons, quarks, diquarks and baryons,~respectively.
		\label{fig:5}}
\end{figure}

The resulting Feynman diagrams for the selfenergy contributions are given in Figure \ref{fig:5}.

Note that it is immediately plain from this formulation that in the situation of confinement, when~the propagators belonging to colored excitations (quarks and diquarks) and thus to states that could not be populated would be cancelled, the system simplifies considerably. When all closed loop diagrams containing quarks and diquarks are neglected, this system reduces to a meson-baryon system. 
Out of the five closed-loop diagrams of Figure \ref{fig:4} remains then only the rightmost one from which 
the two selfenergy diagrams in Figure \ref{fig:5} emerge that contain only meson and baryon lines.

\subsection{Relativistic Density Functional Approach to Nuclear Matter}

\begin{figure}[H]
	\centering
	\parbox[h]{1cm}{
	$\mathcal{U} = ~~$}
\parbox[h]{4cm}{
\includegraphics[scale=0.25]{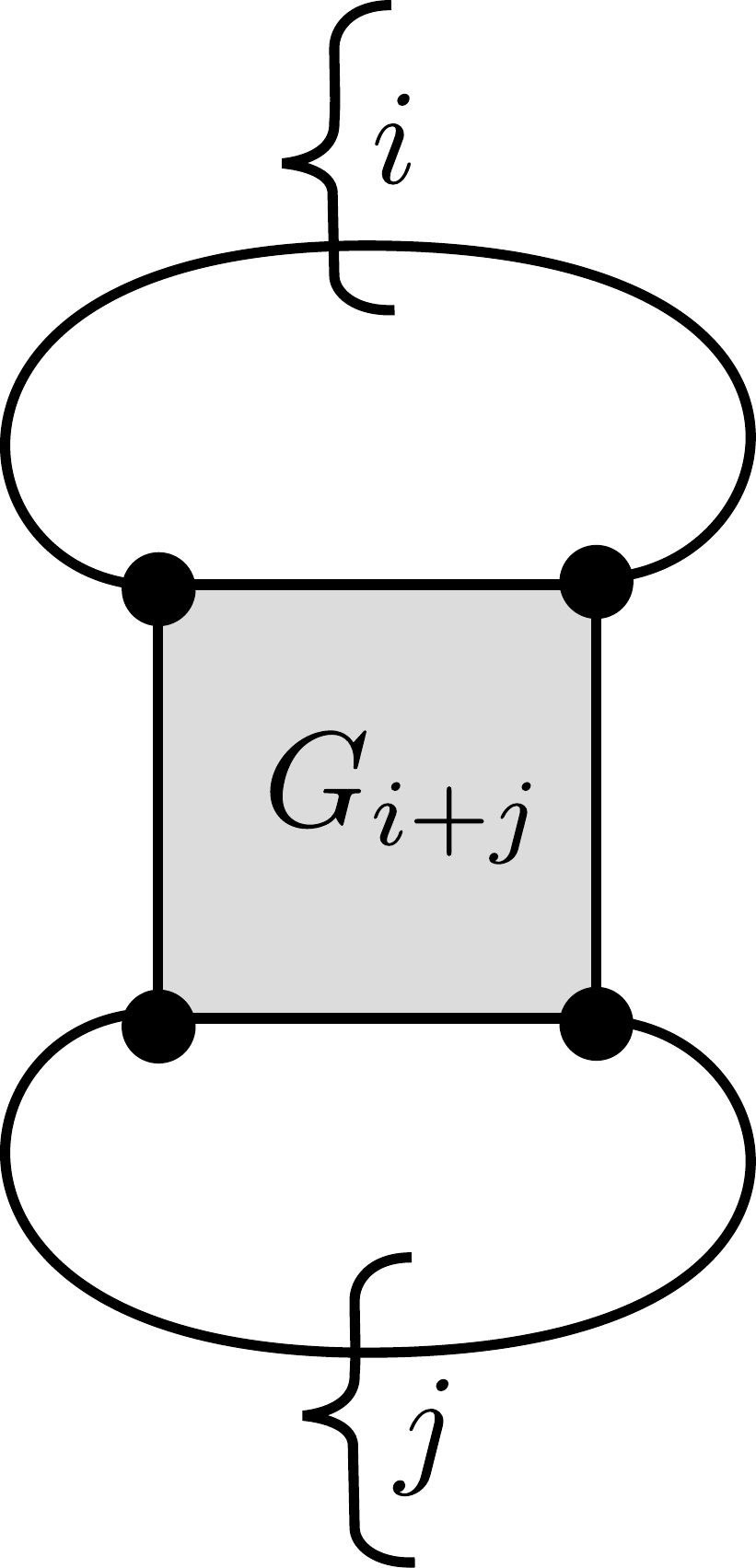}}
	\caption{Effective density-functional for the interaction of species $i$ and $j$ given by a density-dependent coupling $G_{i+j}$ as the local limit of a $\Phi$ functional.}
	\label{U}
\end{figure}

In the limit of quark (and gluon) confinement, the meson-baryon system can be further reduced when the mesonic degrees of freedom are not considered as dynamic ones but just as their meanfield values coupled to the baryon degrees of freedom with effective, possibly density- and temperature-dependent couplings, as sketched in Figure \ref{U}.
In this case, the $\Phi-$derivable approach reduces to a selfconsistent relativistic meanfield theory of nuclear matter, 
\begin{eqnarray}
	\Omega = T\sum_{i=n, p, \Lambda, ...}c_i \left[\mathrm{Tr} \ln S_{i, qu}^{-1} + \sum_{j=S,V} n_{i,j}\Sigma_{i,j} \right] +U\left[\left\{n_{i,S},n_{i,V}\right\}\right]~,
	\label{OmRDF}
\end{eqnarray}
where the functional $U\left[\left\{n_{i,j}\right\}\right]$ stands for the interaction in the system and is expressed in terms of the couplings of bilinears of the baryon spinors that define their scalar and vector densities $n_{i,S}$ and $n_{i,V}$, resp.
The thermodynamic potential (\ref{OmRDF}), where the role of the $\Phi$ functional is now played by the $U$ functional that depends not on the propagators but on the densities and is therefore called a density functional.
The thermodynamic potential (\ref{OmRDF}) shares with the $\Phi-$derivable approach the fulfillment of the conditions of stationarity
\begin{eqnarray}
	\frac{\partial \Omega}{\partial n_{i,S}}=\frac{\partial \Omega}{\partial n_{i,V}}=0~,~~i=n,p,\Lambda,\dots~,
\end{eqnarray} 
and selfconsistency, which is provided by the fact that the (now real) selfenergies in the Dyson equations are obtained as derivatives of the $U$ functional
\begin{eqnarray}
	\frac{\partial U}{\partial n_{i,S}}=\Sigma_{i,S}~,~~\frac{\partial U}{\partial n_{i,V}}=\Sigma_{i,V}~. 
\end{eqnarray}

The baryon quasiparticle propagators $ S_{i, qu}$ fulfill the Dyson equations
$S_{i,qu}^{{-1}}=S_{i,0}^{-1}-\Sigma_{i,S}-\Sigma_{i,V}$, diagrammatically given as

\begin{equation}
\parbox[h][0.6cm][c]{0.1\linewidth}{\includegraphics[scale=0.23]{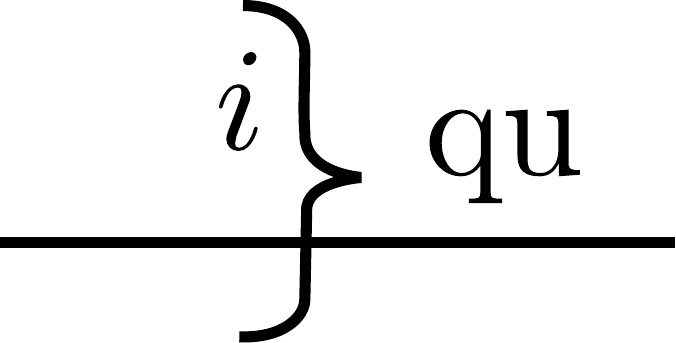}}%
= \;%
\parbox[h][0.6cm][c]{0.1\linewidth}{\includegraphics[scale=0.23]{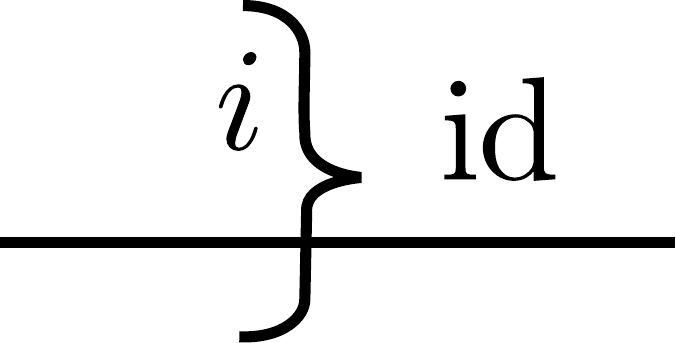}}%
+ \;%
\parbox[h][2.2cm][c]{
	0.3\linewidth}{\includegraphics[scale=0.23]{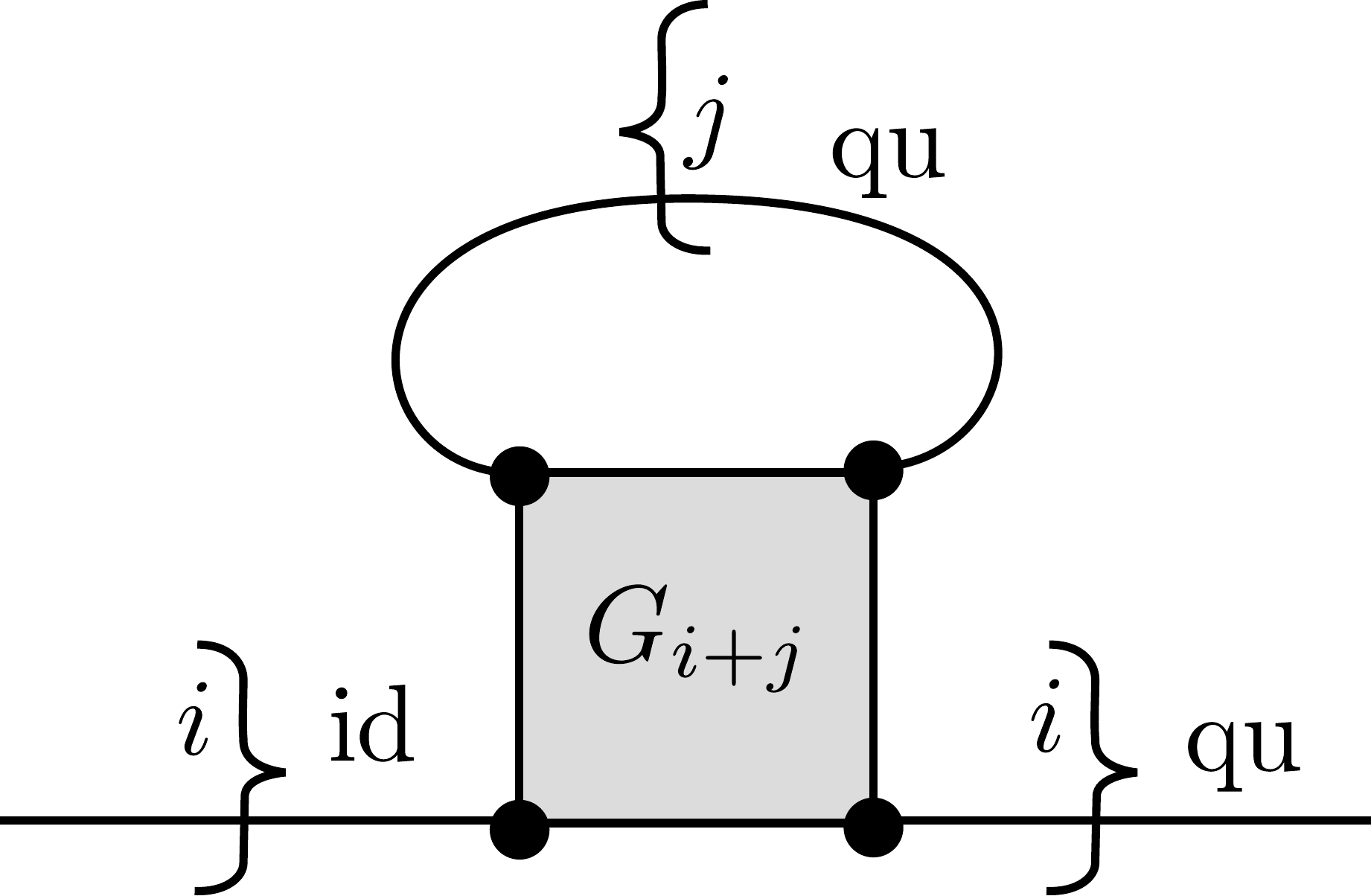}}%
\parbox[h]{1cm}{.}
\label{clusterDSE}
\end{equation}
\vspace{5mm}
 
To this class of relativistic density functional models for baryonic matter belong the density-dependent relativistic meanfield models known as, e.g., DD2, NL3, KVOR, TM1. 
Their nonrelativistic relatives are the density functional models of the Skyrme type.

Recently, these models have been augmented with the inclusion of hadronic excluded volume effects. For an elaborate version of such corrections, see \cite{Typel:2016srf}. The origin of the excluded volume effects is the quark substructure of the baryons which entails the quark exchange effects between baryons that are a consequence of the Pauli principle on the quark level of description.  

\subsection{Quark Pauli Blocking in Hadronic Matter}

The step to cancel all diagrams that contain propagators of colored excitations with the argument of confinement may be a too drastic step when we want to describe matter with a high density so that phase space occupation effects shall become important and the hadronic bound states ``remember'' their quark substructure.
Formally these effects are included into the selfconsistent description of quark-hadron matter within the $\Phi-$derivable approach outlined above in Equation (\ref{Om2}) with the $\Phi$ functional given by the diagrams in Figure \ref{fig:4}. 
In the limit of the confined phase, however, it is important to restore these quark substructure effects as they will drive the system towards deconfinement for sufficiently large densities. 
This can be accomplished by including selfenergy diagrams containing quark and diquark lines in Figure \ref{fig:5} in a perturbative manner by considering their propagators not fully selfconsistently, but in first order with respect to their selfenergies.

\begin{equation}
\label{quark-SE}
\parbox[h][0.6cm][c]{0.1\linewidth}{\includegraphics[scale=0.13]{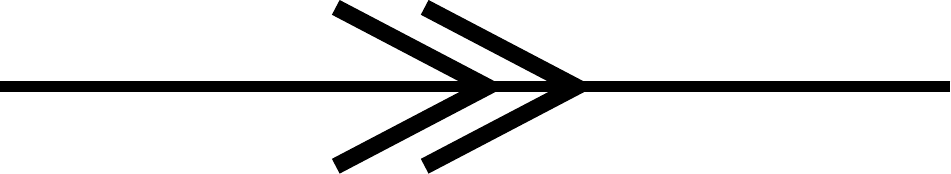}}%
= \;%
\parbox[h][0.6cm][c]{0.1\linewidth}{\includegraphics[scale=0.13]{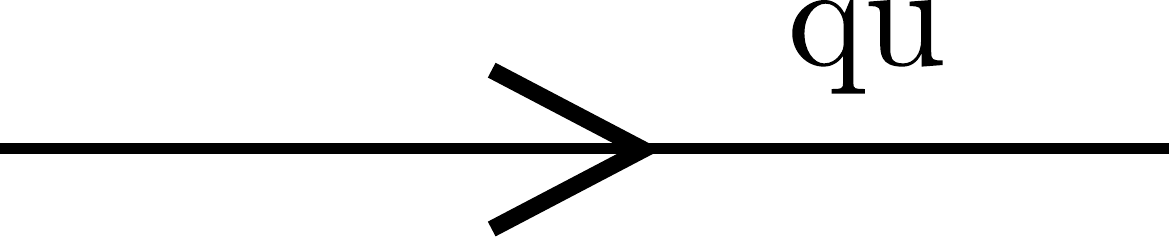}}%
+ \;%
\parbox[h][2.2cm][c]{0.2\linewidth}{\includegraphics[scale=0.13]{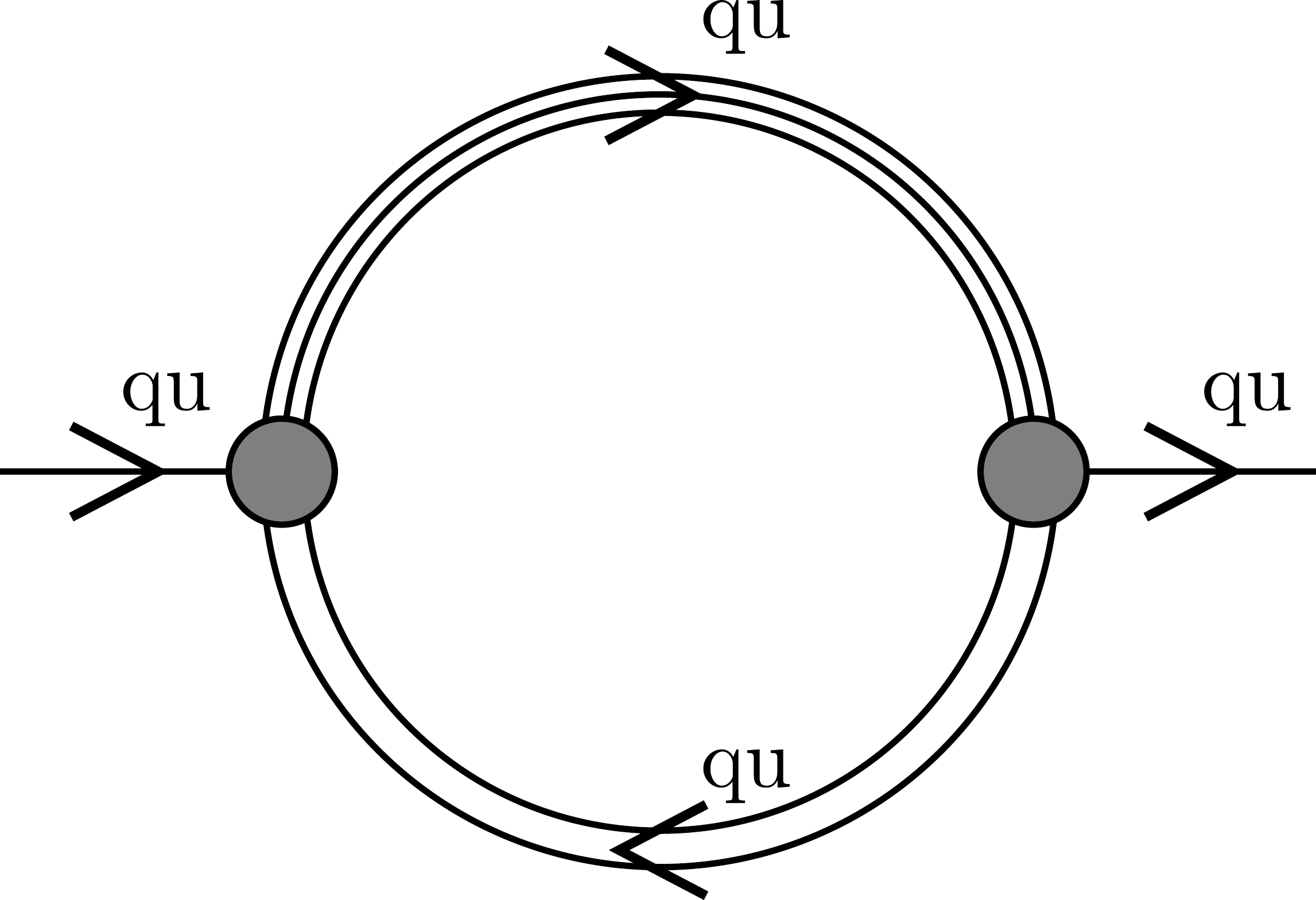}}%
+ %
\mathcal{O} (\Sigma^{(2)})
\end{equation}

\begin{equation}
\label{diquark-SE}
	\parbox[h][0.6cm][c]{0.1\linewidth}{\includegraphics[scale=0.13]{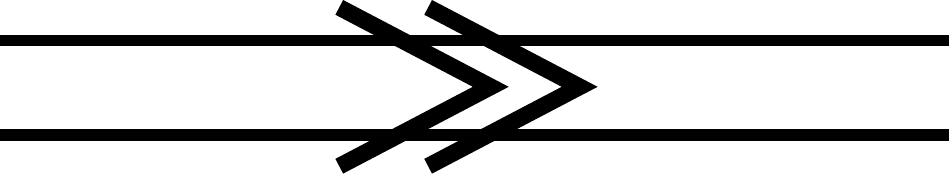}}%
	= \;%
	\parbox[h][0.6cm][c]{0.1\linewidth}{\includegraphics[scale=0.13]{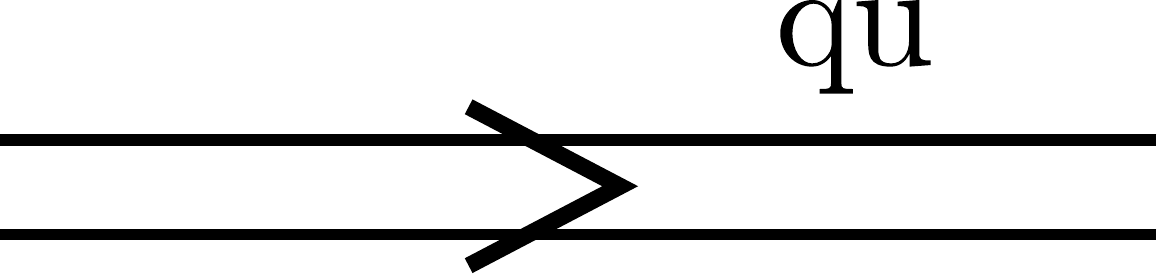}}%
	+ \;%
	\parbox[h][2.2cm][c]{0.2\linewidth}{\includegraphics[scale=0.13]{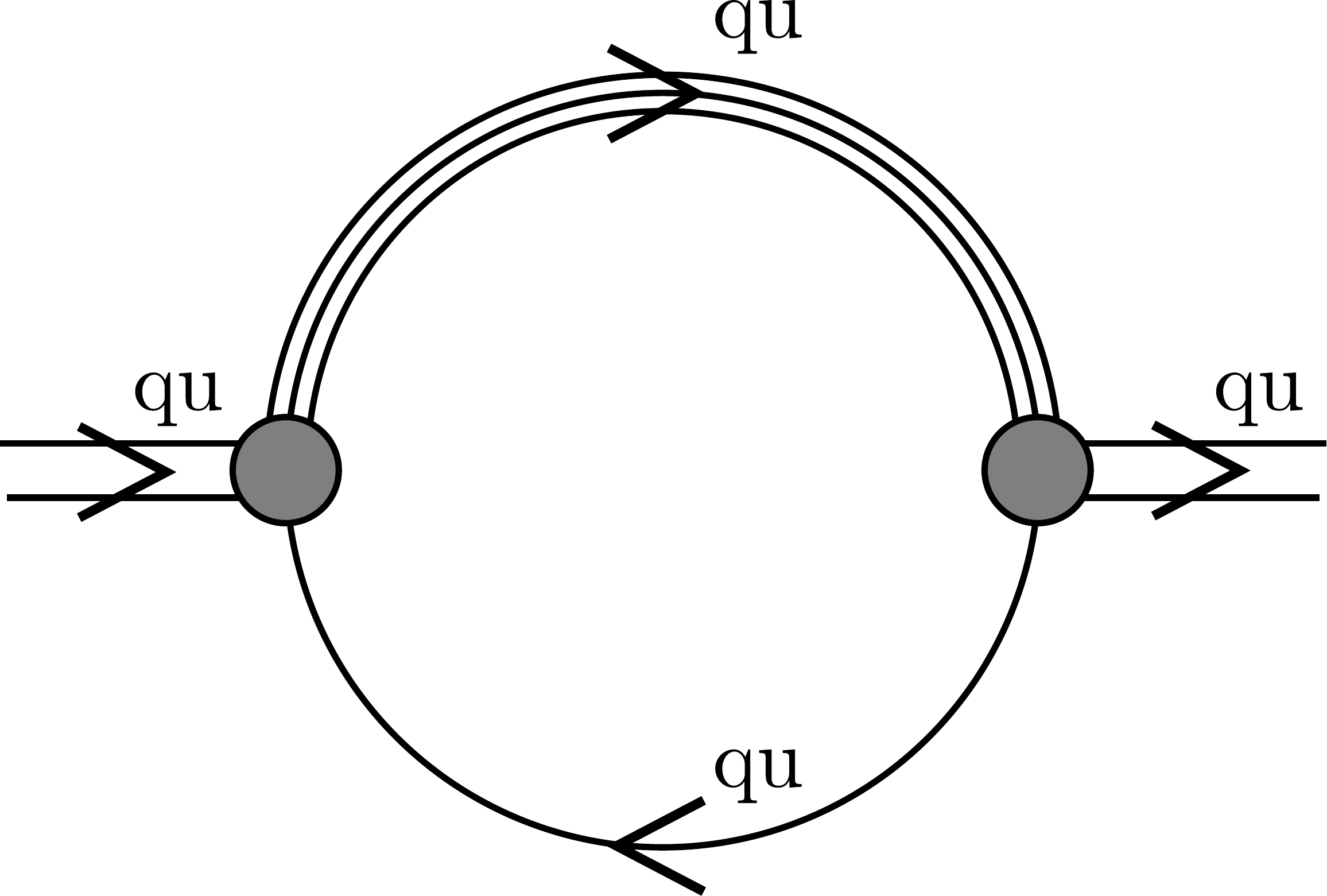}}%
	+ %
	\mathcal{O} (\Sigma^{(2)})
\end{equation}

Here the quark and diquark quasiparticle propagator lines in Equations~(\ref{quark-SE}) and (\ref{diquark-SE}) are defined by a Dyson-Schwinger equation as given generically in  Equation (\ref{clusterDSE}), but with a density functional for the effective interaction that is appropriate for the decription os quark matter and will be discussed in detail in Section~\ref{ssec:quarkRDF}. 

In such a way two quark-diquark substructure contributions to the baryon selfenergy appear due to the corrections (\ref{quark-SE}) and (\ref{diquark-SE}) beyond the quasiparticle approximation for  quark and diquark propagators
\begin{equation}
	\label{Quark-Diquark-Pauli}
	\parbox[h][0.6cm][c]{0.22\linewidth}{\includegraphics[scale=0.13]{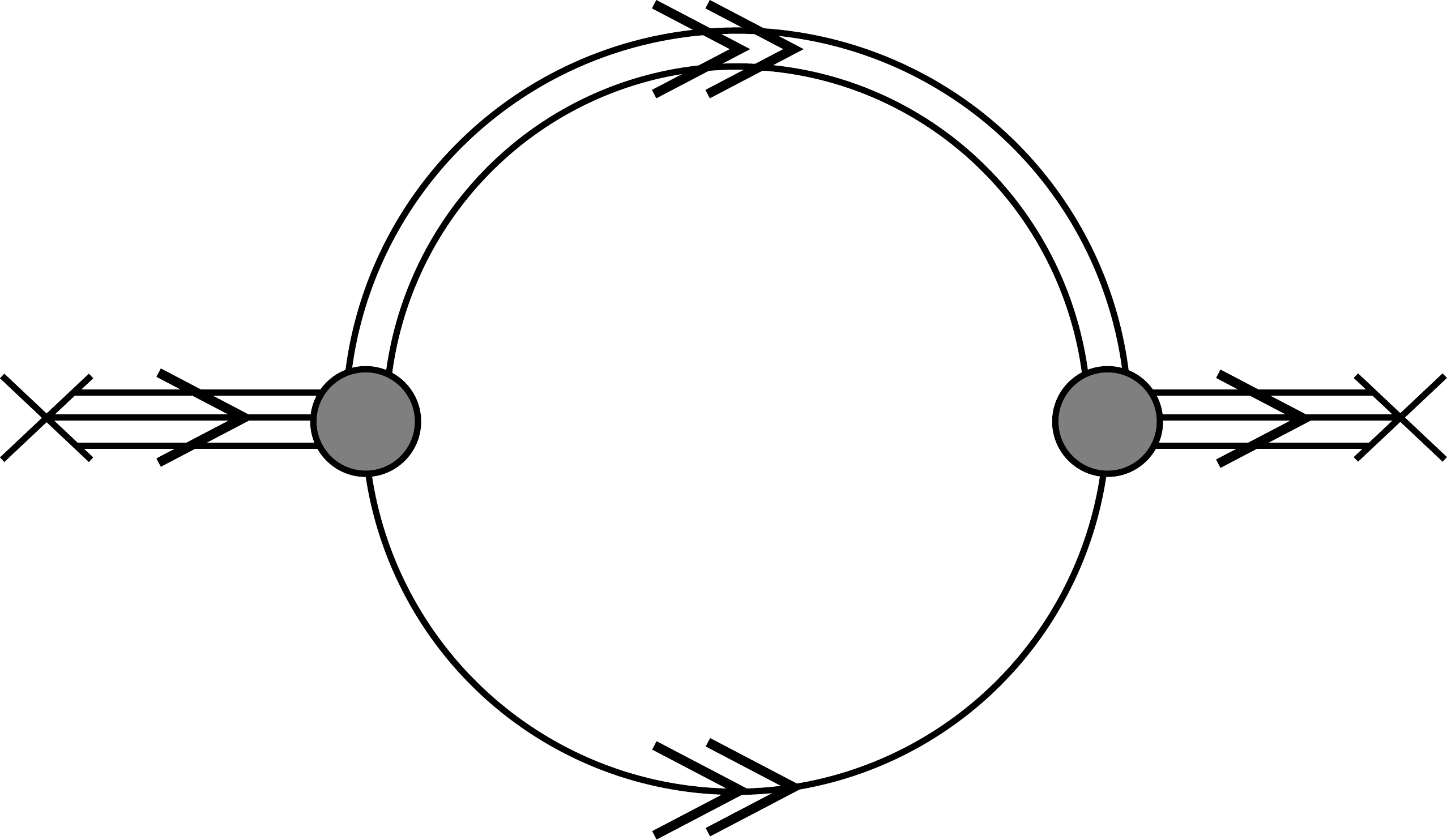}}%
	= \;%
	\Omega (\Sigma^{(0)})
	+ \;%
	\parbox[h][2.2cm][c]{0.2\linewidth}{\includegraphics[scale=0.13]{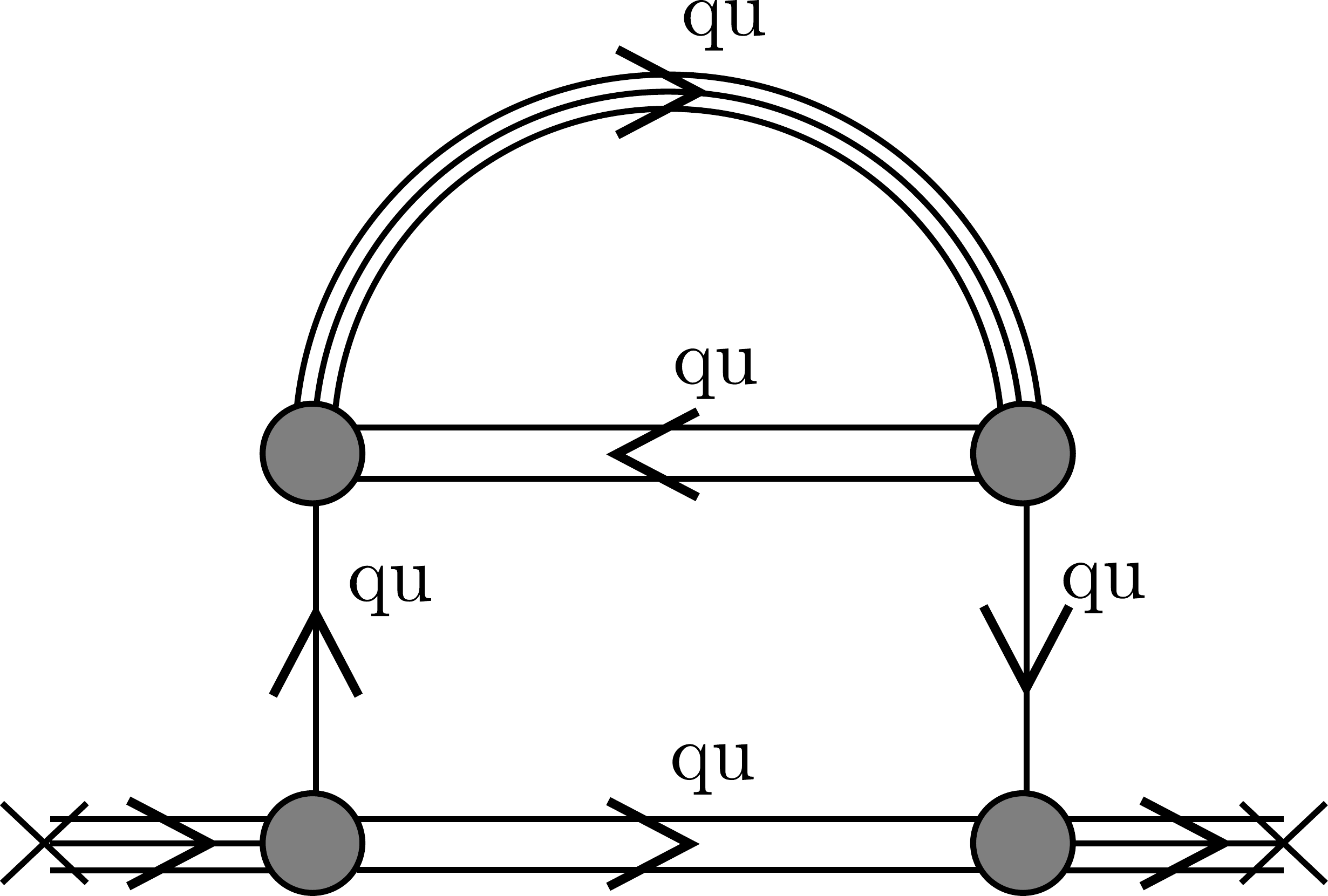}}%
	+ \;%
\parbox[h][0.6cm][c]{0.2\linewidth}{\includegraphics[scale=0.13]{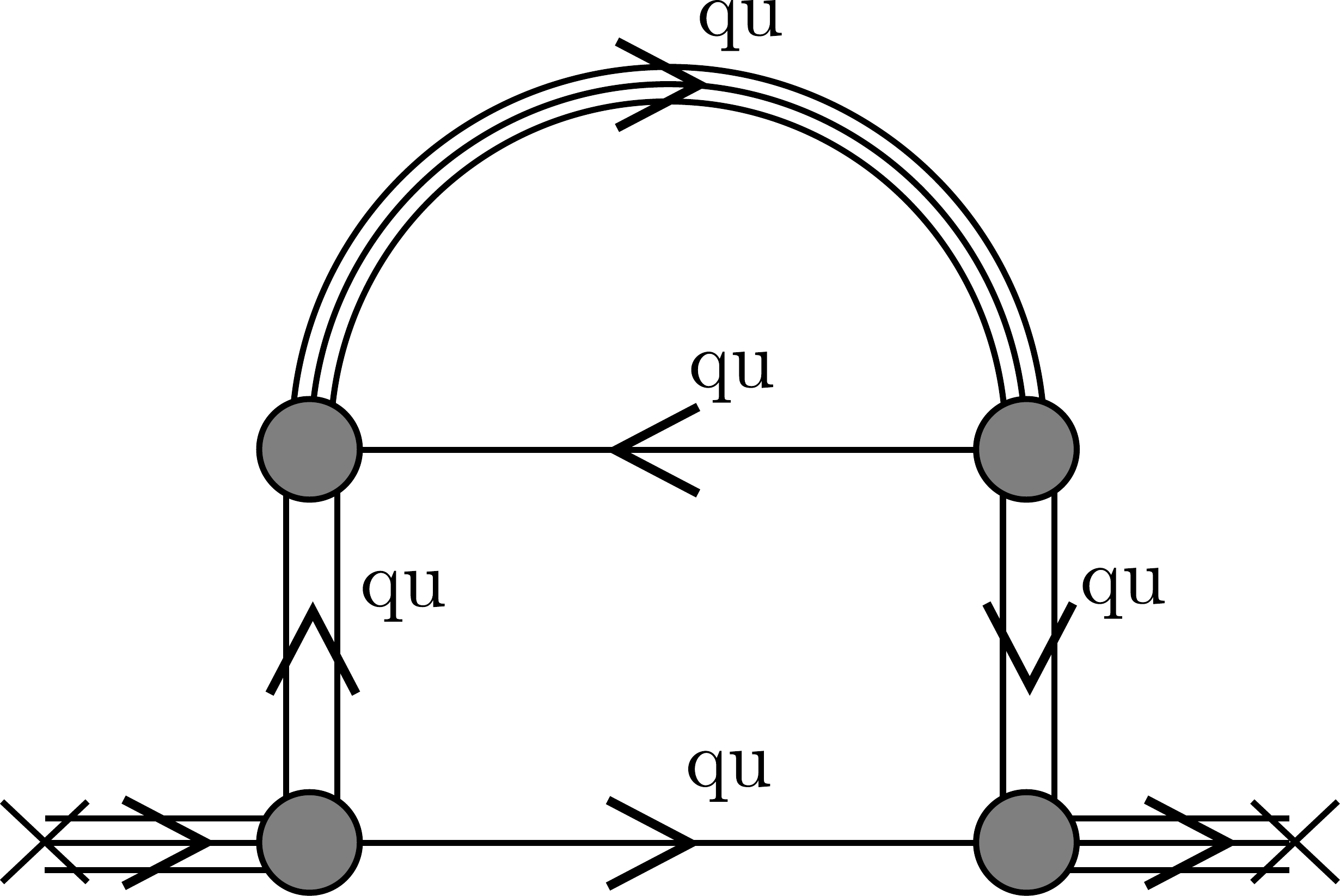}}%
	+ %
	\mathcal{O} (\Sigma^{(2)})~.
\end{equation}

They contain one closed baryon line and are therefore of first order in the baryon density. 
By~functional derivative w.r.t. the baryon propagator line (cutting) an effective quark and diquark exchange interaction can be obtained from those contributions to the baryon selfenergy shown in Equation (\ref{Quark-Diquark-Pauli}).
The diagram for the quark exchange interaction between baryons resulting from cutting the baryon line in the first of the two diagrams in Equation (\ref{Quark-Diquark-Pauli}) is shown in (\ref{qex}) in two forms which are topologically equivalent,
\begin{equation}
\label{qex}
	\parbox[h][0.6cm][c]{0.2\linewidth}{\includegraphics[scale=0.13]{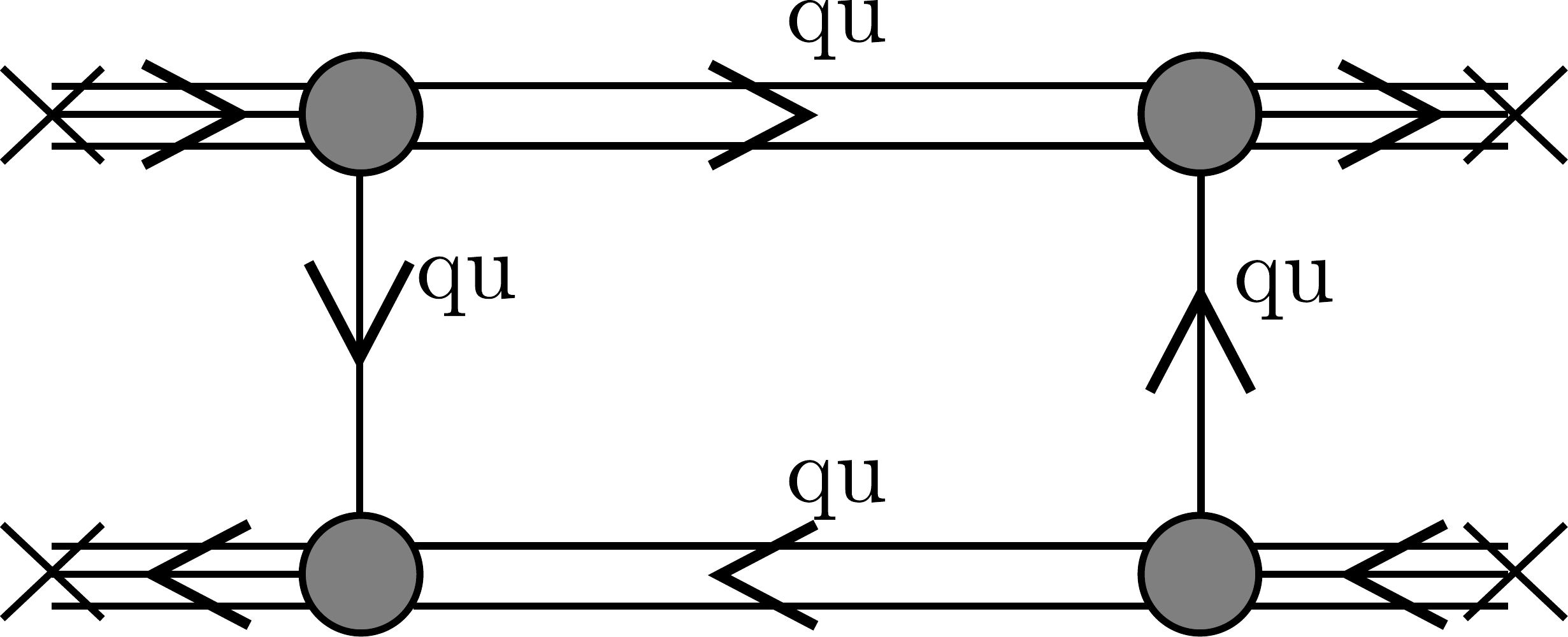}}%
	\Leftrightarrow \;%
	\parbox[h][2.2cm][c]{0.2\linewidth}{\includegraphics[scale=0.13]{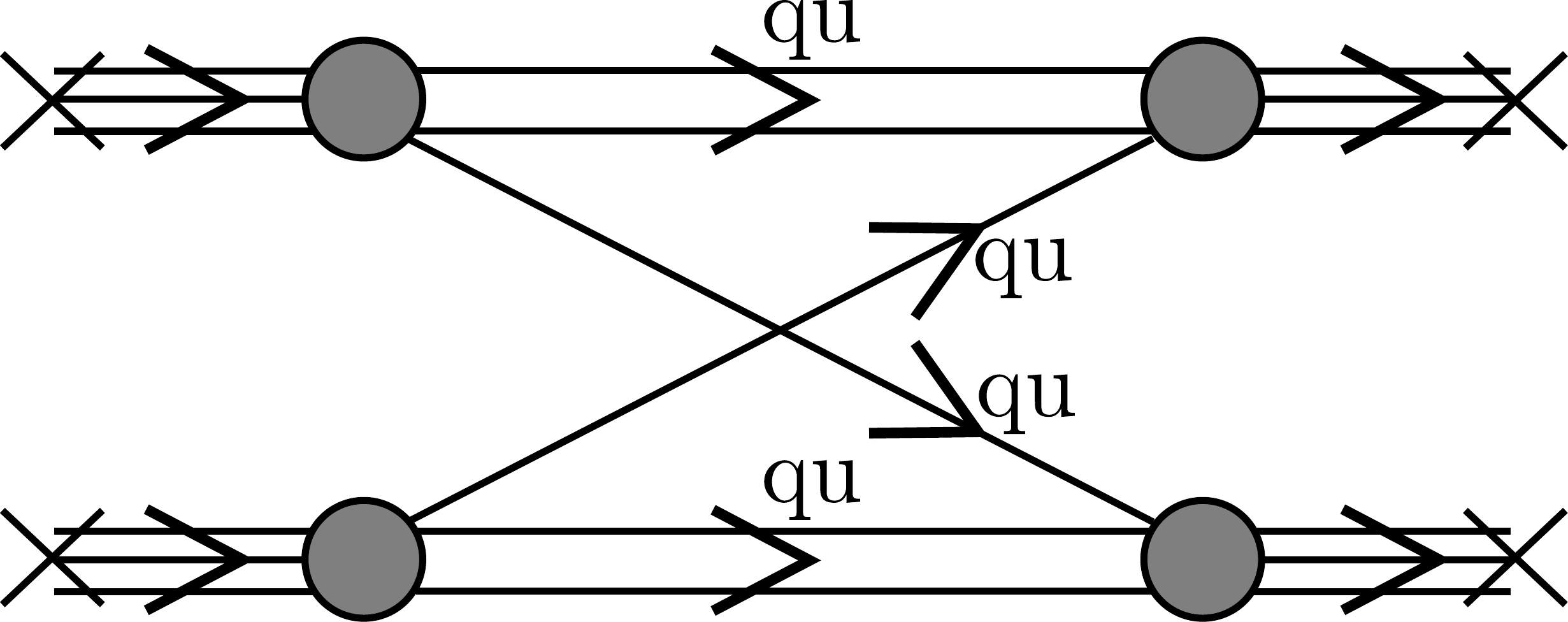}}~.
\end{equation}

Analogously, the diquark exchange interaction is obtained by cutting the baryon line in the second of the two diagrams in Equation (\ref{Quark-Diquark-Pauli}).

The quark Pauli blocking effects in nuclear matter have been evaluated in a nonrelativistic approximation with a confining potential model in Refs.~\cite{Ropke:1986qs,Blaschke:1988zt} where it was found that the result for the repulsive density-dependent nucleon-nucleon interaction corresponds well to the repulsive part of the effective Skyrme interaction functional in Ref.~\cite{Vautherin:1971aw}.
Note that the resulting EoS has been successfully applied in predicting massive hybrid stars with quark matter cores \cite{Blaschke:1989nn}.
A flaw of the nonrelativistic calculations of the quark Pauli blocking effect is that the quark mass is a fixed parameter so that partial chiral symmetry restoration in dense hadronic matter as a selfenergy effect on the quark propagator (consistent with the quark exchange) is not accounted for. This question has recently been taken up by Blaschke, Grigorian and R\"opke who demonstrated that a chirally improved calculation of the quark Pauli blocking effect results in an EoS for nuclear matter that is similar to the DD2 model with excluded volume \cite{Typel:2016srf}.    

Strange quarks as well as strange hadrons belong to the system of our model, as it is formulated for any flavour.
Interesting new aspects, which are expected from this approach concern, for instance, the Pauli blocking between baryons, including hyperons, in dense matter.
This shall be of relevance for the discussion of the hyperon puzzle in compact stars \cite{Klaehn:2017mux,Bastian:2017fzo}.

The quark Pauli blocking effect applies also to mesons and corresponding expressions for selfenergy effects can be extracted from the $\Phi-$derivable approach in a similar manner as for the baryons. This has been outlined in Ref.~\cite{Dubinin:2017iqc}.
In a nonrelativistic potential model calculation, an~effective quark exchange potential for the $\pi$-$\pi$ interaction has been derived \cite{Blaschke:1992qa} which reproduces the scattering length of the pion interaction in the isospin = 2 channel, see also 
\cite{Barnes:1991em}. 

Let us now turn to the other limit of the $\Phi-$derivable approach to a unified EoS for quark-hadron matter, the case of deconfined quarks. In this case, also chiral symmetry is restored, and due to the resulting lowering of the mass threshold the meson and baryon states become unbound (Mott effect). 
Their contribution to the thermodynamics as captured in the corresponding phase shift functions is gradually vanishing at high temperatures and chemical potentials with just chiral quark matter remaining asymptotically.
As a paradigmatic example for the treatment of Mott dissociation of hadrons in hot, dense matter let us consider the case of pion dissociation in hot quark matter.

\subsection{Mott Dissociation of Pions in Quark Matter}
\begin{figure}[H]
	\begin{minipage}[t]{0.33\textwidth}
		\centering
		\includegraphics[scale=0.2]{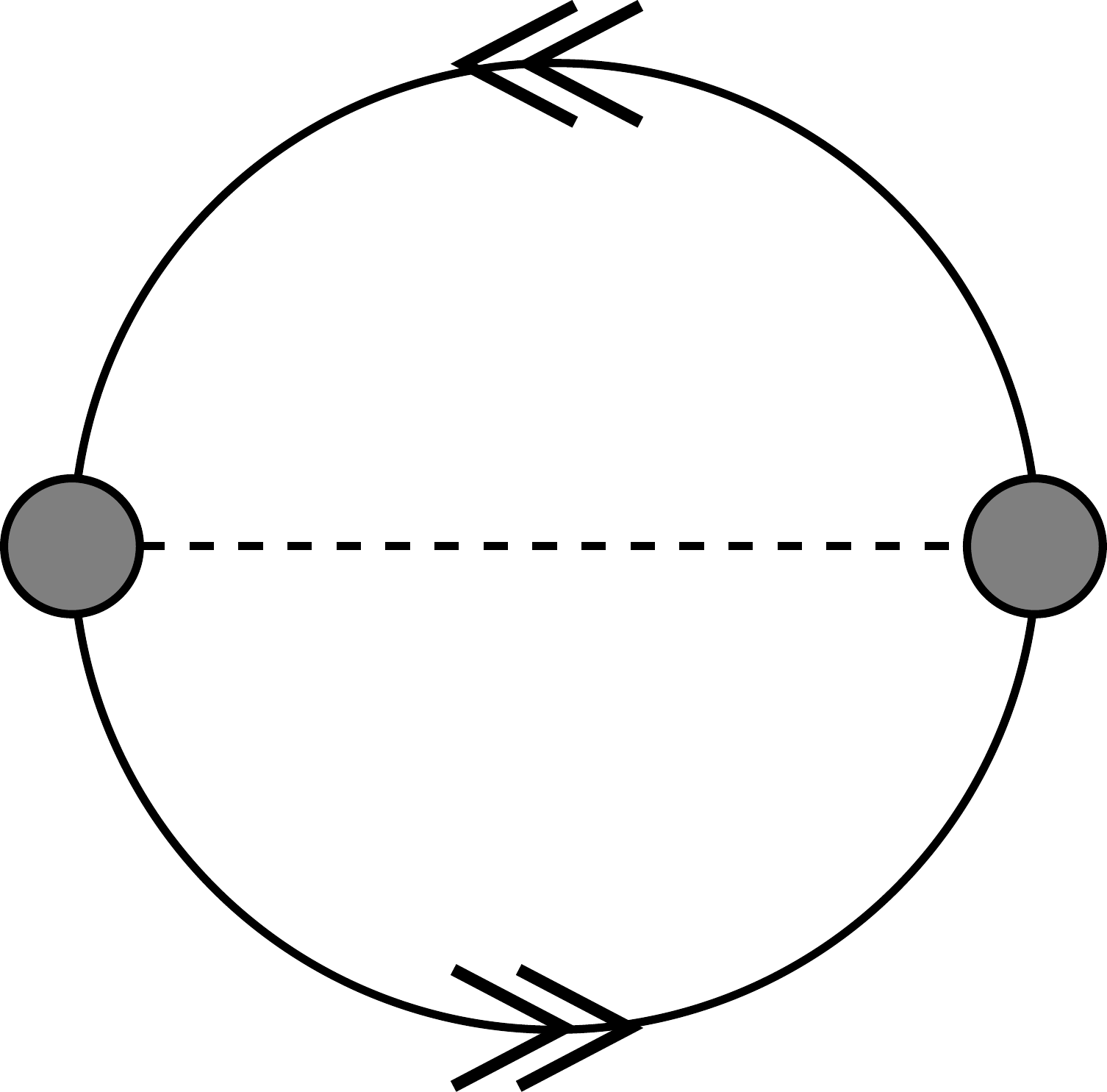}
	\end{minipage}%
	\begin{minipage}[t]{0.33\textwidth}
		\centering
		\includegraphics[scale=0.2]{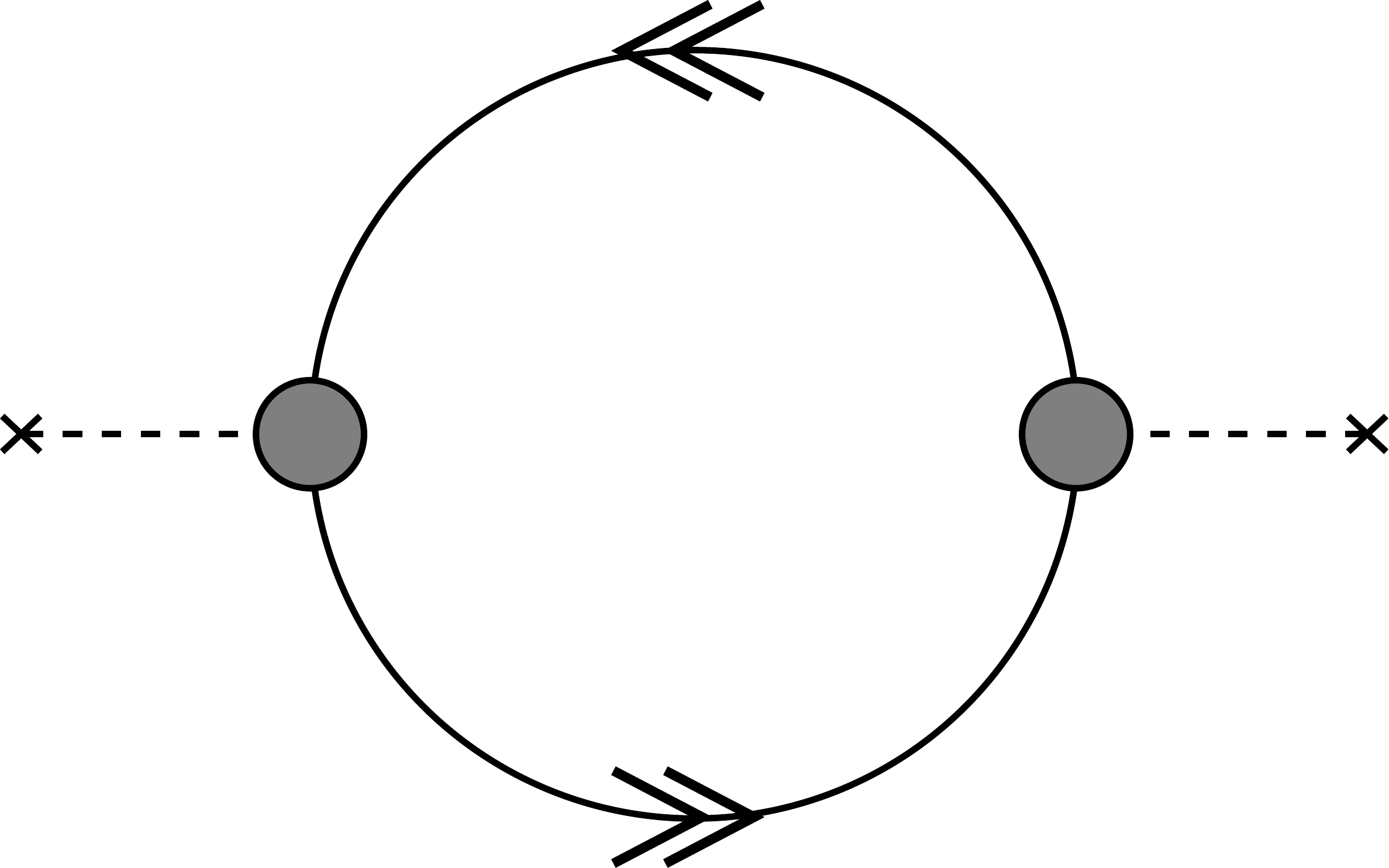}
	\end{minipage}%
	\begin{minipage}[t]{0.33\textwidth}
		\centering
		\includegraphics[scale=0.2]{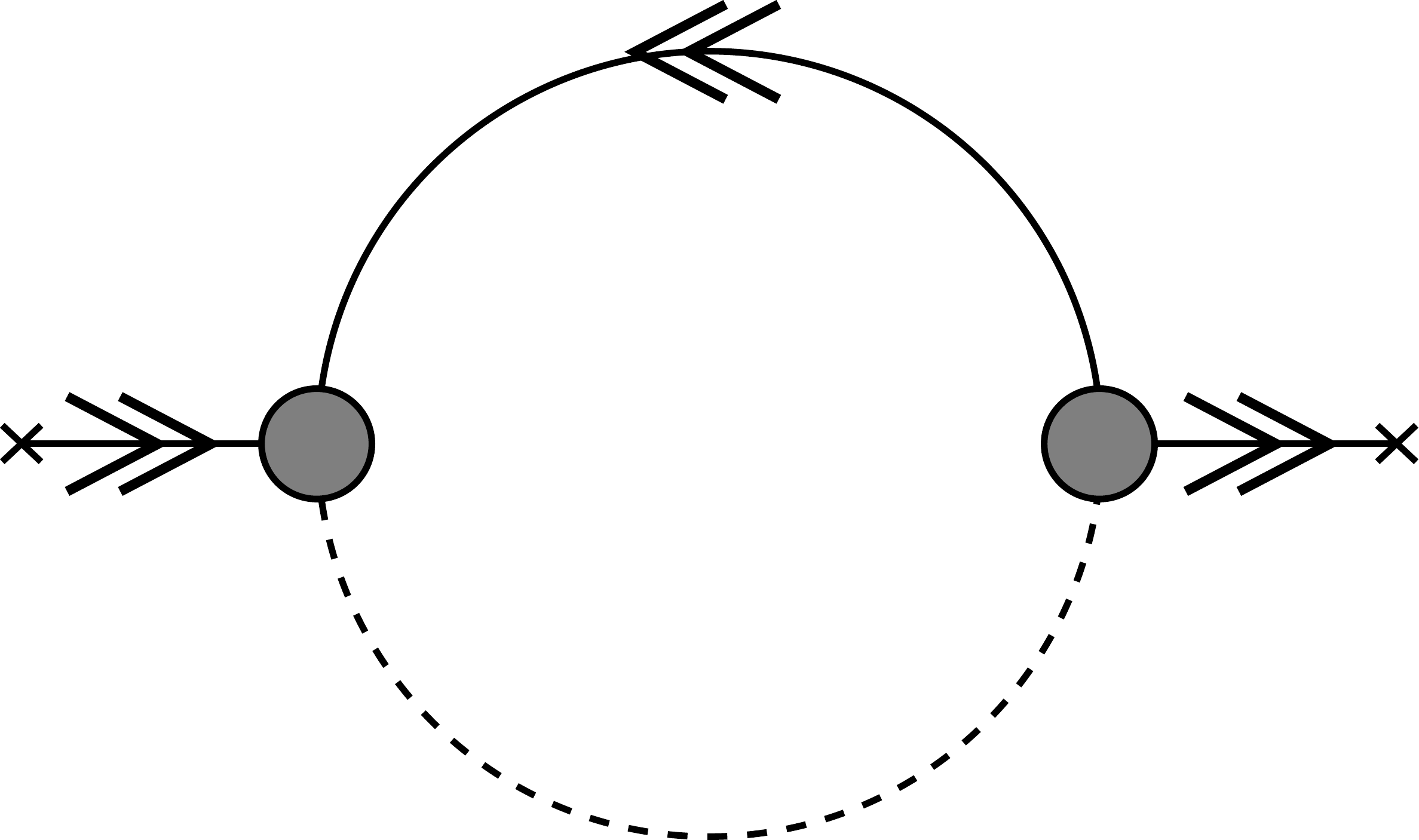}
	\end{minipage}
\caption{The $\Phi$ functional (left panel) for the case of mesons in quark matter, where the bosonic meson propagator is defined by the dashed line and the fermionic quark propagators are shown by the solid lines with arrows.  The corresponding meson and quark selfenergies are shown in the middle and right panels, respectively.} 
\label{fig:11}
\end{figure}

In order to describe the problem of mesons in quark matter within the $\Phi-$derivable approach we define the $\Phi$ functional and the corresponding selfenergy in Figure \ref{fig:11}.

\begin{figure}[H]
	\centering
	\includegraphics[scale=0.4]{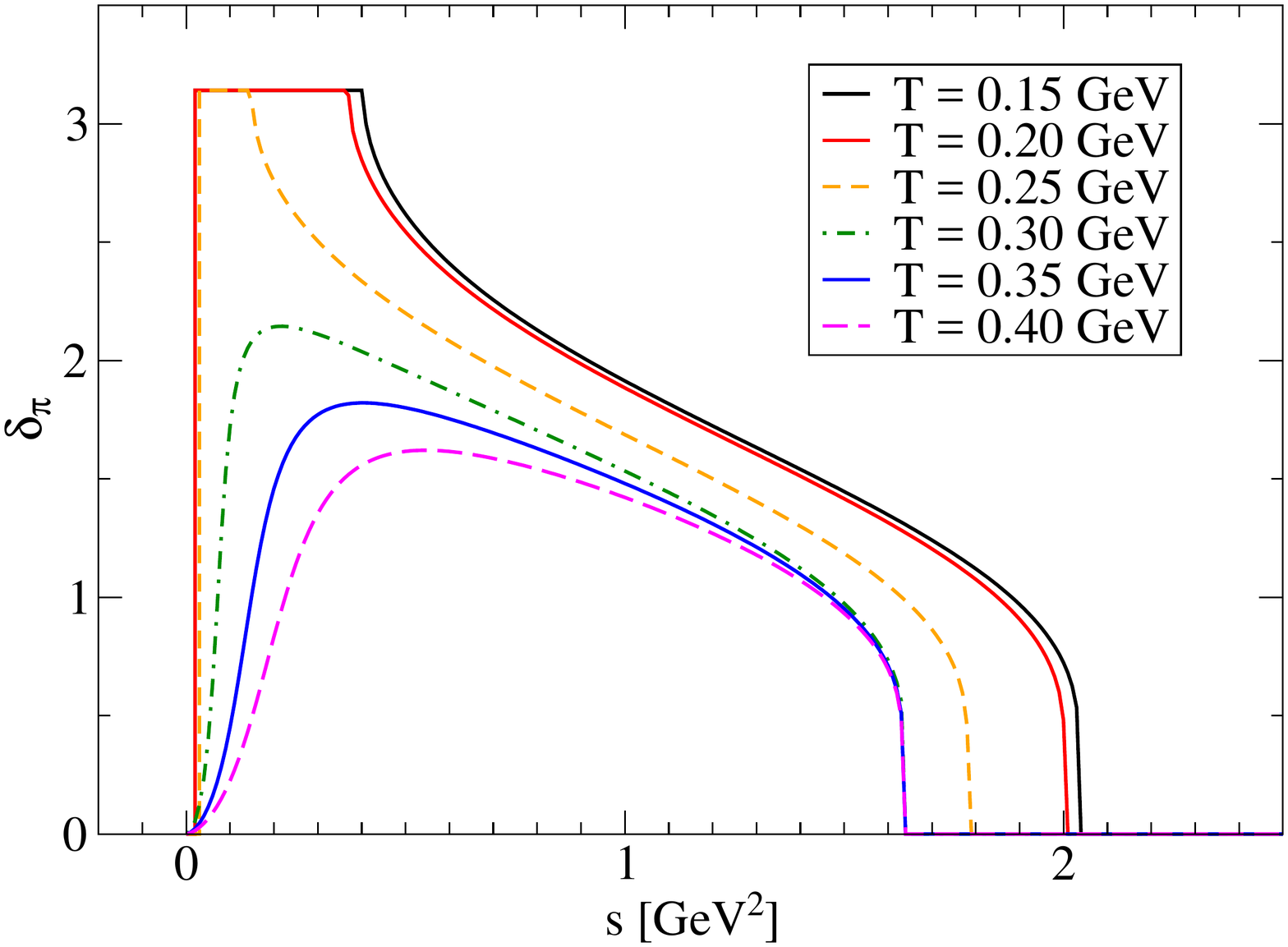}
	\caption{Phase shift of the pion as a quark-antiquark state for different temperatures, below and above the Mott dissociation temperature, from Ref.~\cite{Blaschke:2014zsa,Blaschke:2013zaa}.}
	\label{fig:pion}
\end{figure}

The meson polarization loop $\Pi_M(q,z)$ in the middle panel of Figure \ref{fig:11} enters the definition of the meson T matrix (often called propagator) 
\begin{equation}
	T^{-1}_M(q,\omega+i\eta) = G_S^{-1} - \Pi_M(q,\omega+i\eta)
	=|T_M(q,\omega)|^{-1}{\rm e}^{- i \delta_M(q,\omega)}~,
\end{equation}
which in the polar representation introduces a phase shift 
$\delta_M(q,\omega)=\arctan (\Im T_M/\Re T_M)$, that results in a generalized Beth-Uhlenbeck equation of state for the thermodynamics of the consistently coupled quark-meson system \cite{Blaschke:2013zaa}.
\begin{equation}
	\Omega = \Omega_{\rm MF} + \Omega_M ~, 
\end{equation}
where the selfconsistent quark meanfield contribution is
\begin{equation}
	\Omega_{\rm MF} =
	\frac{\sigma^2_{\rm MF}}{4G_S} - 2N_c N_f \int\frac{d^3p}{(2\pi)^3} 
	\left[ E_p 
	+T \ln\left(1+{\rm e}^{-(E_p-\Sigma_+-\mu)/T} \right) 
	+T \ln\left(1+{\rm e}^{-(E_p+\Sigma_-+\mu)/T} \right) 
\right]
 ~, 
\end{equation}
with the quasiparticle energy shift for quarks (antiquarks) due to mesonic correlations given by 
\mbox{$\Sigma_\pm=\sum_{M=\pi,\sigma} {\rm tr}_D \left[ \Sigma_M\Lambda_\pm \gamma_0\right]/2$} and 
the positive (negative) energy projection operators \mbox{$\Lambda_\pm=(1\pm \gamma_0)/2$}.
The mesonic contribution to the thermodynamics is
\begin{equation}
	\Omega_M = d_M \int \frac{d^3 k}{(2\pi)^3} \int\frac{d\omega}{2\pi}
	\left\{
	\omega + 2 T \ln \left[1 - {\rm e}^{-\omega/T} \right] \right\}2\sin^2\delta_M(k,\omega) ~
	\frac{\delta_M(k,\omega)}{d\omega}~,
\end{equation}
where similar to the case of deuterons in nuclear matter the factor $2\sin^2\delta_M$ accounts for the 
fact that mesonic correlations in the continuum are partly already accounted for by the selfenergies
$\Sigma_M$ defining the improved selfconsistent quasiparticle picture.  
In the previous works of  Refs.~\cite{Hufner:1994ma,Zhuang:1994dw,Blaschke:2013zaa,Yamazaki:2012ux,Wergieluk:2012gd} on this topic, however, the effect of the backreaction from mesonic correlations on the quark meanfield thermodynamics had been disregarded. 
In Figure \ref{fig:pion} we show the phase shift of the pion as a quark-antiquark state for different temperatures, below and above the Mott dissociation temperature. The shape of these functions and their evolution with increasing temperature over the Mott dissociation resembles the similar behaviour of the deuteron phase shift in nuclear matter at increasing density, see~Figure \ref{Phaseshifts}.
\vspace{-18pt}

Here we note from the $\Phi-$derivable approach that for consistency the quark propagator in the quark meanfield thermodynamic potential shall contain effects from the selfenergy $\Sigma_M$ due to the coupling to the mesonic correlations as in the right panel of Figure \ref{fig:11}. 
This total quark selfenergy is the given by $\Sigma({\bf p},p_0)=\sigma_{\rm MF} + \Sigma_M({\bf p},p_0)$,
where for a local NJL model with scalar coupling constant $G_S$ the meanfield contribution is 
\begin{eqnarray}
	\sigma_\mathrm{MF}= 2 N_f N_c G_S \int \frac{d^3 p}{(2\pi)^3} \frac{m}{E_p}[1 - f(E_p-\mu) - f(E_p + \mu)] ~,
\end{eqnarray} 
and the contribution due to scalar/pseudoscalar mesons (corresponding to the diagram shown in the rightmost panel of Figure \ref{fig:11}) is given by \cite{Kitazawa:2014sga}
\begin{align}\label{SigmaM}
\begin{split}
	\Sigma_M({\bf 0},p_0)= d_M\int \frac{d^4q}{(2\pi)^4}\pi \varrho_M({\bf q},q_0) 
	\left\{\frac{(\gamma_0 +m/E_q)[1+g(q_0)-f(E_q - \mu)]}{q_0-p_0+E_q-\mu-i\eta}\right.
\\	+ \left.\frac{(\gamma_0 -m/E_q)[g(q_0)+f(E_q + \mu)]}{q_0-p_0-E_q-\mu-i\eta} \right\}~,
\end{split}
\end{align} 
where $\varrho_M=(-1/\pi)\Im T_M({\bf q}, \omega+i\eta)$ is the meson spectral density and 
$E_q=\sqrt{q^2+m^2}$ is the quark dispersion law with the quark mass $m=m_0+\sigma_{\rm MF}$.
One can observe the similarity of this result (\ref{SigmaM}) with that for a Dirac fermion coupled to a 
pointlike scalar meson, as given in \cite{Blaizot:1991kh}.

Finally, let us consider the aspect of quark deconfinement in dense matter within a relativistic density functional approach which can be considered as a local limit of the $\Phi-$derivable approach to quark-hadron matter (\ref{Om2}) that obtains a form similar to the relativistic density functional theory for hadronic matter. The important difference lies in the density functional that in 
the case of quark matter shall account for confining effects, see \cite{Ropke:1986qs,Kaltenborn:2017}. 
In the final part of this section, we outline the recently developed relativistic version of the so-called string-flip model that proved rather successful for phenomenological applications to compact stars, supernovae and neutron star mergers. 
 
\subsection{Relativistic Density Functional Approach to Quark Matter}
\label{ssec:quarkRDF}
Hence, one obtains a contribution to the energy density functional of quark matter correspondingly~\cite{Khvorostukin:2006aw}.~In analogy to the Walecka model of nuclear matter~\cite{Kapusta:1989tk}, the relativistic density-functional approach to interacting quark matter can be obtained from the path integral approach based on the partition function \cite{Kaltenborn:2017},
\begin{equation}
	\label{eq:Z}
	\mathcal{Z} = \int \mathcal{D}\bar{q}\mathcal{D}q \exp\left\{\int_0^\beta d\tau\int_Vd^3 x \left[\mathcal{L}_{\rm eff} + \bar{q}\gamma_0 \hat{\mu} q\right]\right\}~,
\end{equation}
with $q = (q_\mathrm u, q_\mathrm d)^\mathrm T$, $\hat\mu = \operatorname{diag} (\mu_\mathrm u, \mu_\mathrm d)$ and effective Lagrangian density $\mathcal{L}_{\textrm{eff}}=\mathcal{L}_{\textrm{free}}- U(\bar{q}q, \bar{q}\gamma_0q)$. The~interaction is given by the potential $U(\bar{q}q, \bar{q}\gamma_0q)$, which is a nonlinear functional of the scalar and vector quark field-currents.
In the mean-field approximation, this potential can be expanded around the expectation values of the field currents, $n_\mathrm{s} = \langle\bar{q}q\rangle$ and $n_\mathrm{v} = \langle\bar{q}\gamma_0q\rangle$ respectively,
\begin{equation}
	U(\bar{q}q, \bar{q}\gamma_0 q) = U(n_{\rm s}, n_{\rm v}) + (\bar{q}q -  n_{\rm s})\Sigma_\mathrm{s} + (\bar{q}\gamma_0q -  n_{\rm v}) \Sigma_\mathrm{v} +\ldots~,
\end{equation}
with scalar and vector self-energies, $\Sigma_{\rm s}$ and $\Sigma_{\rm v}$.
By appropriate rearranging of the quantities and performing the path integrals of Equation \eqref{eq:Z} one gets the thermodynamic potential 
\begin{align}
	\Omega &= -T \ln \mathcal Z = \Omega^\mathrm{quasi} + U(n_{\rm s}, n_{\rm v}) -  n_\mathrm s\Sigma_\mathrm{s} -  n_\mathrm v\Sigma_\mathrm{v}\;.
\end{align}

The quasi-particle term (for the case of isospin symmetry and degenerate flavors)
\begin{eqnarray}
\Omega^\mathrm{quasi}=-2N_c N_f T \int \frac{d^3p}{(2\pi)^3}
\left\{\ln\left[1+\mathrm{e}^{-\beta(E^*-\mu^*)}\right] + \ln\left[1+\mathrm{e}^{-\beta(E^*+\mu^*)}\right] \right\}
\end{eqnarray}
can be calculated by using the ideal Fermi gas distribution for quarks with the quasiparticle energy $E^*=\sqrt{p^2+M^2}$, the effective mass $M = m + \Sigma_{s}$ and effective chemical potential $\mu^* = \mu - \Sigma_{v}$.
The~self energies are determined by the density derivations

\begin{align}
	\Sigma_{\mathrm s} &= \frac{\partial U(n_{\rm s}, n_{\rm v})}{\partial n_{\mathrm s}}\;,\quad\mathrm{and}
\\	\Sigma_{\mathrm v} &= \frac{\partial U(n_{\rm s}, n_{\rm v})}{\partial n_{\mathrm v}}\;.
\end{align}

In this approach the stationarity of the thermodynamical potential
\begin{align}
	0 &= \frac{\partial\Omega}{\partial n_{\mathrm s}} = \frac{\partial\Omega}{\partial n_{\mathrm v}}
\end{align}
is always fulfilled.
For the case of isospin asymmetry, see Ref.~\cite{Kaltenborn:2017}.

In the mean-field approximation, the correlation energy can be obtained by folding the string-length distribution function for a given density with some interaction potential~\cite{Horowitz:1991fn,Ropke:1986qs,Horowitz:1991ux}.
Moreover, the average string length between quarks in uniform matter is related to the scalar density, $n_\mathrm s$, being proportional to ${n_\mathrm s^{-1/3}}$.
To capture this phenomenology, the following density functional of the interaction is adopted,
\begin{equation}
	\label{eq:potential}
	U({n}_\mathrm s,{n}_\mathrm v) = D({n}_\mathrm v){n}_\mathrm s^{2/3} +a {n}_\mathrm v^2 + \frac{b {n}_\mathrm v^4}{1+c {n}_\mathrm v^2}~.
\end{equation}

The first term captures aspects of (quark) confinement through the density dependent scalar self-energy, 
\begin{equation}
\Sigma_\mathrm s = \frac{2}{3}D({n}_\mathrm v){n}_\mathrm s^{-1/3}~,
\end{equation}
defining the effective quark mass $M = m + \Sigma_\mathrm s$.
We also employ higher-order quark interactions~\cite{Benic:2014jia}, by inclusion of the third term in Equation \eqref{eq:potential}, for the description of hybrid stars (neutron stars with a quark matter core) in order to obey the observational constraint of $2~$M$_\odot$.
To this end, the denominator in the last term of Equation \eqref{eq:potential} guarantees that for the appropriate choice of the parameters $b$ and $c$, causality is not violated (i.e., the speed of sound $c_\mathrm s=\sqrt{\partial P/\partial \varepsilon}$ does not exceed the speed of light).
All terms in Equation \eqref{eq:potential} that contain the vector density contribute to the shift defining the effective chemical potentials $\mu^* = \mu - \Sigma_\mathrm V$, where
\begin{equation}
\Sigma_\mathrm{v} = 2an_\mathrm{v} + \frac{4b n_\mathrm{v}^3}{1+c n_\mathrm{v}^2} - \frac{2 b c n_\mathrm{v}^5}{(1+c n_\mathrm{v}^2)^2} + \frac{\partial D(n_\mathrm{v})}{\partial n_\mathrm{v}}{n}_s^{2/3}~.
\end{equation} %

The SFM modification takes into account the effective reduction of the in-medium string tension, $D(n_\mathrm v) = D_0 \phi(n_{\rm v};\alpha)$.
It is understood as a consequence of the modification of the pressure on the color field lines by the dual Meissner effect, since the reduction of the available volume corresponds to the reduction of the non-perturbative dual superconductor QCD vacuum that determines the strength of the confining potential between the quarks.
The reduction of the string tension is modeled via a Gaussian function of the baryon density $n_{\rm v}$, 
\begin{equation}
\label{avail}
\phi(n_\mathrm v ; \alpha)=\exp\left[- \alpha (n_\mathrm v\cdot \mathrm{fm}^3)^2\right]~,
\end{equation}
similar the available volume fraction in dense nuclear matter (see \cite{Typel:2016srf} and Section~\ref{ssec:NMhigh} below), as it shall be related to the available volume of 
nonperturbative QCD vacuum that according to the dual superconductor model is responsible for the formation of stringy color field configurations because of the dual Meissner effect.
A detailed discussion of the role of the parameters $a,b,c$ and $\alpha$ is given in Ref.~\cite{Kaltenborn:2017}.

\section{Applications in Core-Collapse Supernovae and Neutron Stars}
\label{sec:appl-astro}
In the recent years the equation of state (EoS) has been significantly constraint, in particular at densities around and in excess of nuclear saturation density ($\rho_0$).{}~At densities $\rho\leq\rho_0$, chiral~effective field theory is the ab-initio approach to the nuclear many-body problem of dilute neutron \mbox{matter~\cite{Hebeler:2010a,Hebeler:2010b,Holt:2012a,Sammarruca:2012,Tews:2013,Krueger:2013,Coraggio:2013}.}
Moreover, the high-precision observation of massive neutron stars of about 2~M$_\odot$~\cite{Demorest:2010,Antoniadis:2013,Fonseca:2016} constraints the supersaturation-density EoS, i.e., sufficient stiffness is required.
The latter aspect challenges the appearance of additional particle degrees of freedom, e.g., hyperons and quarks, which tend to soften the EoS at $\rho>\rho_0$. 

While neutron stars feature matter in $\beta$-equilibrium at zero temperature, the challenge lays in the development of EoS for core-collapse supernova applications, which cover a large domain of temperature, density and isospin asymmetry (cf. Figure 1a of Ref.~\cite{Fischer:2017}).~At $T\leq0.5$~MeV, time-dependent processes determine the nuclear composition, where heavy nuclei dominate.
With increasing temperature, towards $T\simeq 0.5$~MeV, complete chemical and thermal equilibrium known as NSE (nuclear statistical equilibrium) is achieved, where, the nuclear composition is determined from the three independent variables: $T$, $\rho$ (or $n_\mathrm B)$, $y_\mathrm C$ (baryonic charge fraction).
Note that there is a strong density dependence, i.e., these heavy nuclei, originally belonging to the iron group with $A\gtrsim 56$, become increasingly heavier with increasing density.
This phenomenon is well known also from the neutron star crust, where due to the low temperatures only a single nucleus exists for at given density, instead of a (broad) distribution as in the supernova case.
At $\rho=\rho_0$ and at temperatures above $T=5-10$~MeV, nuclei dissolve at the liquid-gas phase transition into homogeneous nuclear matter composed of (quasi-free) nucleons~\cite{Typel:2009sy,Hempel:2011,Ropke:2012qv}. 

It becomes evident that first-principle calculations covering the entire domain are presently inexistent.~Instead, model EOS are being developed for astrophysical applications.~These combine several domains with different degrees of freedom, i.e., heavy nuclei at low temperatures, inhomogeneous nuclear matter with light and heavy nuclei together with the free nucleons (mean field), and homogeneous matter at high temperatures and densities. The latter has long been subject to investigations of a possible phase transition to the quark-gluon plasma. 

The role of the EOS in simulations of core-collapse supernovae was explored within the failed scenario and consequently the formation of black holes, with focus on the dynamics and the neutrino signal~\cite{Sumiyoshi:2006id,Fischer:2009,OConnor:2011,Steiner:2013}.
In the multi-dimensional framework, neutrino-driven supernova explosions were the subjects of investigation~\cite{Marek:2008qi,Suwa:2013,Nagakura:2017}, where it was found that such explosions are favored for soft EOS~\cite{Lattimer:1991nc} with an earlier onset of shock revival and generally higher explosion energies, in comparison to stiff EOS~\cite{Shen:1998gg}.
Moreover, the role of the nuclear symmetry energy has been reviewed~\cite{Fischer:2014}.
This~an important nuclear matter parameter has a strong density dependence and becomes more tightly constrained by experiments, nuclear theory and observations~\cite{Lattimer:2013,Kolomeitsev:2016}.

\subsection{Heavy Nuclear Clusters with $A\gtrsim 56$}
At low temperatures ($T<0.5$~MeV) time-dependent nuclear processes determine the evolution, which corresponds to the outer core of the stellar progenitor, with the nuclear composition of dominantly silicon, sulfur as well as carbon and oxygen. In some cases, even parts of the hydrogen-rich helium envelope are taken into account, e.g., in simulations of supernova explosions following the shock evolution for tens of seconds through parts of the  stellar envelope. Therefore, small nuclear reaction networks are commonly employed in supernova studies~\cite{Thielemann:2004,Fischer:2009af}. They are sufficient for the nuclear energy generation. 

At $T\gtrsim 0.5$~MeV, NSE is fulfilled and the relation $\mu_{(A,Z)}=Z\mu_p+(A-Z)\mu_n$ between the chemical potential of nuclei $\mu_{(A,Z)}$, with atomic mass $A$ and charge $Z$, and the chemical potentials of neutron $\mu_n$ and proton $\mu_p$ holds.
The NSE conditions found in the collapsing stellar core feature a broad distribution of nuclei with a pronounced peak around the iron-group, at low densities (see Figure 2 of Ref.~\cite{Fischer:2017}). In simulations of supernovae, this nuclear distribution is classified by the NSE average, including nuclear shell effects as discussed~\cite{Hempel:2009mc}, which extends beyond the commonly used single-nucleus approximation. This is important for the consideration of weak processes with heavy nuclei. In particular, rates for electron captures on protons bound in these heavy nuclei~\cite{Juodagalvis:2010} are averaged over the NSE composition and provided to the community as a table. In addition, coherent neutrino-nucleus scattering is considered~\cite{Bruenn:1985en}, where even inelastic contributions are take into account~\cite{Langanke:2007ua}, as well as nuclear (de)excitations~\cite{Fuller:1991,Fischer:2013}. The profound understanding of weak processes associated with nuclear transitions is also important for the understanding of the physics of the neutron star crust (cf. Ref.~\cite{Schatz:2014} and references therein), in particular for accreting neutron stars leading to the phenomenon of deep-crustal heating~\cite{Haensel:1990,Brown:1998,Haensel:2008}.

\subsection{Light Nuclear Clusters with $A\leq4$}
In the domain corresponding to NSE, there is a rather narrow density domain where light nuclear clusters such as $^2$H, $^3$H, $^3$He and $^4$He can exist, at finite temperatures on the order of few MeV~\cite{Typel:2009sy}.  There are two aspects related to the presence of these light nuclear clusters, modification of the nuclear EOS due to these degrees of freedom, and the neutrino response due to the inclusion of a large variety of weak processes (cf. Table~1 in Ref.~\cite{Fischer:2016d}). Common model supernova EOS with (light) clusters are based on the modified NSE~\cite{Hempel:2009mc}. However, the dissolving of the clusters towards high density is mimicked by the geometric excluded volume approach, as well as by hand with increasing temperature. This applies equally for light and heavy nuclear clusters. In comparison with the state-of-the-art quantum statistical approach~\cite{Roepke:2009,Roepke:2011} the deficits of NSE are revealed~\cite{Fischer:2017}, while the cluster-virial EOS can provide the constraint at low densities~ \cite{Ropke:2012qv}. It relates to the overestimation of the abundance of light clusters and in particular the too late dissolving of clusters into homogeneous matter.

Besides the NSE approach with 'all' (light) clusters included, in simulations of core-collapse supernovae the simplified nuclear composition $(n,p,\alpha,\langle A,Z \rangle)$, with only $^4$He as representative light cluster, has long been employed~\cite{Lattimer:1991nc,Shen:1998gg}. It leads to an overestimate of the abundances of the unbound baryons and $^4$He. This has important consequences for the supernova results, since such simplistic approach overestimates the neutrino response with the  neutrons and protons, which in turn changes the supernova neutrino fluxes and spectra, however, with only a mild impact on the overall supernova dynamics~\cite{Fischer:2017}.

In supernova simulations the temperatures are generally too high for any significant abundance of any light cluster. Only when the supernova explosion onset has been launched and the remnant central proto-neutron star deleptonizes and cools via the emission of neutrinos, light clusters can start to play a role. However, it has been demonstrated that neutral-current reactions are dominated by scattering on free neutrons, which is the most abundant nuclear species due to the generally large neutron excess (>95\%) of supernova matter. Scattering reactions with light clusters play a negligible role. On the other hand, the neutrino response for light clusters is dominated by charged-current (break-up) reactions involving deuteron, triton and helium-3. However, the $\nu_e$ charged-current opacity is dominated by absorption on free neutrons. The situation is different for $\bar\nu_e$, since it was shown that protons and light clusters have similar abundances~\cite{Fischer:2016d,Fischer:2017}. Taking properly into medium modifications for the cross sections and the proper phase-space of the contributing particles, the final rates are generally small. They never reach values as for the standard rates with protons, and hence the impact from charged-current weak processes with light clusters was found to be negligible~\cite{Fischer:2016d,Fischer:2017}. Note that this study was based on the NSE approach which generally overestimates the abundance of light clusters. Hence, any improved treatment of the nuclear composition will most likely even reduce the impact of light clusters and the associated weak processes.

\subsection{Homogeneous Matter at Supersaturation Density and Phase Transition to Quark Matter}
\label{ssec:NMhigh}
With increasing densities, around $\rho_0$ (depending on the temperature), the transition to homogeneous nuclear matter proceeds where the EOS becomes less and less constrained by nuclear physics.~As discussed above, the quark substructure of baryons shall become apparent at supersaturation densities and manifest itself by a nucleonic hard core repulsion due to quark Pauli blocking. This effect is likely to be enhanced by chiral symmetry restoration at high densities. To~explore its role for the nuclear EOS at supersaturation densities the geometric excluded volume mechanism can be employed~\cite{Typel:2016srf}, where the available volume of the nucleons, $V_N = V\,\phi(\rho;\rm v)$, is~proportional to the total volume $V$ of the system where as proportionality factor occurs the available volume fraction 
\begin{equation}
\label{phi-v}
\phi(\rho;\mathrm{v})=\exp \left[-\mathrm{|v|v} \left( \rho - \rho_0 \right)^2 \right]~.
\end{equation}

This is a density functional taken here according to~\cite{Typel:2016srf} in a Gaussian form similar to (\ref{avail}).
Depending on the sign of the excluded volume parameter, v, it allows to model both,  stiffening and softening of the supersaturation-density EOS based on some reference model. This approach has been applied to confirm the independence of supernova simulations, e.g., the supernova shock dynamics as well as the evolution of neutrino luminosities and average energies, on the supersaturation-density EOS~\cite{Fischer:2016a}.~In studies of neutron stars, this approach results in significantly altered neutron star properties, e.g., maximum masses and radii~\cite{Benic:2014jia,Kaltenborn:2017}. The latter property is currently constraint only poorly, from the analysis of observations of low-mass x-ray binary systems~\cite{Steiner:2010,Suleimanov11,Steiner:2013b}.

Another uncertain aspect of the supersaturation-density EoS is the possibility of a phase transition from nuclear matter, with hadrons as degrees of freedom, to the deconfined quark gluon plasma with quarks and gluons as the new degrees of freedom. This has long been explored in the context of cold neutron stars. 
Unfortunately, the evaluation of the partition function of Quantum Chromodynamics (QCD)---the theory of strongly interacting matter---is possible only at vanishing baryon density by means of large-scale numerical simulations of this gauge field theory in a representation on space-time lattices~\cite{Fodor:2004nz,Ratti:2005jh}. 
These numerical ab-initio solutions predict a smooth cross-over transition at a temperature of $T=154\pm9$~MeV at $\mu_B\simeq 0$, see Refs.~\cite{,Katz:2012JHEP,Laermann:2012PRL,Katz:2014PhLB}. 
Consequently, to study the role of quark degrees of freedom at high baryon densities, 
effective models for low-energy QCD have to be employed. 
Generally, the nuclear and quark matter phases are modeled separately and a phase transition construction is employed. 
This so-called {\it two-phase approach} results in a first-order phase transition by design. 
Note further that perturbative QCD, which is valid in the limit of asymptotic freedom, where the smallness of the coupling between quarks allows the usage of perturbative methods and corrections to the behaviour of an ideal gas of ultra-relativistic particles~\cite{Kurkela:2014vha} are small, become applicable only at extremely high temperatures and densities exceeding by far the values attainable in compact stars or ultrarelativistic heavy-ion collisions. 
Instead, for studies of neutron stars and supernovae, effective quark matter models have been commonly employed, such as the thermodynamic bag model~\cite{Farhi:1984qu}, models based on the Nambu-Jona-Lasinio (NJL) type~\cite{Nambu:1961tp,Klevansky:1992qe,Buballa:2003qv}, the recently developed vector-interaction enhanced chiral bag model~\cite{Klaehn:2015,Klaehn:2017a} and in particular the density-functional-based DD2-SFM hybrid EoS approach \cite{Kaltenborn:2017} that has been outlined above.
With the finite-temperature extension of the DD2F-SFM EoS it was possible to show that the deconfinement phase transition may serve as an explosion mechanism for massive ($\sim$50 M$_\odot$) blue-supergiant stars~\cite{Fischer:2017lag} long sought-for.
At the same time, it explains the occurrence of a population of neutron stars born with high masses.

It has been realized that repulsive interactions are the necessary ingredient to provide a sufficient stiffness for the EOS at high densities in order to yield massive neutron stars with quark-matter core (known as hybrid stars) in agreement with the current constraint of $2$~M$_\odot$. 
Moreover, higher-order vector repulsion terms~\cite{Benic:2014jia,Kaltenborn:2017} can lead to the 'twin' phenomenon~(cf. \cite{Haensel:2007yy,Read:2008iy,Zdunik:2012dj,Alford:2013aca} and references therein). It relates to the existence of compact stellar objects with similar-to-equal masses but different radii, to a strong phase transition with a large latent heat. 
As a consequence, the hybrid stars for such an EoS appear in the mass-radius diagram on a disconnected ``third family'' branch, separated from the branch of neutron stars (second family) by a sequence of unstable configurations. 
As has been demonstrated impressively for the DD2-SFM hybrid EoS \cite{Kaltenborn:2017},
by varying only the available volume parameter $\alpha$ describing the screening of the confining interaction in dense matter, the twin phenomenon can be obtained at high masses 
of $\sim$2~M$_\odot$ as well as for typical pulsar masses of $1.3$--$1.4$ M$_\odot$ or even below.  
This feature allows to discuss such hybrid stars in the context of the first multi-messenger observation of the binary compact star merger GW170817 
\cite{TheLIGOScientific:2017qsa}, where such a scenario appears as an alternative to the conservative one of a binary neutron star merger with a relatively soft EoS
\cite{Ayriyan:2017nby,Paschalidis:2017qmb,Blaschke:2018mqw}. 
In this interesting situation the NICER (Neutron Star Interior Composition Explorer)\footnote{\url{https://www.nasa.gov/nicer}} NASA mission has the potential to rule out the soft EoS scenario, when it would measure for its primary target, the~nearest millisecond pulsar PSR J0437-4715 with a mass of $1.44\pm 0.07$ M$_\odot$ a large radius of, say, $14$~km with the expected high precision of 0.5~km.
Such a measurement would contradict the constraint on compactness of neutron stars extracted from GW170817 in the double neutron star merger scenario that constrains the radius in the mass range of $1.4$ M$_\odot$ to $R<13.4$ km, see~\cite{Annala:2017llu}.
Such a measurement would imply the discovery of the third family of compact stars 
\cite{Blaschke:2018mqw} with an onset mass in the mass range $1.16$--$1.60$ M$_\odot$ extracted from the gravitational wave signal of binary inspiral \cite{TheLIGOScientific:2017qsa}. 

\section{Cluster Formation and Quark Deconfinement Transition in Heavy-Ion Collisions}\label{sec:appl-hic}

\subsection{Light Cluster Formation and Symmetry Energy in Low-Energy Heavy-Ion Collisions}

The only possibility to probe the properties of hot and dense nuclear matter in the laboratory are 
heavy ion collisions (HIC). The fragment distributions, their energy spectra and correlations 
measured in the detectors are used to infer the properties of the initial state (``fireball'')
produced in HIC. To~reconstruct the initial state, one has to model the time evolution
of the expanding hot and dense matter, including the formation of correlations and clusters.
	
A strict quantum statistical approach to this nonequilibrium process is not available at present.
A possible method is the Zubarev nonequilibrium statistical operator which is able 
to describe the formation of correlations in the expanding hot and dense matter \cite{Ropke:2017zts,Ropke:2017own}.
First steps to describe cluster formation ($A < 4$) in expanding matter have been performed 
\cite{Kuhrts:2000zs},
but have to be worked out further, in particular to include larger clusters ($A \ge 4$). Kinetic codes
based on a single-particle description, but also QMD and AMD codes which simulate cluster formation
using the coalescence model and simpler concepts, have to be improved to obtain a microscopic description of 
cluster formation. Work~in this direction is in progress 
\cite{Bastian:2016xna,Bastian:2017rhb,Roepke:2017ohb}.
Alternatively, the freeze-out concept has been used to model the expansion process of the ``fireball''.
Within the chemical freeze-out approach, it is assumed that at a certain instant of the expanding
and down-cooling fireball, the composition is frozen because the ``chemical'' reactions become slow
so that the composition remains unchanged. This concept of a sudden freeze-out has been applied 
to HIC not only at moderate energies, but also at very high energies
\cite{Roepke:2017ohb}.
We give a short summary of some results obtained from HIC experiments at moderate energies ($\approx$35 MeV/\emph{A}), 
where the production of light clusters was measured and interpreted within a freeze-out model.
The main issue of these investigation was to show the relevance of clustering in nuclear systems 
and the necessity to describe medium effects. In particular, the equation of state of nuclear matter is of interest, 
at moderate temperatures ($T \le 20$ MeV) and at subsaturation densities.

A first series of experiments was related to the symmetry energy \cite{Kowalski:2006ju,Natowitz:2010ti,Wada:2011qm,Hagel:2014wja,Typel:2013zna}.

In contrast to the standard treatment of symmetry energy within mean-field approaches,
see, for instance, \cite{Li:2008gp},
it does not vanish in the zero density limit, but is significantly determined by cluster formation
in the low-density region. This is a very obvious result, and more or less trivial in the density region 
where medium effects can be neglected. However, going to higher densities, medium effects, 
in particular the dissolution of bound states, must be included so that a smooth transition
to the near-saturation density region is expected, where mean-field approaches can be applied.
It has been shown in \cite{Kowalski:2006ju,Natowitz:2010ti,Wada:2011qm,Hagel:2014wja,Typel:2013zna} 
that general expressions for the symmetry energy can be obtained 
which reproduce the results of experiments at low densities, determined by cluster formation,
but agree with mean-field approaches at high densities. A cluster-virial approach may improve
the description in the intermediate region.

The direct observation of medium effects in the nuclear matter EoS from HIC experiments is
more involved. The composition, in particular the yield of light clusters ($A \le 4$), is obtained in the low-density
limit from a mass-action law using the binding energies of free nuclei. 
We~discussed this approach above, Section \ref{sec:qs-nucl}, as nuclear statistical equilibrium (NSE).
A special ratio of cluster yields $Y_i$,
the so-called chemical constant $K_A=Y_A/(Y_p^Z Y_n^N)$,
 can be considered which in this low-density limit is solely a function of the 
temperature and the volume, but not the chemical potentials.
Because in chemical equilibrium the chemical potentials of the cluster $A$, consisting
of $Z$ protons and \mbox{$N=A-Z$} neutrons, are related as $\mu_A=Z \mu_p+N \mu_n$,
the chemical potentials cancel in the low-density limit.
This~simple approach, neglecting any density effects, has been disproved by 
experiments with HIC~\cite{Qin:2011qp}.
The reason is the modification of the binding energies if the density is increasing.
In~particular, self-energy shifts and Pauli blocking lead to the reduction of the binding energy and
the dissolution at the so called Mott density. The measured chemical constants can be used
for the experimental determination of in-medium cluster binding energies
and Mott points in nuclear matter~\cite{Hagel:2011ws}.

We can discus these experimental results as clear indication for the need to consider medium effects. 
Within a QS  approach
including quasiparticle shifts and correlations in the continuum \cite{Ropke:2014fia}, 
it~was possible to reproduce the data for the chemical constants
of the light elements $d, t, h, \alpha$ obtained from the cluster yields. More simple 
models for medium corrections such as the semiempirical excluded volume concept \cite{Hempel:2009mc}
can be adapted to reproduce the data \cite{Hempel:2015yma}.

Also for these experiments, a nuclear matter EoS is needed which describes cluster formation
and the medium modification, as well as the treatment of continuum correlations. 
A cluster-virial approach may improve the calculation of the EoS in a wide region of the phase diagram.

In conclusion, medium effects, in particular self-energy shifts and Pauli blocking for light clusters, are 
verified by recent HIC experiments. An improved cluster viral approach should be worked out to 
describe adequately the contributions of correlations in the continuum, expressed by in-medium scattering
phase shifts between the different constituents.~The freeze-out model to describe the expanding fireball 
may be considered as an approximation to treat the nonequilibrium process. 
Kinetic approaches, for instance transport codes 
allowing for cluster production (such as QMD or AMD), may~be developed further to include in-medium 
correlation effects. This would allow for a systematic and consistent treatment of HIC experiments.
Nevertheless, the correct description of the thermodynamic equilibrium, in particular 
the  cluster viral approach, is a benchmark for all nonequilibrium approaches, and should be advanced in the future.	

\subsection{Deconfinement Transition in Relativistic Heavy-Ion Collisions}

The main goal of the theoretical developments towards a unified EoS for quark-hadron matter is to achieve a most reliable prediction for the behavior of warm, dense strongly interacting matter including the deconfinement transition, which is described here as a Mott dissociation of baryonic and mesonic bound states, triggered by the restoration of the dynamically broken chiral restoration.
Dynamical chiral symmetry breaking is a rather robust phenomenon that can be described in a broad variety of chiral quark models of the Nambu-Jona-Lasinio type, i.e., with a four-fermion interaction of the current-current form with a sufficiently strong coupling to allow for a nontrivial solution of the gap equation for the quark mass. This is a nonperturbative effect that cannot be obtained in any finite order of perturbation theory
and is nowadays an obligatory element of modern descriptions of quark matter.
At finite temperatures, however, such models often fail to provide a sensible description of the EoS of QCD matter since the quark matter pressure dominates over the hadronic one already at unphysically low temperatures, due to the lack of a quark confinement mechanism in those models.
A simple way out is provided by adopting a bag pressure for mimicking confinement. 
This is a too simple concept and spoils the beauty of a dynamical description. 
While the details of the confinement mechanism in QCD are still debated, a viable compromise is provided by the concept of a confining density functional that is based on string-type interactions between color charges and, together with the concept of saturation of color interactions within nearest neighbors (string-flip model) allows for the treatment of such confining interactions in quark matter within a relativistic, selfconsistent quasiparticle model. 

The success of the DD2(DD2F)-SFM hybrid EoS for astrophysical applications has been summarized in the previous section. The applications for heavy-ion collisions, in the isospin-symmetric case, are still under way. A necessary requirement for a sensible description of the complex system of an ultrarelativistic heavy-ion collision calls for a numerical simulation code like THESEUS, the~three-fluid hydrodynamics-based event simulator extended by UrQMD simulations of final-state interactions \cite{Batyuk:2016qmb}. 
Such a description is most appropriate for the investigation of possible effects of a phase transition in the baryon stopping regime, i.e., when the projectile and target fluids collide and form a highly compressed baryonic matter system, i.e., in the collision energy range $\sqrt{s_{NN}}=2 - 20$ which is covered by the THESEUS program and by 
the high-statistics collison experiments like NA61, RHIC BES or RHIC FXT and the upcoming NICA and FAIR experiments.
Therefore, we have chosen THESEUS as the tool for identifying the QCD phase transition
and for investigating the effects of a first-order phase transition on heavy-ion collision observables.
Previous studies of this question have been performed with THESEUS \cite{Batyuk:2017sku} and with the three-fluid hydrodynamics code in it \cite{Ivanov:2013wha,Ivanov:2013yqa,Ivanov:2013yla} using model EoS of three kinds: purely hadronic, crossover and with a strong first-order transition
\footnote{Recently, a thermodynamically consistent generalization of the excluded-volume improved RDF approach to the hadronic EoS has been suggested which employs a density- and temperature-dependent excluded volume parameter. Within this setting, a second first-order phase transition with a critical endpoint in the QCD phase diagram has been obtained \cite{Typel:2017vif}. Such a formulation may be most convenient, e.g., for Bayesian studies of the structure of the QCD phase diagram to be extracted from data of heavy-ion collision experiments.}. 
The upgrade of the EoS with the  DD2(DD2F)-SFM hybrid EoS as described in this work is under way. In particular, one expects modifications from the lower density region of the phase transition and its temperature dependence (for a direct comparison, see the right panel of Figure 2 in \cite{Bastian:2015avq}) and a better description of flow observables due to the increased stiffening of the high-density part of the new hybrid EoS when compared to the one in Ref.~\cite{Batyuk:2016qmb}.

A main goal of the unified approach to the quark-hadron EoS as outlined in this work is the possibility to obtain an EoS with a second critical endpoint in the QCD phase diagram associated with the chiral/deconfinement transition. The first one, corresponding to the liquid-gas phase transition in nuclear matter is already an integral part of the RDF description of nuclear matter within the DD2(DD2F) part of the present approach.      
We want to point out that with this approach one may have achieved a systematic formulation of a theory for quark-hadron matter that allows to address also the presently puzzling questions of chemical freezeout of hadrons and nuclei like: 
\begin{enumerate}[leftmargin=*,labelsep=4.9mm]
	\item Can the success of the thermal statistical model in describing the production of nuclear clusters as measured by the ALICE experiment at LHC \cite{Andronic:2017pug} be interpreted so that they freeze out directly when hadronizing the QGP so that they may be viewed as preformed multiquark systems already in the QGP?
	\item What are the necessary ingredients to understand chemical freezeout of hadrons and clusters kinetically \cite{Blaschke:2017lvd}?
\end{enumerate}

With these prospects for the development of a unified quark-hadron matter EoS we want to conclude the present work.
\section{Conclusions}
\label{sec:conclusions}

We have outlined an approach to a unified equation of state for quark-hadron matter on the basis of a $\Phi-$derivable approach to the generalized Beth-Uhlenbeck equation of state for a cluster decomposition of thermodynamic quantities like the density. 
To this end we have summarized the cluster virial expansion for nuclear matter and demonstrated the equivalence of the Green's function approach to the $\Phi-$derivable formulation. 
As an example, the formation and dissociation of deuterons in nuclear matter was discussed. 
We have formulated the cluster $\Phi-$derivable approach to quark-hadron matter which allows to take into account the specifics of chiral symmetry restoration and deconfinement in triggering the Mott-dissociation of hadrons.
Applications to the phenomenology of nuclear clusters and quark deconfinement in the astrophysics of supernovae and compact stars as well as in heavy-ion collisions are outlined. 

This approach unifies the description of a strongly coupled quark-gluon plasma with that of a medium-modified hadron resonance gas description which are shown to be its limiting cases. 
The developed formalism shall replace the common two-phase approach to the description of the deconfinement and chiral phase transition, where separately developed equations of state for hadronic and quark matter are matched with Gibbs conditions of phase equilibrium.
Roughly speaking, one would develop a Ginzburg-Landau-type density functional which allows first, second and higher-order transitions, including crossovers.
Examples are the van-der-Waals EoS which has a first-order transition with critical endpoint or the RMF models of nuclear matter for the liquid-gas transition.
The cluster virial expansion shall allow a formulation of the quark-hadron transition in a similar way.

\vspace{6pt}

\authorcontributions{NUFB as first author is responsible for the final form and content of the whole paper, integrating the contributions of the coauthors, who provided Sections \ref{sec:qs-nucl}, \ref{sec:appl-hic} (GR), \ref{sec:phi-nucl}, \ref{sec:phi-quarks} (DB) and \ref{sec:appl-astro} (TF).}

\acknowledgments{%
We acknowledge support by the Polish National Science Center (NCN) under grant number UMO-2016/23/B/ST2/00720 (TF) and grant number UMO-2014/13/B/ST9/02621 (NUFB) and by the Bogoliubov-Infeld program (NUFB, TF) for collaboration between JINR Dubna and Polish Institutes, as well as the Heisenberg-Landau program for collaboration between JINR Dubna and German Institutes (GR).
GR and DB are grateful for support within the MEPhI Academic Excellence programme under contract no. 02.a03.21.0005 for their work in Sections 2 and 3, respectively.
The work of DB was supported in part by a grant from the Russian Science Foundation under contract number 17-12-01427.
This work was supported by the COST Actions CA15213 ``THOR'', CA16117 ``ChETEC'' and CA16214 ``PHAROS''.
}
\conflictsofinterest{The authors declare no conflict of interest.}

\begin{thebibliography}{999}

\bibitem[Natowitz \em{et~al.}(2002)Natowitz, Hagel, Ma, Murray, Qin, Wada, and
  Wang]{Natowitz:2002nw}
Natowitz, J.B.; Hagel, K.; Ma, Y.; Murray, M.; Qin, L.; Wada, R.; Wang, J.
\newblock {Limiting temperatures and the equation of state of nuclear matter}.
\newblock {\em Phys. Rev. Lett.} {\bf 2002}, {\em 89},~212701.

\bibitem[Bazavov \em{et~al.}(2012)Bazavov et~al.]{Bazavov:2011nk}
Bazavov, A.; Bhattacharya, T.; Cheng, M.; DeTar, C.; Ding, H.T.; Gottlieb, S.; Gupta, R.; Hegde, P.; Heller,~U.M.; Karsch, F.; et al.~\newblock {The chiral and deconfinement aspects of the QCD transition}.
\newblock {\em Phys. Rev. D} {\bf 2012}, {\em 85},~054503.

\bibitem[Baym and Kadanoff(1961)]{Baym:1961zz}
Baym, G.; Kadanoff, L.P.
\newblock {Conservation Laws and Correlation Functions}.
\newblock {\em Phys. Rev.} {\bf 1961}, {\em 124},~287--299.

\bibitem[Baym(1962)]{Baym:1962sx}
Baym, G.~{Selfconsistent approximation in many body systems}.~{\em Phys.~Rev.}~{\bf 1962},~{\em 127},~1391--1401. 

\bibitem[Kraeft \em{et~al.}(1986)Kraeft, Kremp, Ebeling, and
  R\"{o}pke]{KKER1986}
Kraeft, W.D.; Kremp, D.; Ebeling, W.; R\"{o}pke, G.
\newblock {\em Quantum Statistics of Charged Particle Systems}; Springer: Berlin, Germany,   1986.


\bibitem[Weinhold \em{et~al.}(1998)Weinhold, Friman, and
  N{\"o}renberg]{Weinhold:1997ig}
Weinhold, W.; Friman, B.; N{\"o}renberg, W.~{Thermodynamics of Delta resonances}.
\newblock {\em Phys. Lett. B} {\bf 1998}, {\em 433}, 236--242.

\bibitem[Weinhold(1998)]{Weinhold:1998tha}
Weinhold, W.
\newblock {Thermodynamik mit Resonanzzuständen}.
\newblock Ph.D. Thesis, Technischen Universität Darmstadt, Darmstadt, Germany, 1998.

\bibitem[Zimmermann and Stolz(1985)]{Zimmermann:1985ji}
Zimmermann, R.; Stolz, H.
\newblock The Mass Action Law in Two-Component Fermi Systems Revisited Excitons
  and Electron-Hole Pairs.
\newblock {\em Phys. Status Solidi } {\bf 1985}, {\em 131},~151--164.

\bibitem[R{\"o}pke \em{et~al.}(1982{\natexlab{a}})R{\"o}pke, M{\"u}nchow, and
  Schulz]{Ropke:1982vzx}
R{\"o}pke, G.; M{\"u}nchow, L.; Schulz, H.
\newblock {On the phase stability of hot nuclear matter and the applicability
  of detailed balance equations}.
\newblock {\em Phys. Lett. B} {\bf 1982}, {\em 110},~21--24.

\bibitem[R{\"o}pke \em{et~al.}(1982{\natexlab{b}})R{\"o}pke, M{\"u}nchow, and
  Schulz]{Ropke:1982ino}
R{\"o}pke, G.; M{\"u}nchow, L.; Schulz, H.
\newblock {Particle clustering and Mott transitions in nuclear matter at finite
  temperature}.
\newblock {\em Nucl. Phys. A} {\bf 1982}, {\em 379},~536--552. 

\bibitem[R{\"o}pke \em{et~al.}(1983)R{\"o}pke, Schmidt, M{\"u}nchow, and
  Schulz]{Ropke:1983lbc}
R{\"o}pke, G.; Schmidt, M.; M{\"u}nchow, L.; Schulz, H.
\newblock {Particle clustering and Mott transition in nuclear matter at finite
  temperature (II)}.
\newblock {\em Nucl. Phys. A} {\bf 1983}, {\em 399},~587--602. 

\bibitem[Schmidt \em{et~al.}(1990)Schmidt, R{\"o}pke, and
  Schulz]{Schmidt:1990zz}
Schmidt, M.; R{\"o}pke, G.; Schulz, H.
\newblock Generalized Beth-Uhlenbeck approach for hot nuclear matter.
\newblock {\em Ann. Phys.} {\bf 1990}, {\em 202},~57--99.

\bibitem[R{\"o}pke(2017{\natexlab{b}})]{Ropke:2017own} 
R{\"o}pke, G.
\newblock {Correlations and Clustering in Dilute Matter}. In {\em Nuclear
  Particle Correlations and Cluster Physics}; Schröder, W., Ed.; World Scientific: Singapore,  2017; pp.
  31--69.

\bibitem[R{\"o}pke \em{et~al.}(2018)R{\"o}pke, Blaschke, Ivanov, Karpenko,
  Rogachevsky, and Wolter]{Roepke:2017ohb}
R{\"o}pke, G.; Blaschke, D.; Ivanov, Y.B.; Karpenko, I.; Rogachevsky, O.V.;
  Wolter, H.H.
\newblock {Medium effects on freeze-out of light clusters at NICA energies}.
\newblock {\em Phys. Part.  Nucl. Lett.} {\bf 2018}, {\em
  15},~225--229.

\bibitem[Beth and Uhlenbeck(1937)]{Beth:1937zz}
Beth, E.; Uhlenbeck, G.
\newblock {The quantum theory of the non-ideal gas. II. Behaviour at low
  temperatures}.
\newblock {\em Physica} {\bf 1937}, {\em 4},~915--924. 

\bibitem[R{\"o}pke \em{et~al.}(2013)R{\"o}pke, Bastian, Blaschke, Kl{\"a}hn,
  Typel, and Wolter]{Ropke:2012qv}
R{\"o}pke, G.; Bastian, N.U.; Blaschke, D.; Kl{\"a}hn, T.; Typel, S.; Wolter,
  H.H.
\newblock {Cluster virial expansion for nuclear matter within a quasiparticle
  statistical approach}.
\newblock {\em Nucl. Phys. A} {\bf 2013}, {\em 897},~70--92.

\bibitem[R{\"o}pke(2015)]{Ropke:2014fia}
R{\"o}pke, G.
\newblock {Nuclear matter equation of state including two-, three-, and
  four-nucleon correlations}.
\newblock {\em Phys. Rev. C} {\bf 2015}, {\em 92},~054001.

\bibitem[R{\"o}pke(2014)]{Ropke:2014mwa}
\textls[-25]{R{\"o}pke, G.
\newblock {Clustering in nuclear environment}.
\newblock {\em J. Phys. } {\bf 2014}, {\em 569},~012031. } 

\bibitem[Horowitz and Schwenk(2006)]{Horowitz:2005nd}
Horowitz, C.J.; Schwenk, A.
\newblock {Cluster formation and the virial equation of state of low-density
  nuclear matter}.
\newblock {\em Nucl. Phys. A} {\bf 2006}, {\em 776},~55--79. 

\bibitem[Blaschke(2015)]{Blaschke:2015bxa}
Blaschke, D.
\newblock {Cluster virial expansion for quark and nuclear matter}.
\newblock {\em PoS BaldinISHEPPXXII} {\bf 2015}, 113

\bibitem[Vanderheyden and Baym(1998)]{Vanderheyden:1998ph}
Vanderheyden, B.; Baym, G.
\newblock {Selfconsistent approximations in relativistic plasmas: Quasiparticle
  analysis of the thermodynamic properties}.
\newblock {\em J. Stat. Phys.} {\bf 1998}, {\em 93},~843--861.

\bibitem[Blaizot \em{et~al.}(2001)Blaizot, Iancu, and Rebhan]{Blaizot:2000fc}
Blaizot, J.P.; Iancu, E.; Rebhan, A.
\newblock {Approximately selfconsistent resummations for the thermodynamics of
  the quark gluon plasma. 1. Entropy and density}.
\newblock {\em Phys. Rev. D} {\bf 2001}, {\em 63},~065003.

\bibitem[Uhlenbeck and Beth(1936)]{UHLENBECK1936729}
Uhlenbeck, G.E.; Beth, E.
\newblock The quantum theory of the non-ideal gas I. Deviations from the
  classical theory.
\newblock {\em Physica} {\bf 1936}, {\em 3},~729--745.

\bibitem[Typel(2016)]{Typel:2016srf}
\textls[-35]{Typel, S.
\newblock {Variations on the excluded-volume mechanism}.
\newblock {\em Eur. Phys. J. A} {\bf 2016}, {\em 52},~16. } 

\bibitem[R{\"o}pke \em{et~al.}(1986)R{\"o}pke, Blaschke, and
  Schulz]{Ropke:1986qs}
R{\"o}pke, G.; Blaschke, D.; Schulz, H.
\newblock {Pauli Quenching Effects in a Simple String Model of Quark/Nuclear
  Matter}.
\newblock {\em Phys. Rev. D} {\bf 1986}, {\em 34},~3499--3513.

\bibitem[Blaschke and R{\"o}pke(1988)]{Blaschke:1988zt}
Blaschke, D.; R{\"o}pke, G.
\newblock {Pauli Quenching for Hadrons in Nuclear Matter: A Quark Substructure
  Effect}.
\newblock{\em Dubna Preprint} {\bf 1988} JINR-E2-88-77 (unpublished).

\bibitem[Vautherin and Brink(1972)]{Vautherin:1971aw}
Vautherin, D.; Brink, D.M.
\newblock {Hartree-Fock calculations with Skyrme's interaction. 1. Spherical
  nuclei}.
\newblock {\em Phys.~Rev. C} {\bf 1972}, {\em 5},~626--647. 

\bibitem[Blaschke \em{et~al.}(1990)Blaschke, Tovmasian, and
  K{\"a}mpfer]{Blaschke:1989nn}
Blaschke, D.; Tovmasian, T.; K{\"a}mpfer, B.
\newblock {Predicting Stable Quark Cores in Neutron Stars From a Unified
  Description of Quark---Hadron Matter}.
\newblock {\em Sov. J. Nucl. Phys.} {\bf 1990}, {\em 52},~675--678.

\bibitem[Kl{\"a}hn and Blaschke(2017)]{Klaehn:2017mux}
  Kl{\"a}hn, T.; Blaschke, D.
  \newblock {Strange Matter in Compact Stars}.
  \newblock {\em EPJ Web Conf.} {\bf 2018}, {\em 171}, 08001

\bibitem[Bastian \em{et~al.}(2017)Bastian, Blaschke, Cierniak, Fischer,
  Kaltenborn, Marczenko, and Typel]{Bastian:2017fzo}
Bastian, N.U.F.; Blaschke, D.B.; Cierniak, M.; Fischer, T.; Kaltenborn, M.A.R.;
  Marczenko, M.; Typel, S.
\newblock {Strange matter prospects within the string-flip model}.
  \newblock {\em EPJ Web Conf.} {\bf 2018}, {\em 171}, 20002

\bibitem[Dubinin \em{et~al.}(2017)Dubinin, Blaschke, Friesen, and
  Turko]{Dubinin:2017iqc}
Dubinin, A.; Blaschke, D.; Friesen, A.; Turko, L.~{Pauli Blocking Effect Within the Relativistic Pion Gas}.
\newblock {\em Acta~Phys. Pol. Suppl.} {\bf 2017}, {\em 10},~903. 

\bibitem[Blaschke and R{\"o}pke(1993)]{Blaschke:1992qa}
Blaschke, D.; R{\"o}pke, G.
\newblock {Quark exchange contribution to the effective meson meson interaction
  potential}.
\newblock {\em Phys. Lett. B} {\bf 1993}, {\em 299},~332--337.

\bibitem[Barnes and Swanson(1992)]{Barnes:1991em}
Barnes, T.; Swanson, E.S.
\newblock {A Diagrammatic approach to meson meson scattering in the
  nonrelativistic quark potential model}.
\newblock {\em Phys. Rev. D} {\bf 1992}, {\em 46},~131--159.

\bibitem[Blaschke \em{et~al.}(2014)Blaschke, Buballa, Dubinin, R{\"o}pke, and
  Zablocki]{Blaschke:2013zaa}
Blaschke, D.; Buballa, M.; Dubinin, A.; R{\"o}pke, G.; Zablocki, D.
\newblock {Generalized Beth---Uhlenbeck approach to mesons and diquarks in hot,
  dense quark matter}.
\newblock {\em Ann. Phys.} {\bf 2014}, {\em 348},~228--255.

\bibitem[H{\"u}fner \em{et~al.}(1994)H{\"u}fner, Klevansky, Zhuang, and
  Voss]{Hufner:1994ma}
H{\"u}fner, J.; Klevansky, S.P.; Zhuang, P.; Voss, H.
\newblock {Thermodynamics of a quark plasma beyond the mean field: A
  generalized Beth-Uhlenbeck approach}.
\newblock {\em Ann. Phys.} {\bf 1994}, {\em 234},~225--244.

\bibitem[Zhuang \em{et~al.}(1994)Zhuang, H{\"u}fner, and
  Klevansky]{Zhuang:1994dw}
Zhuang, P.; H{\"u}fner, J.; Klevansky, S.P.
\newblock {Thermodynamics of a quark---Meson plasma in the Nambu-Jona-
Lasinio
  model}.
\newblock {\em Nucl. Phys. A} {\bf 1994}, {\em 576},~525--552.

\bibitem[Yamazaki and Matsui(2013)]{Yamazaki:2012ux}
Yamazaki, K.; Matsui, T.
\newblock {Quark-Hadron Phase Transition in the PNJL model for interacting
  quarks}.
\newblock {\em Nucl. Phys. A} {\bf 2013}, {\em 913},~19--50.

\bibitem[Wergieluk \em{et~al.}(2013)Wergieluk, Blaschke, Kalinovsky, and
  Friesen]{Wergieluk:2012gd}
Wergieluk, A.; Blaschke, D.; Kalinovsky, {Y}.L.; Friesen, A.
\newblock {Pion dissociation and Levinson`s theorem in hot PNJL quark matter}.
\newblock {\em Phys. Part. Nucl. Lett.} {\bf 2013}, {\em 10},~660--668.

\bibitem[Blaschke \em{et~al.}(2015)Blaschke, Dubinin, and
  Buballa]{Blaschke:2014zsa}
Blaschke, D.; Dubinin, A.; Buballa, M.
\newblock {Polyakov-loop suppression of colored states in a quark-meson-diquark
  plasma}.
\newblock {\em Phys. Rev. D} {\bf 2015}, {\em 91},~125040.

\bibitem[Kitazawa \em{et~al.}(2014)Kitazawa, Kunihiro, and
  Nemoto]{Kitazawa:2014sga}
Kitazawa, M.; Kunihiro, T.; Nemoto, Y.
\newblock {Emergence of soft quark excitations by the coupling with a soft mode
  of the QCD critical point}.
\newblock {\em Phys. Rev. D} {\bf 2014}, {\em 90},~116008.

\bibitem[Blaizot(1992)]{Blaizot:1991kh}
Blaizot, J.P.
\newblock {Quantum fields at finite temperature and density}.
\newblock {\em J. Korean Phys. Soc.} {\bf 1992}, {\em 25},~S65--S98.

\bibitem[{Kaltenborn} \em{et~al.}(2017){Kaltenborn}, {Bastian}, and
  {Blaschke}]{Kaltenborn:2017}
{Kaltenborn}, M.A.R.; {Bastian}, N.U.F.; {Blaschke}, D.B.
\newblock {Quark-Nuclear Hybrid Equation of State with Excluded Volume Effects}.
\newblock {\em \prd} {\bf 2017}, {\em 96},~056024.

\bibitem[Khvorostukin \em{et~al.}(2006)Khvorostukin, Skokov, Toneev, and
  Redlich]{Khvorostukin:2006aw}
Khvorostukin, A.S.; Skokov, V.V.; Toneev, V.D.; Redlich, K.
\newblock {Lattice QCD constraints on the nuclear equation of state}.
\newblock {\em Eur. Phys. J. C} {\bf 2006}, {\em 48},~531--543.

\bibitem[Kapusta(1989)]{Kapusta:1989tk}
Kapusta, J.I.
\newblock {{Finite Temperature Field Theory}}.  In {\em Cambridge
  Monographs on Mathematical Physics}; Cambridge University Press: Cambridge, UK,   1989.
\bibitem[Horowitz and Piekarewicz(1992)]{Horowitz:1991fn}
Horowitz, C.J.; Piekarewicz, J.
\newblock {Quark models of nuclear matter: 1. Basic models and ground state
  properties}.
\newblock {\em Nucl. Phys. A} {\bf 1992}, {\em 536},~669--696.

\bibitem[Horowitz and Piekarewicz(1991)]{Horowitz:1991ux}
Horowitz, C.J.; Piekarewicz, J.
\newblock {Nuclear to quark matter transition in the string flip model}.
\newblock {\em Phys. Rev. C} {\bf 1991}, {\em 44},~2753--2764.

\bibitem[Benic \em{et~al.}(2015)Benic, Blaschke, Alvarez-Castillo, Fischer, and
  Typel]{Benic:2014jia}
Benic, S.; Blaschke, D.; Alvarez-Castillo, D.E.; Fischer, T.; Typel, S.
\newblock {A new quark-hadron hybrid equation of state for astrophysics---I.
  High-mass twin compact stars}.
\newblock {\em Astron. Astrophys.} {\bf 2015}, {\em 577},~A40.

\bibitem[{Hebeler} and {Schwenk}(2010)]{Hebeler:2010a}
{Hebeler}, K.; {Schwenk}, A.
\newblock {Chiral three-nucleon forces and neutron matter}.
\newblock {\em Phys. Rev. C} {\bf 2010}, {\em 82},~014314.

\bibitem[{Hebeler} \em{et~al.}(2010){Hebeler}, {Lattimer}, {Pethick}, and
  {Schwenk}]{Hebeler:2010b}
{Hebeler}, K.; {Lattimer}, J.M.; {Pethick}, C.J.; {Schwenk}, A.
\newblock {Constraints on Neutron Star Radii Based on Chiral Effective Field
  Theory Interactions}.
\newblock {\em Phys. Rev. L} {\bf 2010}, {\em 105},~161102.

\bibitem[{Holt} \em{et~al.}(2012){Holt}, {Kaiser}, and {Weise}]{Holt:2012a}
{Holt}, J.W.; {Kaiser}, N.; {Weise}, W.
\newblock {Chiral nuclear dynamics with three-body forces}.
\newblock {\em Prog. Part. Nucl. Phys.} {\bf 2012}, {\em
  67},~353--358.

\bibitem[{Sammarruca} \em{et~al.}(2012){Sammarruca}, {Chen}, {Coraggio},
  {Itaco}, and {Machleidt}]{Sammarruca:2012}
{Sammarruca}, F.; {Chen}, B.; {Coraggio}, L.; {Itaco}, N.; {Machleidt}, R.
\newblock {Dirac-Brueckner-Hartree-Fock versus chiral effective field theory}.
\newblock {\em Phys. Rev. C} {\bf 2012}, {\em 86},~054317.

\bibitem[{Tews} \em{et~al.}(2013){Tews}, {Kr{\"u}ger}, {Hebeler}, and
  {Schwenk}]{Tews:2013}
{Tews}, I.; {Kr{\"u}ger}, T.; {Hebeler}, K.; {Schwenk}, A.
\newblock {Neutron Matter at Next-to-Next-to-Next-to-Leading Order in Chiral
  Effective Field Theory}.
\newblock {\em \prl} {\bf 2013}, {\em 110},~032504.

\bibitem[{Kr{\"u}ger} \em{et~al.}(2013){Kr{\"u}ger}, {Tews}, {Hebeler}, and
  {Schwenk}]{Krueger:2013}
{Kr{\"u}ger}, T.; {Tews}, I.; {Hebeler}, K.; {Schwenk}, A.
\newblock {Neutron matter from chiral effective field theory interactions}.
\newblock {\em Phys. Rev. C} {\bf 2013}, {\em 88},~025802.

\bibitem[{Coraggio} \em{et~al.}(2013){Coraggio}, {Holt}, {Itaco}, {Machleidt},
  and {Sammarruca}]{Coraggio:2013}
{Coraggio}, L.; {Holt}, J.W.; {Itaco}, N.; {Machleidt}, R.; {Sammarruca}, F.
\newblock {Reduced regulator dependence of neutron-matter predictions with
  perturbative chiral interactions}.
\newblock {\em Phys. Rev. C} {\bf 2013}, {\em 87},~014322.

\bibitem[{Demorest} \em{et~al.}(2010){Demorest}, {Pennucci}, {Ransom},
  {Roberts}, and {Hessels}]{Demorest:2010}
{Demorest}, P.B.; {Pennucci}, T.; {Ransom}, S.M.; {Roberts}, M.S.E.; {Hessels},
  J.W.T.
\newblock {A two-solar-mass neutron star measured using Shapiro delay}.
\newblock {\em Nature} {\bf 2010}, {\em 467},~1081--1083.

\bibitem[{Antoniadis} \em{et~al.}(2013){Antoniadis}, {Freire}, {Wex}, {Tauris},
  {Lynch}, {van Kerkwijk}, {Kramer}, {Bassa}, {Dhillon}, {Driebe}, {Hessels},
  {Kaspi}, {Kondratiev}, {Langer}, {Marsh}, {McLaughlin}, {Pennucci}, {Ransom},
  {Stairs}, {van Leeuwen}, {Verbiest}, and {Whelan}]{Antoniadis:2013}
{Antoniadis}, J.; {Freire}, P.C.C.; {Wex}, N.; {Tauris}, T.M.; {Lynch}, R.S.;
  {van Kerkwijk}, M.H.; {Kramer}, M.; {Bassa}, C.; {Dhillon}, V.S.; {Driebe},
  T.; et al.
\newblock {A Massive Pulsar in a Compact Relativistic Binary}.
\newblock {\em Science} {\bf 2013}, {\em 340},~448.

\bibitem[{Fonseca} \em{et~al.}(2016){Fonseca}, {Pennucci}, {Ellis}, {Stairs},
  {Nice}, {Ransom}, {Demorest}, {Arzoumanian}, {Crowter}, {Dolch}, {Ferdman},
  {Gonzalez}, {Jones}, {Jones}, {Lam}, {Levin}, {McLaughlin}, {Stovall},
  {Swiggum}, and {Zhu}]{Fonseca:2016}
{Fonseca}, E.; {Pennucci}, T.T.; {Ellis}, J.A.; {Stairs}, I.H.; {Nice}, D.J.;
  {Ransom}, S.M.; {Demorest}, P.B.; {Arzoumanian},~Z.; {Crowter}, K.; {Dolch},
  T.;  et al.
\newblock {The NANOGrav Nine-year Data Set: Mass and Geometric Measurements of
  Binary Millisecond Pulsars}.
\newblock {\em Astrophys. J.} {\bf 2016}, {\em 832},~167.

\bibitem[{Fischer} \em{et~al.}(2017){Fischer}, {Bastian}, {Blaschke},
  {Cerniak}, {Hempel}, {Kl{\"a}hn}, {Mart{\'{\i}}nez-Pinedo}, {Newton},
  {R{\"o}pke}, and {Typel}]{Fischer:2017}
\textls[-15]{{Fischer}, T.; {Bastian}, N.U.; {Blaschke}, D.; {Cerniak}, M.; {Hempel}, M.;
  {Kl{\"a}hn}, T.; {Mart{\'{\i}}nez-Pinedo}, G.; {Newton},~W.G.; {R{\"o}pke},
  G.; {Typel}, S.
 \newblock {The state of matter in simulations of core-collapse supernovae---Reflections and recent developments}.
 \newblock{\em Publ.\ Astron.\ Soc.\ Austral.} \textbf{2017}, {\em 34},~67 }

\bibitem[Typel \em{et~al.}(2010)Typel, R{\"o}pke, Kl{\"a}hn, Blaschke, and
  Wolter]{Typel:2009sy}
Typel, S.; R{\"o}pke, G.; Kl{\"a}hn, T.; Blaschke, D.; Wolter, H.H.
\newblock {Composition and thermodynamics of nuclear matter with light
  clusters}.
\newblock {\em Phys. Rev. C} {\bf 2010}, {\em81},~015803.

\bibitem[{Hempel} \em{et~al.}(2011){Hempel}, {Schaffner-Bielich}, {Typel}, and
  {R{\"o}pke}]{Hempel:2011}
{Hempel}, M.; {Schaffner-Bielich}, J.; {Typel}, S.; {R{\"o}pke}, G.
\newblock {Light clusters in nuclear matter: Excluded volume versus quantum
  many-body approaches}.
\newblock {\em \prc} {\bf 2011}, {\em 84},~055804.

\bibitem[Sumiyoshi \em{et~al.}(2006)Sumiyoshi, Yamada, Suzuki, and
  Chiba]{Sumiyoshi:2006id}
Sumiyoshi, K.; Yamada, S.; Suzuki, H.; Chiba, S.
\newblock {Neutrino signals from the formation of black hole: A probe of
  equation of state of dense matter}.
\newblock {\em \prl} {\bf 2006}, {\em 97},~091101.

\bibitem[{Fischer} \em{et~al.}(2009){Fischer}, {Whitehouse}, {Mezzacappa},
  {Thielemann}, and {Liebend{\"o}rfer}]{Fischer:2009}
{Fischer}, T.; {Whitehouse}, S.C.; {Mezzacappa}, A.; {Thielemann}, F.K.;
  {Liebend{\"o}rfer}, M.
\newblock {The neutrino signal from protoneutron star accretion and black hole
  formation}.
\newblock {\em Astron.  Astrophys. } {\bf 2009}, {\em 499},~1--15.

\bibitem[{O'Connor} and {Ott}(2011)]{OConnor:2011}
{O'Connor}, E.; {Ott}, C.D.
\newblock {Black Hole Formation in Failing Core-Collapse Supernovae}.
\newblock {\em \apj} {\bf 2011}, {\em 730},~70.

\bibitem[{Steiner} \em{et~al.}(2013){Steiner}, {Hempel}, and
  {Fischer}]{Steiner:2013}
{Steiner}, A.W.; {Hempel}, M.; {Fischer}, T.
\newblock {Core-collapse Supernova Equations of State Based on Neutron Star
  Observations}.
\newblock {\em \apj} {\bf 2013}, {\em 774},~17.

\bibitem[{Marek} \em{et~al.}(2009){Marek}, {Janka}, and
  {M{\"u}ller}]{Marek:2008qi}
{Marek}, A.; {Janka}, H.T.; {M{\"u}ller}, E.
\newblock {Equation-of-state dependent features in shock-oscillation modulated
  neutrino and gravitational-wave signals from supernovae}.
\newblock {\em \aap} {\bf 2009}, {\em 496},~475--494.

\bibitem[{Suwa} \em{et~al.}(2013){Suwa}, {Takiwaki}, {Kotake}, {Fischer},
  {Liebend{\"o}rfer}, and {Sato}]{Suwa:2013}
{Suwa}, Y.; {Takiwaki}, T.; {Kotake}, K.; {Fischer}, T.; {Liebend{\"o}rfer},
  M.; {Sato}, K.
\newblock {On the Importance of the Equation of State for the Neutrino-driven
  Supernova Explosion Mechanism}.
\newblock {\em \apj} {\bf 2013}, {\em 764},~99.

\bibitem[{Nagakura} \em{et~al.}(2017){Nagakura}, {Iwakami}, {Furusawa},
  {Okawa}, {Harada}, {Sumiyoshi}, {Yamada}, {Matsufuru}, and
  {Imakura}]{Nagakura:2017}
{Nagakura}, H.; {Iwakami}, W.; {Furusawa}, S.; {Okawa}, H.; {Harada}, A.;
  {Sumiyoshi}, K.; {Yamada}, S.; {Matsufuru}, H.; {Imakura}, A.
\newblock {Simulations of Core-Collapse Supernovae in Spatial Axisymmetry with Full
  Boltzmann Neutrino Transport}.
 \newblock {\em Astrophys.\ J.} {\bf 2018}, {\em 854}, no. 2, 136

\bibitem[Lattimer and Swesty(1991)]{Lattimer:1991nc}
Lattimer, J.M.; Swesty, F.
\newblock {A Generalized equation of state for hot, dense matter}.
\newblock {\em \npa} {\bf 1991}, {\em 535},~331--376.

\bibitem[Shen \em{et~al.}(1998)Shen, Toki, Oyamatsu, and
  Sumiyoshi]{Shen:1998gg}
Shen, H.; Toki, H.; Oyamatsu, K.; Sumiyoshi, K.
\newblock {Relativistic equation of state of nuclear matter for supernova and
  neutron star}.
\newblock {\em \npa} {\bf 1998}, {\em 637},~435--450.

\bibitem[{Fischer} \em{et~al.}(2014){Fischer}, {Hempel}, {Sagert}, {Suwa}, and
  {Schaffner-Bielich}]{Fischer:2014}
{Fischer}, T.; {Hempel}, M.; {Sagert}, I.; {Suwa}, Y.; {Schaffner-Bielich}, J.
\newblock {Symmetry energy impact in simulations of core-collapse supernovae}.
\newblock {\em \epja} {\bf 2014}, {\em 50},~46.

\bibitem[{Lattimer} and {Lim}(2013)]{Lattimer:2013}
{Lattimer}, J.M.; {Lim}, Y.
\newblock {Constraining the Symmetry Parameters of the Nuclear Interaction}.
\newblock {\em \apj} {\bf 2013}, {\em 771},~51.

\bibitem[{Tews} \em{et~al.}(2016){Tews}, {Lattimer}, {Ohnishi}, and
  {Kolomeitsev}]{Kolomeitsev:2016}
{Tews}, I.; {Lattimer}, J.M.; {Ohnishi}, A.; {Kolomeitsev}, E.E.
\newblock {Symmetry Parameter Constraints from A Lower Bound on the Neutron-Matter Energy}.
\newblock{\em Astrophys.\ J.} {\bf 2017} {\em 848}, no. 2, 105

\bibitem[Thielemann \em{et~al.}(2004)Thielemann, Brachwitz, Höflich,
  Martinez-Pinedo, and Nomoto]{Thielemann:2004}
Thielemann, F.K.; Brachwitz, F.; Höflich, P.; Martinez-Pinedo, G.; Nomoto, K.
\newblock The physics of type Ia supernovae.
\newblock {\em New Astron. Rev.} {\bf 2004}, {\em 48},~605--610.

\bibitem[Fischer \em{et~al.}(2010)Fischer, Whitehouse, Mezzacappa, Thielemann,
  and Liebend{\"o}rfer]{Fischer:2009af}
Fischer, T.; Whitehouse, S.; Mezzacappa, A.; Thielemann, F.K.;
  Liebend{\"o}rfer, M.
\newblock {Protoneutron star evolution and the neutrino driven wind in general
  relativistic neutrino radiation hydrodynamics simulations}.
\newblock {\em \aap} {\bf 2010}, {\em 517},~A80.

\bibitem[Hempel and Schaffner-Bielich(2010)]{Hempel:2009mc}
Hempel, M.; Schaffner-Bielich, J.
\newblock {Statistical Model for a Complete Supernova Equation of State}.
\newblock {\em Nucl. Phys. A} {\bf 2010}, {\em 837},~210--254.

\bibitem[{Juodagalvis} \em{et~al.}(2010){Juodagalvis}, {Langanke}, {Hix},
  {Mart{\'{\i}}nez-Pinedo}, and {Sampaio}]{Juodagalvis:2010}
{Juodagalvis}, A.; {Langanke}, K.; {Hix}, W.R.; {Mart{\'{\i}}nez-Pinedo}, G.;
  {Sampaio}, J.M.
\newblock {Improved estimate of electron capture rates on nuclei during stellar
  core collapse}.
\newblock {\em \npa} {\bf 2010}, {\em 848},~454--478.

\bibitem[Bruenn(1985)]{Bruenn:1985en}
Bruenn, S.W.
\newblock {Stellar core collapse: Numerical model and infall epoch}.
\newblock {\em \apjs} {\bf 1985}, {\em 58},~771--841.

\bibitem[Langanke \em{et~al.}(2008)Langanke, Martinez-Pinedo, M{\"u}ller,
  Janka, Marek, et~al.]{Langanke:2007ua}
Langanke, K.; Martinez-Pinedo, G.; M{\"u}ller, B.; Janka, H.T.; Marek, A.
\newblock {Effects of Inelastic Neutrino-Nucleus Scattering on Supernova
  Dynamics and Radiated Neutrino Spectra}.
\newblock {\em \prl} {\bf 2008}, {\em 100},~011101.

\bibitem[{Fuller} and {Meyer}(1991)]{Fuller:1991}
{Fuller}, G.M.; {Meyer}, B.S.
\newblock {High-temperature neutrino-nucleus processes in stellar collapse}.
\newblock {\em \apj} {\bf 1991}, {\em 376},~701--716.

\bibitem[{Fischer} \em{et~al.}(2013){Fischer}, {Langanke}, and
  {Mart{\'{\i}}nez-Pinedo}]{Fischer:2013}
{Fischer}, T.; {Langanke}, K.; {Mart{\'{\i}}nez-Pinedo}, G.
\newblock {Neutrino-pair emission from nuclear de-excitation in core-collapse
  supernova simulations}.
\newblock {\em \prc} {\bf 2013}, {\em 88},~065804.

\bibitem[{Schatz} \em{et~al.}(2014){Schatz}, {Gupta}, {M{\"o}ller}, {Beard},
  {Brown}, {Deibel}, {Gasques}, {Hix}, {Keek}, {Lau}, {Steiner}, and
  {Wiescher}]{Schatz:2014}
{Schatz}, H.; {Gupta}, S.; {M{\"o}ller}, P.; {Beard}, M.; {Brown}, E.F.;
  {Deibel}, A.T.; {Gasques}, L.R.; {Hix}, W.R.; {Keek}, L.; {Lau},~R.;
  et al.
\newblock {Strong neutrino cooling by cycles of electron capture and
  {$\beta$}$^{-}$ decay in neutron star crusts}.
\newblock {\em \nat} {\bf 2014}, {\em 505},~62--65.

\bibitem[{Haensel} and {Zdunik}(1990)]{Haensel:1990}
{Haensel}, P.; {Zdunik}, J.L.
\newblock {Non-equilibrium processes in the crust of an accreting neutron
  star}.
\newblock {\em \aap} {\bf 1990}, {\em 227},~431--436.

\bibitem[{Brown} \em{et~al.}(1998){Brown}, {Bildsten}, and
  {Rutledge}]{Brown:1998}
{Brown}, E.F.; {Bildsten}, L.; {Rutledge}, R.E.
\newblock {Crustal Heating and Quiescent Emission from Transiently Accreting
  Neutron Stars}.
\newblock {\em \apjl} {\bf 1998}, {\em 504},~L95--L98.

\bibitem[{Haensel} and {Zdunik}(2008)]{Haensel:2008}
{Haensel}, P.; {Zdunik}, J.L.
\newblock {Models of crustal heating in accreting neutron stars}.
\newblock {\em \aap} {\bf 2008}, {\em 480},~459--464.

\bibitem[{Fischer} \em{et~al.}(2016){Fischer}, {Mart{\'{\i}}nez-Pinedo},
  {Hempel}, {Huther}, {R{\"o}pke}, {Typel}, and {Lohs}]{Fischer:2016d}
{Fischer}, T.; {Mart{\'{\i}}nez-Pinedo}, G.; {Hempel}, M.; {Huther}, L.;
  {R{\"o}pke}, G.; {Typel}, S.; {Lohs}, A.
\newblock {Expected impact from weak reactions with light nuclei in
  corecollapse supernova simulations}. In  Proceedings of the 13th International Symposium on Origin of Matter and Evolution of Galaxies, Beijing, China, 24--27 June 2015.

\bibitem[{R{\"o}pke}(2009)]{Roepke:2009}
{R{\"o}pke}, G.
\newblock {Light nuclei quasiparticle energy shifts in hot and dense nuclear
  matter}.
\newblock {\em \prc} {\bf 2009}, {\em 79},~014002.

\bibitem[{R{\"o}pke}(2011)]{Roepke:2011}
{R{\"o}pke}, G.
\newblock {Parametrization of light nuclei quasiparticle energy shifts and
  composition of warm and dense nuclear matter}.
\newblock {\em \npa} {\bf 2011}, {\em 867},~66--80.

\bibitem[{Fischer}(2016)]{Fischer:2016a}
{Fischer}, T.~{Constraining the supersaturation density equation of state from
  core-collapse supernova simulations---Excluded volume extension of the
  baryons }.
\newblock {\em \epja} {\bf 2016}, {\em 52},~54.

\bibitem[{Steiner} \em{et~al.}(2010){Steiner}, {Lattimer}, and
  {Brown}]{Steiner:2010}
{Steiner}, A.W.; {Lattimer}, J.M.; {Brown}, E.F.
\newblock {The Equation of State from Observed Masses and Radii of Neutron
  Stars}.
\newblock {\em \apj} {\bf 2010}, {\em 722},~33--54.

\bibitem[Suleimanov \em{et~al.}(2011)Suleimanov, Poutanen, Revnivtsev, and
  Werner]{Suleimanov11}
Suleimanov, V.; Poutanen, J.; Revnivtsev, M.; Werner, K.
\newblock Neutron star stiff equation of state derived from cooling phases of
  the X-ray burster 4U 1724-307.
\newblock {\em \apj} {\bf 2011}, {\em 742},~122.

\bibitem[{Steiner} \em{et~al.}(2013){Steiner}, {Lattimer}, and
  {Brown}]{Steiner:2013b}
{Steiner}, A.W.; {Lattimer}, J.M.; {Brown}, E.F.
\newblock {The Neutron Star Mass-Radius Relation and the Equation of State of
  Dense Matter}.
\newblock {\em \apjl} {\bf 2013}, {\em 765},~L5.

\bibitem[Fodor and Katz(2004)]{Fodor:2004nz}
Fodor, Z.; Katz, S.
\newblock {Critical point of QCD at finite T and mu, lattice results for
  physical quark masses}.
\newblock {\em J. High Energy Phys.} {\bf 2004}, {\em 0404},~050.

\bibitem[Ratti \em{et~al.}(2006)Ratti, Thaler, and Weise]{Ratti:2005jh}
Ratti, C.; Thaler, M.A.; Weise, W.
\newblock {Phases of QCD: Lattice thermodynamics and a field theoretical
  model}.
\newblock {\em \prd} {\bf 2006}, {\em 73},~014019.

\bibitem[{Bors{\'a}nyi} \em{et~al.}(2012){Bors{\'a}nyi}, {Fodor}, {Katz},
  {Krieg}, {Ratti}, and {Szab{\'o}}]{Katz:2012JHEP}
{Bors{\'a}nyi}, S.; {Fodor}, Z.; {Katz}, S.D.; {Krieg}, S.; {Ratti}, C.;
  {Szab{\'o}}, K.
\newblock {Fluctuations of conserved charges at finite temperature from lattice
  QCD}.
\newblock {\em J. High Energy Phys.} {\bf 2012}, {\em 1},~138.

\bibitem[{Bazavov} \em{et~al.}(2012){Bazavov}, {Ding}, {Hegde}, {Kaczmarek},
  {Karsch}, {Laermann}, {Mukherjee}, {Petreczky}, {Schmidt}, {Smith},
  {Soeldner}, and {Wagner}]{Laermann:2012PRL}
{Bazavov}, A.; {Ding}, H.T.; {Hegde}, P.; {Kaczmarek}, O.; {Karsch}, F.;
  {Laermann}, E.; {Mukherjee}, S.; {Petreczky}, P.; {Schmidt}, C.; {Smith}, D.;
  et al.
\newblock {Freeze-Out Conditions in Heavy Ion Collisions from QCD
  Thermodynamics}.
\newblock {\em \prl} {\bf 2012}, {\em 109},~192302.

\bibitem[{Bors{\'a}nyi} \em{et~al.}(2014){Bors{\'a}nyi}, {Fodor}, {Hoelbling},
  {Katz}, {Krieg}, and {Szab{\'o}}]{Katz:2014PhLB}
{Bors{\'a}nyi}, S.; {Fodor}, Z.; {Hoelbling}, C.; {Katz}, S.D.; {Krieg}, S.;
  {Szab{\'o}}, K.K.
\newblock {Full result for the QCD equation of state with 2 + 1 flavors}.
\newblock {\em Phys. Lett. B} {\bf 2014}, {\em 730},~99--104.

\bibitem[Kurkela \em{et~al.}(2014)Kurkela, Fraga, Schaffner-Bielich, and
  Vuorinen]{Kurkela:2014vha}
Kurkela, A.; Fraga, E.S.; Schaffner-Bielich, J.; Vuorinen, A.
\newblock {Constraining neutron star matter with Quantum Chromodynamics}.
\newblock {\em \apj} {\bf 2014}, {\em 789},~27.

\bibitem[Farhi and Jaffe(1984)]{Farhi:1984qu}
Farhi, E.; Jaffe, R.
\newblock {Strange Matter}.
\newblock {\em \prd} {\bf 1984}, {\em 30},~2379.

\bibitem[Nambu and Jona-Lasinio(1961)]{Nambu:1961tp}
Nambu, Y.; Jona-Lasinio, G.
\newblock {Dynamical Model of Elementary Particles Based on an Analogy with
  Superconductivity. 1.}
\newblock {\em \pr} {\bf 1961}, {\em 122},~345--358.

\bibitem[Klevansky(1992)]{Klevansky:1992qe}
Klevansky, S.
\newblock {The Nambu-Jona-Lasinio model of quantum chromodynamics}.
\newblock {\em Rev. Mod. Phys.} {\bf 1992}, {\em 64},~649--708.


\bibitem[Buballa(2005)]{Buballa:2003qv}
Buballa, M.
\newblock {NJL model analysis of quark matter at large density}.
\newblock {\em Phys.Rept.} {\bf 2005}, {\em 407},~205--376.

\bibitem[{Kl{\"a}hn} and {Fischer}(2015)]{Klaehn:2015}
{Kl{\"a}hn}, T.; {Fischer}, T.
\newblock {Vector Interaction Enhanced Bag Model for Astrophysical
  Applications}.
\newblock {\em \apj} {\bf 2015}, {\em 810},~134.

\bibitem[{Kl{\"a}hn} \em{et~al.}(2017){Kl{\"a}hn}, {Fischer}, and
  {Hempel}]{Klaehn:2017a}
{Kl{\"a}hn}, T.; {Fischer}, T.; {Hempel}, M.
\newblock {Simultaneous chiral symmetry restoration and deconfinement -
  Consequences for the QCD phase diagram}.
\newblock {\em \apj} {\bf 2017}, {\em 836},~89.

\bibitem[Fischer \em{et~al.}(2017)Fischer, Bastian, Wu, Typel, Klähn, and
  Blaschke]{Fischer:2017lag}
Fischer, T.; Bastian, N.U.F.; Wu, M.R.; Typel, S.; Klähn, T.; Blaschke, D.B.
\newblock {High-density phase transition paves the way for supernova explosions
  of massive blue-supergiant stars}. \emph{arXiv} {\bf 2017}, arXiv:1712.08788.

\bibitem[{Haensel} \em{et~al.}(2007){Haensel}, {Potekhin}, and
  {Yakovlev}]{Haensel:2007yy}
{Haensel}, P.; {Potekhin}, A.Y.; {Yakovlev}, D.G.
\newblock {{Neutron Stars 1: Equation of State and Structure}}.
  In~{\em Astrophysics and Space Science Library}; Springer: Berlin, Germany,  2007.

\bibitem[{Read} \em{et~al.}(2009){Read}, {Lackey}, {Owen}, and
  {Friedman}]{Read:2008iy}
{Read}, J.S.; {Lackey}, B.D.; {Owen}, B.J.; {Friedman}, J.L.
\newblock {Constraints on a phenomenologically parametrized neutron-star
  equation of state}.
\newblock {\em \prd} {\bf 2009}, {\em 79},~124032.

\bibitem[{Zdunik} and {Haensel}(2013)]{Zdunik:2012dj}
{Zdunik}, J.L.; {Haensel}, P.
\newblock {Maximum mass of neutron stars and strange neutron-star cores}.
\newblock {\em \aap} {\bf 2013}, {\em 551},~A61.

\bibitem[{Alford} \em{et~al.}(2013){Alford}, {Han}, and
  {Prakash}]{Alford:2013aca}
{Alford}, M.G.; {Han}, S.; {Prakash}, M.
\newblock {Generic conditions for stable hybrid stars}.
\newblock {\em \prd} {\bf 2013}, {\em 88},~083013.

\bibitem[Abbott \em{et~al.}(2017)Abbott et~al.]{TheLIGOScientific:2017qsa}
Abbott, B.;  Abbott, R.; Abbott, T.D.; Acernese, F.; Ackley, K.; Adams, C.; Adams, T.; Addesso, P.; Adhikari,~R.X.; Adya, V.B.; et al.
\newblock {GW170817: Observation of Gravitational Waves from a Binary Neutron
  Star Inspiral}.
\newblock {\em Phys. Rev. Lett.} {\bf 2017}, {\em 119},~161101.

\bibitem[Ayriyan \em{et~al.}(2018)Ayriyan, Bastian, Blaschke, Grigorian,
  Maslov, and Voskresensky]{Ayriyan:2017nby}
Ayriyan, A.; Bastian, N.U.; Blaschke, D.; Grigorian, H.; Maslov, K.;
  Voskresensky, D.N.
\newblock {How robust is a third family of compact stars against pasta phase
  effects?}
\newblock {\em Phys. Rev. C} {\bf 2018}, {\em 97},~045802.

\bibitem[Paschalidis \em{et~al.}(2018)Paschalidis, Yagi, Alvarez-Castillo,
  Blaschke, and Sedrakian]{Paschalidis:2017qmb}
Paschalidis, V.; Yagi, K.; Alvarez-Castillo, D.; Blaschke, D.B.; Sedrakian, A.
\newblock Implications from GW170817 and I-Love-Q relations for relativistic
  hybrid stars.
\newblock {\em Phys. Rev. D} {\bf 2018}, {\em 97},~084038.

\bibitem[Blaschke and Chamel(2018)]{Blaschke:2018mqw}
Blaschke, D.; Chamel, N.
\newblock {Phases of dense matter in compact stars}. \emph{arXiv} \textbf{2018}, arXiv:1803.01836.

\bibitem[Annala \em{et~al.}(2017)Annala, Gorda, Kurkela, and
  Vuorinen]{Annala:2017llu}
Annala, E.; Gorda, T.; Kurkela, A.; Vuorinen, A.
\newblock {Gravitational-wave constraints on the neutron-star-matter Equation
  of State}. \emph{Phys. Rev. Lett.} {\bf 2017}, \emph{120}, 17270.

\bibitem[R{\"o}pke(2017{\natexlab{a}})]{Ropke:2017zts}
R{\"o}pke, G.
\newblock {Nuclear matter EoS including few-nucleon correlations}.
\newblock {\em Nuovo Cim. C} {\bf 2017}, {\em 39},~392.

\bibitem[Kuhrts \em{et~al.}(2001)Kuhrts, Beyer, Danielewicz, and
  R{\"o}pke]{Kuhrts:2000zs}
Kuhrts, C.; Beyer, M.; Danielewicz, P.; R{\"o}pke, G.
\newblock {Medium corrections in the formation of light charged particles in
  heavy ion reactions}.
\newblock {\em Phys. Rev. C} {\bf 2001}, {\em 63},~034605.

\bibitem[Bastian \em{et~al.}(2016)Bastian, Batyuk, Blaschke, Danielewicz,
  Ivanov, Karpenko, R{\"o}pke, Rogachevsky, and Wolter]{Bastian:2016xna}
Bastian, N.U.; Batyuk, P.; Blaschke, D.; Danielewicz, P.; Ivanov, {Y}.B.; Karpenko, I.; R{\"o}pke, G.; Rogachevsky,~O.; Wolter, H.H.
\newblock {Light cluster production at NICA}.
\newblock {\em Eur. Phys. J. A} {\bf 2016}, {\em 52},~244.

\bibitem[Bastian \em{et~al.}(2017)Bastian, Blaschke, and
  R{\"o}pke]{Bastian:2017rhb}
Bastian, N.U.; Blaschke, D.; R{\"o}pke, G.
\newblock {Light cluster production at NICA}.
\newblock {\em Acta Phys. Pol. Suppl.} {\bf 2017}, {\em 10},~899.

\bibitem[Kowalski \em{et~al.}(2007)Kowalski et~al.]{Kowalski:2006ju}
Kowalski, S.; Natowitz, J.B.; Shlomo, S.; Wada, R.; Hagel, K.; Wang, J.; Materna, T.; Chen, Z.; Ma, Y.G.; Qin,~L.; et al.
\newblock {Experimental determination of the symmetry energy of a low density
  nuclear gas}.
\newblock {\em Phys. Rev. C} {\bf 2007}, {\em 75},~014601.

\bibitem[Natowitz \em{et~al.}(2010)Natowitz et~al.]{Natowitz:2010ti}
Kowalski, S.; Natowitz, J.B.; Shlomo, S.; Wada, R.; Hagel, K.; Wang, J.; Materna, T.; Chen, Z.; Ma, Y.G.; Qin,~L.; et al.
\newblock {Symmetry energy of dilute warm nuclear matter}.
\newblock {\em Phys. Rev. Lett.} {\bf 2010}, {\em 104},~202501.

\bibitem[Wada \em{et~al.}(2012)Wada et~al.]{Wada:2011qm}
Wada, R.
\newblock {The Nuclear Matter Symmetry Energy at $0.03\leq \rho/\rho_0\leq
  0.2$}.
\newblock {\em Phys. Rev. C} {\bf 2012}, {\em 85},~064618.

\bibitem[Hagel \em{et~al.}(2014)Hagel, Natowitz, and R{\"o}pke]{Hagel:2014wja}
Hagel, K.; Natowitz, J.B.; R{\"o}pke, G.
\newblock {The equation of state and symmetry energy of low density nuclear
  matter}.
\newblock {\em Eur. Phys. J. A} {\bf 2014}, {\em 50},~39.

\bibitem[Typel \em{et~al.}(2014)Typel, Wolter, R{\"o}pke, and
  Blaschke]{Typel:2013zna}
Typel, S.; Wolter, H.H.; R{\"o}pke, G.; Blaschke, D.
\newblock {Effects of the liquid-gas phase transition and cluster formation on
  the symmetry energy}.
\newblock {\em Eur. Phys. J. A} {\bf 2014}, {\em 50},~17.

\bibitem[Li \em{et~al.}(2008)Li, Chen, and Ko]{Li:2008gp}
Li, B.A.; Chen, L.W.; Ko, C.M.
\newblock {Recent Progress and New Challenges in Isospin Physics with Heavy-Ion
  Reactions}.
\newblock {\em Phys. Rept.} {\bf 2008}, {\em 464},~113--281.

\bibitem[Qin \em{et~al.}(2012)Qin et~al.]{Qin:2011qp}
Qin, L.
\newblock {Laboratory Tests of Low Density Astrophysical Equations of State}.
\newblock {\em Phys. Rev. Lett.} {\bf 2012}, {\em 108},~172701.

\bibitem[Hagel \em{et~al.}(2012)Hagel et~al.]{Hagel:2011ws}
Hagel, K.; Wada, R.; Qin, L.; Natowitz, J.B.; Shlomo, S.; Bonasera, A.; Röpke, G.; Typel, S.; Chen, Z.; Huang,~M.; et al.
\newblock {Experimental Determination of In-Medium Cluster Binding Energies and
  Mott Points in Nuclear Matter}.
\newblock {\em Phys. Rev. Lett.} {\bf 2012}, {\em 108},~062702.

\bibitem[Hempel \em{et~al.}(2015)Hempel, Hagel, Natowitz, R{\"o}pke, and
  Typel]{Hempel:2015yma}
Hempel, M.; Hagel, K.; Natowitz, J.; R{\"o}pke, G.; Typel, S.
\newblock {Constraining supernova equations of state with equilibrium constants
  from heavy-ion collisions}.
\newblock {\em Phys. Rev. C} {\bf 2015}, {\em 91},~045805.

\bibitem[{Batyuk} \em{et~al.}(2016){Batyuk}, {Blaschke}, {Bleicher}, {Ivanov},
  {Karpenko}, {Merts}, {Nahrgang}, {Petersen}, and
  {Rogachevsky}]{Batyuk:2016qmb}
{Batyuk}, P.; {Blaschke}, D.; {Bleicher}, M.; {Ivanov}, Y.B.; {Karpenko}, I.;
  {Merts}, S.; {Nahrgang}, M.; {Petersen}, H.; {Rogachevsky}, O.
\newblock {Event simulation based on three-fluid hydrodynamics for collisions
  at energies available at the Dubna Nuclotron-based Ion Collider Facility and
  at the Facility for Antiproton and Ion Research in Darmstadt}.
\newblock {\em \prc} {\bf 2016}, {\em 94},~044917.

\bibitem[Batyuk \em{et~al.}(2017)Batyuk, Blaschke, Bleicher, Ivanov, Karpenko,
  Malinina, Merts, Nahrgang, Petersen, and Rogachevsky]{Batyuk:2017sku}
\textls[-15]{Batyuk, P.; Blaschke, D.; Bleicher, M.; Ivanov, Y.B.; Karpenko, I.;
  Malinina, L.; Merts, S.; Nahrgang,~M.; Petersen,~H.; Rogachevsky, O.
\newblock {Three-fluid Hydrodynamics-based Event Simulator Extended by UrQMD
  final State interactions (THESEUS) for FAIR-NICA-SPS-BES/RHIC energies}.~{In Proceedings of the 6th International Conference on New Frontiers in Physics (ICNFP
  2017),  Kolymbari,  Greece,  17--26 August  2017}.}

\bibitem[Ivanov(2013{\natexlab{a}})]{Ivanov:2013wha}
Ivanov, {Y}.B.
\newblock {Alternative Scenarios of Relativistic Heavy-Ion Collisions: I.
  Baryon Stopping}.
\newblock {\em Phys. Rev. C} {\bf 2013}, {\em 87},~064904.

\bibitem[Ivanov(2013{\natexlab{b}})]{Ivanov:2013yqa}
Ivanov, {Y}.B.
\newblock {Alternative Scenarios of Relativistic Heavy-Ion Collisions: II.
  Particle Production}.
\newblock {\em Phys. Rev. C} {\bf 2013}, {\em 87},~064905.

\bibitem[Ivanov(2014)]{Ivanov:2013yla}
Ivanov, {Y}.B.
\newblock {Alternative Scenarios of Relativistic Heavy-Ion Collisions: III.
  Transverse Momentum Spectra}.
\newblock {\em Phys. Rev. C} {\bf 2014}, {\em 89},~024903.

\bibitem[Typel and Blaschke(2018)]{Typel:2017vif}
Typel, S.; Blaschke, D.
\newblock {A Phenomenological Equation of State of Strongly Interacting Matter
  with First-Order Phase Transitions and Critical Points}.
\newblock {\em Universe} {\bf 2018}, {\em 4},~32.

\bibitem[Bastian and Blaschke(2016)]{Bastian:2015avq}
Bastian, N.U.; Blaschke, D.
\newblock {Towards a new quark-nuclear matter EoS for applications in
  astrophysics and heavy-ion collisions}.
\newblock {\em J. Phys.} {\bf 2016}, {\em 668},~012042.

\bibitem[Andronic \em{et~al.}(2017)Andronic, Braun-Munzinger, Redlich, and
  Stachel]{Andronic:2017pug}
Andronic, A.; Braun-Munzinger, P.; Redlich, K.; Stachel, J.
\newblock {Decoding the phase structure of QCD via particle production at high
  energy}. \emph{arXiv} {\bf 2017}, arXiv:1710.09425.

\bibitem[Blaschke \em{et~al.}(2017)Blaschke, Jankowski, and
  Naskret]{Blaschke:2017lvd}
Blaschke, D.; Jankowski, J.; Naskret, M.
\newblock {Formation of hadrons at chemical freeze-out}.  \emph{arXiv} {\bf 2017}, arXiv:1705.00169.

\end{thebibliography}
\reftitle{References}

\end{document}